\documentclass[conference]{IEEEtran}
\pdftrailerid{}

\usepackage[T1]{fontenc}
\usepackage[utf8]{inputenc}
\usepackage{amsmath,amssymb,amsthm}
\usepackage{booktabs}
\usepackage{array}      
\usepackage{enumitem}   
\usepackage{graphicx}
\graphicspath{{figures/}}
\usepackage{xcolor}
\usepackage{tikz}       
\usetikzlibrary{positioning,arrows.meta,fit}
\usepackage{pgfplots}
\pgfplotsset{compat=1.18}
\usepgfplotslibrary{groupplots}
\usepackage{multirow}
\usepackage{placeins}   
\usepackage{threeparttable}
\usepackage{cite}       
\usepackage{url}
\usepackage{microtype}
\usepackage{balance}
\usepackage[hidelinks,
            pdftitle={Ranking Is Not Deciding},
            pdfauthor={Pushkal Kumar; Arvin Sahaym},
            pdfkeywords={retrieval-augmented generation; data poisoning;
                         benchmark validity; cost-sensitive classification;
                         adaptive attacks},
            pdfsubject={Preprint}]{hyperref}
\usepackage[capitalise,noabbrev]{cleveref} 
\crefname{figure}{Fig.}{Figs.}
\Crefname{figure}{Fig.}{Figs.}

\theoremstyle{plain}

\newtheorem{proposition}{Proposition}
\newtheorem{lemma}{Lemma}

\theoremstyle{definition}

\theoremstyle{remark}

\crefname{proposition}{Proposition}{Propositions}
\crefname{assumption}{Assumption}{Assumptions}
\crefname{definition}{Definition}{Definitions}
\crefname{theorem}{Theorem}{Theorems}

\newif\ifdraft \drafttrue
\ifdraft
  \newcommand{\res}[1]{\textcolor{red}{[\ifmmode\text{#1}\else#1\fi]}}
  \newcommand{\todo}[1]{\textcolor{orange}{[TODO: \ifmmode\text{#1}\else#1\fi]}}
\else
  \newcommand{\res}[1]{\GenericError{}{Unresolved result placeholder: #1}{}{}}
  \newcommand{\todo}[1]{\GenericError{}{Unresolved TODO: #1}{}{}}
\fi
\newcommand{\inputgen}[1]{\IfFileExists{#1.tex}{\input{#1}}{}}
\makeatletter
\providecommand{\resdef}[2]{%
  \expandafter\gdef\csname res@val@\detokenize{#1}\endcsname{#2}}
\@ifundefined{res@fallback}{\let\res@fallback\res}{}
\renewcommand{\res}[1]{%
  \ifcsname res@val@\detokenize{#1}\endcsname
    \ifmmode
      \text{\csname res@val@\detokenize{#1}\endcsname}%
    \else
      \csname res@val@\detokenize{#1}\endcsname
    \fi
  \else
    \res@fallback{#1}%
  \fi}
\makeatother
\resdef{E0-candidate-jaccard}{0.909--1.000}
\resdef{E0-corpus-table}{\ref{tab:e0-corpus}}
\resdef{E0-pi0}{0.343}
\resdef{E0-pi0-deployed}{0.343}
\resdef{E0-pi1-range}{0.090--0.746}
\resdef{E0-poison-rates}{0.0147--0.0233}
\resdef{E0-train-minutes}{22}
\resdef{E1-AUROC-containonly-lm-financial}{0.995}
\resdef{E1-AUROC-containonly-lm-legal}{0.994}
\resdef{E1-AUROC-containonly-lm-medical}{1.000}
\resdef{E1-AUROC-containonly-lmns-legal}{0.997}
\resdef{E1-AUROC-containonly-lmns-medical}{1.000}
\resdef{E1-AUROC-containonly-lmsh-legal}{0.997}
\resdef{E1-AUROC-containonly-lmsh-medical}{1.000}
\resdef{E1-AUROC-containonly-nat-financial}{0.814}
\resdef{E1-AUROC-containonly-nat-legal}{0.933}
\resdef{E1-AUROC-containonly-nat-medical}{0.653}
\resdef{E1-AUROC-containonly-nat-nq}{0.718}
\resdef{E1-AUROC-containonly-natlm-financial}{0.830}
\resdef{E1-AUROC-containonly-natlm-legal}{0.977}
\resdef{E1-AUROC-containonly-natlm-medical}{0.877}
\resdef{E1-AUROC-containonly-nq-published}{1.000}
\resdef{E1-AUROC-containonly-released}{0.718}
\resdef{E1-AUROC-containonly-tpl-financial}{0.995}
\resdef{E1-AUROC-containonly-tpl-legal}{0.979}
\resdef{E1-AUROC-containonly-tpl-medical}{1.000}
\resdef{E1-AUROC-containonly-tpl-nq}{1.000}
\resdef{E1-AUROC-drs-lm-financial}{0.671}
\resdef{E1-AUROC-drs-lm-legal}{0.688}
\resdef{E1-AUROC-drs-lm-medical}{0.536}
\resdef{E1-AUROC-drs-lmns-legal}{0.836}
\resdef{E1-AUROC-drs-lmns-medical}{0.907}
\resdef{E1-AUROC-drs-lmsh-legal}{0.649}
\resdef{E1-AUROC-drs-lmsh-medical}{0.526}
\resdef{E1-AUROC-drs-medical}{0.882 \ensuremath{\pm} 0.001}
\resdef{E1-AUROC-drs-nat-financial}{0.709}
\resdef{E1-AUROC-drs-nat-legal}{0.776}
\resdef{E1-AUROC-drs-nat-medical}{0.935}
\resdef{E1-AUROC-drs-nat-nq}{0.450}
\resdef{E1-AUROC-drs-natlm-financial}{0.701}
\resdef{E1-AUROC-drs-natlm-legal}{0.694}
\resdef{E1-AUROC-drs-natlm-medical}{0.560}
\resdef{E1-AUROC-drs-nq-published}{0.539}
\resdef{E1-AUROC-drs-released}{0.450}
\resdef{E1-AUROC-drs-tpl-financial}{0.672}
\resdef{E1-AUROC-drs-tpl-legal}{0.690}
\resdef{E1-AUROC-drs-tpl-medical}{0.882}
\resdef{E1-AUROC-drs-tpl-nq}{0.539}
\resdef{E1-AUROC-dup-lm-financial}{N/A}
\resdef{E1-AUROC-dup-lm-legal}{N/A}
\resdef{E1-AUROC-dup-lm-medical}{N/A}
\resdef{E1-AUROC-dup-lmns-legal}{N/A}
\resdef{E1-AUROC-dup-lmns-medical}{N/A}
\resdef{E1-AUROC-dup-lmsh-legal}{N/A}
\resdef{E1-AUROC-dup-lmsh-medical}{N/A}
\resdef{E1-AUROC-dup-nat-financial}{N/A}
\resdef{E1-AUROC-dup-nat-legal}{N/A}
\resdef{E1-AUROC-dup-nat-medical}{N/A}
\resdef{E1-AUROC-dup-nat-nq}{N/A}
\resdef{E1-AUROC-dup-natlm-financial}{N/A}
\resdef{E1-AUROC-dup-natlm-legal}{N/A}
\resdef{E1-AUROC-dup-natlm-medical}{N/A}
\resdef{E1-AUROC-dup-nq-published}{N/A}
\resdef{E1-AUROC-dup-released}{N/A}
\resdef{E1-AUROC-dup-tpl-financial}{N/A}
\resdef{E1-AUROC-dup-tpl-legal}{N/A}
\resdef{E1-AUROC-dup-tpl-medical}{N/A}
\resdef{E1-AUROC-dup-tpl-nq}{N/A}
\resdef{E1-AUROC-lenonly-lm-financial}{0.090}
\resdef{E1-AUROC-lenonly-lm-legal}{0.024}
\resdef{E1-AUROC-lenonly-lm-medical}{0.486}
\resdef{E1-AUROC-lenonly-lmns-legal}{0.030}
\resdef{E1-AUROC-lenonly-lmns-medical}{0.525}
\resdef{E1-AUROC-lenonly-lmsh-legal}{0.025}
\resdef{E1-AUROC-lenonly-lmsh-medical}{0.454}
\resdef{E1-AUROC-lenonly-nat-financial}{0.362}
\resdef{E1-AUROC-lenonly-nat-legal}{0.927}
\resdef{E1-AUROC-lenonly-nat-medical}{1.000}
\resdef{E1-AUROC-lenonly-nat-nq}{0.864}
\resdef{E1-AUROC-lenonly-natlm-financial}{0.269}
\resdef{E1-AUROC-lenonly-natlm-legal}{0.062}
\resdef{E1-AUROC-lenonly-natlm-medical}{0.524}
\resdef{E1-AUROC-lenonly-nq-published}{0.806}
\resdef{E1-AUROC-lenonly-released}{0.864}
\resdef{E1-AUROC-lenonly-tpl-financial}{0.098}
\resdef{E1-AUROC-lenonly-tpl-legal}{0.817}
\resdef{E1-AUROC-lenonly-tpl-medical}{1.000}
\resdef{E1-AUROC-lenonly-tpl-nq}{0.806}
\resdef{E1-AUROC-medical}{$\ge$ 0.993}
\resdef{E1-AUROC-nli-lm-legal}{—}
\resdef{E1-AUROC-nli-lm-medical}{—}
\resdef{E1-AUROC-nli-nat-financial}{—}
\resdef{E1-AUROC-nli-nat-legal}{—}
\resdef{E1-AUROC-nli-nat-medical}{—}
\resdef{E1-AUROC-nli-nat-nq}{—}
\resdef{E1-AUROC-nli-nq-published}{—}
\resdef{E1-AUROC-nli-released}{—}
\resdef{E1-AUROC-nli-tpl-financial}{—}
\resdef{E1-AUROC-nli-tpl-legal}{—}
\resdef{E1-AUROC-nli-tpl-medical}{—}
\resdef{E1-AUROC-nli-tpl-nq}{—}
\resdef{E1-AUROC-nq-natural-max}{0.758}
\resdef{E1-AUROC-nq-published-max}{0.953}
\resdef{E1-AUROC-ppl-lm}{0.981 \ensuremath{\pm} 0.009}
\resdef{E1-AUROC-ppl-lm-financial}{0.660}
\resdef{E1-AUROC-ppl-lm-legal}{0.878}
\resdef{E1-AUROC-ppl-lm-medical}{0.981}
\resdef{E1-AUROC-ppl-lmns}{1.000}
\resdef{E1-AUROC-ppl-lmns-legal}{0.942}
\resdef{E1-AUROC-ppl-lmns-medical}{1.000}
\resdef{E1-AUROC-ppl-lmsh}{1.000}
\resdef{E1-AUROC-ppl-lmsh-legal}{0.948}
\resdef{E1-AUROC-ppl-lmsh-medical}{1.000}
\resdef{E1-AUROC-ppl-nat}{1.000}
\resdef{E1-AUROC-ppl-nat-financial}{0.766}
\resdef{E1-AUROC-ppl-nat-legal}{0.945}
\resdef{E1-AUROC-ppl-nat-medical}{1.000}
\resdef{E1-AUROC-ppl-nat-nq}{0.729}
\resdef{E1-AUROC-ppl-natlm}{0.966}
\resdef{E1-AUROC-ppl-natlm-financial}{0.741}
\resdef{E1-AUROC-ppl-natlm-legal}{0.922}
\resdef{E1-AUROC-ppl-natlm-medical}{0.966}
\resdef{E1-AUROC-ppl-nq-published}{0.773 \ensuremath{\pm} 0.000}
\resdef{E1-AUROC-ppl-released}{0.729 \ensuremath{\pm} 0.000}
\resdef{E1-AUROC-ppl-tpl}{0.993 \ensuremath{\pm} 0.000}
\resdef{E1-AUROC-ppl-tpl-financial}{0.663}
\resdef{E1-AUROC-ppl-tpl-legal}{0.902}
\resdef{E1-AUROC-ppl-tpl-medical}{0.993}
\resdef{E1-AUROC-ppl-tpl-nq}{0.773}
\resdef{E1-AUROC-ragdefender-lm-financial}{N/A}
\resdef{E1-AUROC-ragdefender-lm-legal}{N/A}
\resdef{E1-AUROC-ragdefender-lm-medical}{N/A}
\resdef{E1-AUROC-ragdefender-lmns-legal}{N/A}
\resdef{E1-AUROC-ragdefender-lmns-medical}{N/A}
\resdef{E1-AUROC-ragdefender-lmsh-legal}{N/A}
\resdef{E1-AUROC-ragdefender-lmsh-medical}{N/A}
\resdef{E1-AUROC-ragdefender-nat-financial}{N/A}
\resdef{E1-AUROC-ragdefender-nat-legal}{N/A}
\resdef{E1-AUROC-ragdefender-nat-medical}{N/A}
\resdef{E1-AUROC-ragdefender-nat-nq}{N/A}
\resdef{E1-AUROC-ragdefender-natlm-financial}{N/A}
\resdef{E1-AUROC-ragdefender-natlm-legal}{N/A}
\resdef{E1-AUROC-ragdefender-natlm-medical}{N/A}
\resdef{E1-AUROC-ragdefender-nq-published}{N/A}
\resdef{E1-AUROC-ragdefender-released}{N/A}
\resdef{E1-AUROC-ragdefender-tpl-financial}{N/A}
\resdef{E1-AUROC-ragdefender-tpl-legal}{N/A}
\resdef{E1-AUROC-ragdefender-tpl-medical}{N/A}
\resdef{E1-AUROC-ragdefender-tpl-nq}{N/A}
\resdef{E1-AUROC-raguard-lm}{0.838}
\resdef{E1-AUROC-raguard-lm-financial}{0.930}
\resdef{E1-AUROC-raguard-lm-legal}{0.883}
\resdef{E1-AUROC-raguard-lm-medical}{0.838}
\resdef{E1-AUROC-raguard-lmns}{1.000}
\resdef{E1-AUROC-raguard-lmns-legal}{0.956}
\resdef{E1-AUROC-raguard-lmns-medical}{1.000}
\resdef{E1-AUROC-raguard-lmsh}{1.000}
\resdef{E1-AUROC-raguard-lmsh-legal}{0.964}
\resdef{E1-AUROC-raguard-lmsh-medical}{1.000}
\resdef{E1-AUROC-raguard-nat}{1.000}
\resdef{E1-AUROC-raguard-nat-financial}{0.568}
\resdef{E1-AUROC-raguard-nat-legal}{0.968}
\resdef{E1-AUROC-raguard-nat-medical}{1.000}
\resdef{E1-AUROC-raguard-nat-nq}{0.758}
\resdef{E1-AUROC-raguard-natlm}{0.854}
\resdef{E1-AUROC-raguard-natlm-financial}{0.539}
\resdef{E1-AUROC-raguard-natlm-legal}{0.846}
\resdef{E1-AUROC-raguard-natlm-medical}{0.854}
\resdef{E1-AUROC-raguard-nq-published}{0.953}
\resdef{E1-AUROC-raguard-released}{0.758}
\resdef{E1-AUROC-raguard-tpl}{1.000}
\resdef{E1-AUROC-raguard-tpl-financial}{0.932}
\resdef{E1-AUROC-raguard-tpl-legal}{0.964}
\resdef{E1-AUROC-raguard-tpl-medical}{1.000}
\resdef{E1-AUROC-raguard-tpl-nq}{0.953}
\resdef{E1-AUROC-released-max}{0.758}
\resdef{E1-AUROC-trust-lm-financial}{0.986}
\resdef{E1-AUROC-trust-lm-legal}{0.995 \ensuremath{\pm} 0.001}
\resdef{E1-AUROC-trust-lm-medical}{0.993 \ensuremath{\pm} 0.003}
\resdef{E1-AUROC-trust-lmns-legal}{0.997}
\resdef{E1-AUROC-trust-lmns-medical}{1.000}
\resdef{E1-AUROC-trust-lmsh-legal}{0.998}
\resdef{E1-AUROC-trust-lmsh-medical}{0.998}
\resdef{E1-AUROC-trust-nat-financial}{0.922}
\resdef{E1-AUROC-trust-nat-legal}{0.998 \ensuremath{\pm} 0.000}
\resdef{E1-AUROC-trust-nat-medical}{1.000 \ensuremath{\pm} 0.000}
\resdef{E1-AUROC-trust-nat-nq}{0.997}
\resdef{E1-AUROC-trust-natlm-financial}{0.805}
\resdef{E1-AUROC-trust-natlm-legal}{0.940}
\resdef{E1-AUROC-trust-natlm-medical}{0.940}
\resdef{E1-AUROC-trust-nq-published}{1.000}
\resdef{E1-AUROC-trust-released}{0.997}
\resdef{E1-AUROC-trust-tpl-financial}{0.988}
\resdef{E1-AUROC-trust-tpl-legal}{1.000}
\resdef{E1-AUROC-trust-tpl-medical}{1.000}
\resdef{E1-AUROC-trust-tpl-nq}{1.000}
\resdef{E1-AUROC-trustrag-lm-financial}{N/A}
\resdef{E1-AUROC-trustrag-lm-legal}{N/A}
\resdef{E1-AUROC-trustrag-lm-medical}{N/A}
\resdef{E1-AUROC-trustrag-lmns-legal}{N/A}
\resdef{E1-AUROC-trustrag-lmns-medical}{N/A}
\resdef{E1-AUROC-trustrag-lmsh-legal}{N/A}
\resdef{E1-AUROC-trustrag-lmsh-medical}{N/A}
\resdef{E1-AUROC-trustrag-nat-financial}{N/A}
\resdef{E1-AUROC-trustrag-nat-legal}{N/A}
\resdef{E1-AUROC-trustrag-nat-medical}{N/A}
\resdef{E1-AUROC-trustrag-nat-nq}{N/A}
\resdef{E1-AUROC-trustrag-natlm-financial}{N/A}
\resdef{E1-AUROC-trustrag-natlm-legal}{N/A}
\resdef{E1-AUROC-trustrag-natlm-medical}{N/A}
\resdef{E1-AUROC-trustrag-nq-published}{N/A}
\resdef{E1-AUROC-trustrag-released}{N/A}
\resdef{E1-AUROC-trustrag-tpl-financial}{N/A}
\resdef{E1-AUROC-trustrag-tpl-legal}{N/A}
\resdef{E1-AUROC-trustrag-tpl-medical}{N/A}
\resdef{E1-AUROC-trustrag-tpl-nq}{N/A}
\resdef{E1-AUROC-uniform-lm-financial}{0.981}
\resdef{E1-AUROC-uniform-lm-legal}{0.989 \ensuremath{\pm} 0.002}
\resdef{E1-AUROC-uniform-lm-medical}{0.880 \ensuremath{\pm} 0.010}
\resdef{E1-AUROC-uniform-lmns-legal}{0.992}
\resdef{E1-AUROC-uniform-lmns-medical}{0.999}
\resdef{E1-AUROC-uniform-lmsh-legal}{0.995}
\resdef{E1-AUROC-uniform-lmsh-medical}{0.953}
\resdef{E1-AUROC-uniform-nat-financial}{0.896}
\resdef{E1-AUROC-uniform-nat-legal}{0.996 \ensuremath{\pm} 0.000}
\resdef{E1-AUROC-uniform-nat-medical}{1.000 \ensuremath{\pm} 0.000}
\resdef{E1-AUROC-uniform-nat-nq}{0.990}
\resdef{E1-AUROC-uniform-natlm-financial}{0.735}
\resdef{E1-AUROC-uniform-natlm-legal}{0.888}
\resdef{E1-AUROC-uniform-natlm-medical}{0.503}
\resdef{E1-AUROC-uniform-nq-published}{1.000}
\resdef{E1-AUROC-uniform-released}{0.990}
\resdef{E1-AUROC-uniform-tpl-financial}{0.985}
\resdef{E1-AUROC-uniform-tpl-legal}{0.999}
\resdef{E1-AUROC-uniform-tpl-medical}{1.000}
\resdef{E1-AUROC-uniform-tpl-nq}{1.000}
\resdef{E1-AUROC-valor-lm-financial}{0.981}
\resdef{E1-AUROC-valor-lm-legal}{0.989 \ensuremath{\pm} 0.002}
\resdef{E1-AUROC-valor-lm-medical}{0.880 \ensuremath{\pm} 0.010}
\resdef{E1-AUROC-valor-lmns-legal}{0.992}
\resdef{E1-AUROC-valor-lmns-medical}{0.999}
\resdef{E1-AUROC-valor-lmsh-legal}{0.995}
\resdef{E1-AUROC-valor-lmsh-medical}{0.953}
\resdef{E1-AUROC-valor-nat-financial}{0.896}
\resdef{E1-AUROC-valor-nat-legal}{0.996 \ensuremath{\pm} 0.000}
\resdef{E1-AUROC-valor-nat-medical}{1.000 \ensuremath{\pm} 0.000}
\resdef{E1-AUROC-valor-nat-nq}{0.990}
\resdef{E1-AUROC-valor-natlm-financial}{0.735}
\resdef{E1-AUROC-valor-natlm-legal}{0.888}
\resdef{E1-AUROC-valor-natlm-medical}{0.503}
\resdef{E1-AUROC-valor-nq-published}{1.000}
\resdef{E1-AUROC-valor-released}{0.990}
\resdef{E1-AUROC-valor-tpl-financial}{0.985}
\resdef{E1-AUROC-valor-tpl-legal}{0.999}
\resdef{E1-AUROC-valor-tpl-medical}{1.000}
\resdef{E1-AUROC-valor-tpl-nq}{1.000}
\resdef{E1-ECE-containonly-nq-published}{N/A}
\resdef{E1-ECE-containonly-released}{0.181}
\resdef{E1-ECE-drs-nq-published}{N/A}
\resdef{E1-ECE-drs-released}{N/A}
\resdef{E1-ECE-dup-nq-published}{N/A}
\resdef{E1-ECE-dup-released}{N/A}
\resdef{E1-ECE-financial}{0.016 \ensuremath{\pm} 0.003}
\resdef{E1-ECE-legal}{0.015 \ensuremath{\pm} 0.001}
\resdef{E1-ECE-lenonly-nq-published}{N/A}
\resdef{E1-ECE-lenonly-released}{N/A}
\resdef{E1-ECE-lm-medical}{0.413 \ensuremath{\pm} 0.011}
\resdef{E1-ECE-lmns-medical}{0.474}
\resdef{E1-ECE-lmsh-medical}{0.471}
\resdef{E1-ECE-medical}{0.004 \ensuremath{\pm} 0.000}
\resdef{E1-ECE-nat-medical}{0.012}
\resdef{E1-ECE-natlm-medical}{0.088}
\resdef{E1-ECE-nli-nq-published}{—}
\resdef{E1-ECE-nli-released}{—}
\resdef{E1-ECE-nq}{0.030 \ensuremath{\pm} 0.000}
\resdef{E1-ECE-ppl-nq-published}{N/A}
\resdef{E1-ECE-ppl-released}{N/A}
\resdef{E1-ECE-ragdefender-nq-published}{N/A}
\resdef{E1-ECE-ragdefender-released}{N/A}
\resdef{E1-ECE-raguard-nq-published}{N/A}
\resdef{E1-ECE-raguard-released}{N/A}
\resdef{E1-ECE-tpl-medical}{0.004}
\resdef{E1-ECE-trust-lm-financial}{0.023}
\resdef{E1-ECE-trust-lm-legal}{0.035}
\resdef{E1-ECE-trust-lm-medical}{0.413}
\resdef{E1-ECE-trust-lmns-legal}{0.027}
\resdef{E1-ECE-trust-lmns-medical}{0.474}
\resdef{E1-ECE-trust-lmsh-legal}{0.020}
\resdef{E1-ECE-trust-lmsh-medical}{0.471}
\resdef{E1-ECE-trust-nat-financial}{0.175}
\resdef{E1-ECE-trust-nat-legal}{0.014}
\resdef{E1-ECE-trust-nat-medical}{0.012}
\resdef{E1-ECE-trust-nat-nq}{0.087}
\resdef{E1-ECE-trust-natlm-financial}{0.219}
\resdef{E1-ECE-trust-natlm-legal}{0.197}
\resdef{E1-ECE-trust-natlm-medical}{0.088}
\resdef{E1-ECE-trust-nq-published}{0.030}
\resdef{E1-ECE-trust-released}{0.087}
\resdef{E1-ECE-trust-tpl-financial}{0.016}
\resdef{E1-ECE-trust-tpl-legal}{0.015}
\resdef{E1-ECE-trust-tpl-medical}{0.004}
\resdef{E1-ECE-trust-tpl-nq}{0.030}
\resdef{E1-ECE-trustrag-nq-published}{N/A}
\resdef{E1-ECE-trustrag-released}{N/A}
\resdef{E1-ECE-uniform-lm-financial}{0.023}
\resdef{E1-ECE-uniform-lm-legal}{0.035}
\resdef{E1-ECE-uniform-lm-medical}{0.413}
\resdef{E1-ECE-uniform-lmns-legal}{0.027}
\resdef{E1-ECE-uniform-lmns-medical}{0.474}
\resdef{E1-ECE-uniform-lmsh-legal}{0.020}
\resdef{E1-ECE-uniform-lmsh-medical}{0.471}
\resdef{E1-ECE-uniform-nat-financial}{0.175}
\resdef{E1-ECE-uniform-nat-legal}{0.014}
\resdef{E1-ECE-uniform-nat-medical}{0.012}
\resdef{E1-ECE-uniform-nat-nq}{0.087}
\resdef{E1-ECE-uniform-natlm-financial}{0.219}
\resdef{E1-ECE-uniform-natlm-legal}{0.197}
\resdef{E1-ECE-uniform-natlm-medical}{0.088}
\resdef{E1-ECE-uniform-nq-published}{0.030}
\resdef{E1-ECE-uniform-released}{0.087}
\resdef{E1-ECE-uniform-tpl-financial}{0.016}
\resdef{E1-ECE-uniform-tpl-legal}{0.015}
\resdef{E1-ECE-uniform-tpl-medical}{0.004}
\resdef{E1-ECE-uniform-tpl-nq}{0.030}
\resdef{E1-ECE-valor-lm-financial}{0.023}
\resdef{E1-ECE-valor-lm-legal}{0.035}
\resdef{E1-ECE-valor-lm-medical}{0.413}
\resdef{E1-ECE-valor-lmns-legal}{0.027}
\resdef{E1-ECE-valor-lmns-medical}{0.474}
\resdef{E1-ECE-valor-lmsh-legal}{0.020}
\resdef{E1-ECE-valor-lmsh-medical}{0.471}
\resdef{E1-ECE-valor-nat-financial}{0.175}
\resdef{E1-ECE-valor-nat-legal}{0.014}
\resdef{E1-ECE-valor-nat-medical}{0.012}
\resdef{E1-ECE-valor-nat-nq}{0.087}
\resdef{E1-ECE-valor-natlm-financial}{0.219}
\resdef{E1-ECE-valor-natlm-legal}{0.197}
\resdef{E1-ECE-valor-natlm-medical}{0.088}
\resdef{E1-ECE-valor-nq-published}{0.030}
\resdef{E1-ECE-valor-released}{0.087}
\resdef{E1-ECE-valor-tpl-financial}{0.016}
\resdef{E1-ECE-valor-tpl-legal}{0.015}
\resdef{E1-ECE-valor-tpl-medical}{0.004}
\resdef{E1-ECE-valor-tpl-nq}{0.030}
\resdef{E1-FNR-containonly-lm-financial}{0.009}
\resdef{E1-FNR-containonly-lm-legal}{0.031}
\resdef{E1-FNR-containonly-lm-medical}{0.001}
\resdef{E1-FNR-containonly-lmns-legal}{0.009}
\resdef{E1-FNR-containonly-lmns-medical}{0.000}
\resdef{E1-FNR-containonly-lmsh-legal}{0.023}
\resdef{E1-FNR-containonly-lmsh-medical}{0.000}
\resdef{E1-FNR-containonly-nat-financial}{0.190}
\resdef{E1-FNR-containonly-nat-legal}{0.390}
\resdef{E1-FNR-containonly-nat-medical}{1.000}
\resdef{E1-FNR-containonly-nat-nq}{0.882}
\resdef{E1-FNR-containonly-natlm-financial}{0.174}
\resdef{E1-FNR-containonly-natlm-legal}{0.204}
\resdef{E1-FNR-containonly-natlm-medical}{0.966}
\resdef{E1-FNR-containonly-nq-published}{0.000}
\resdef{E1-FNR-containonly-released}{0.882}
\resdef{E1-FNR-containonly-tpl-financial}{0.009}
\resdef{E1-FNR-containonly-tpl-legal}{0.077}
\resdef{E1-FNR-containonly-tpl-medical}{0.000}
\resdef{E1-FNR-containonly-tpl-nq}{0.000}
\resdef{E1-FNR-drs-lm-financial}{0.838}
\resdef{E1-FNR-drs-lm-legal}{0.857}
\resdef{E1-FNR-drs-lm-medical}{0.988}
\resdef{E1-FNR-drs-lmns-legal}{0.479}
\resdef{E1-FNR-drs-lmns-medical}{0.605}
\resdef{E1-FNR-drs-lmsh-legal}{0.922}
\resdef{E1-FNR-drs-lmsh-medical}{0.990}
\resdef{E1-FNR-drs-medical}{0.700}
\resdef{E1-FNR-drs-nat-financial}{0.889}
\resdef{E1-FNR-drs-nat-legal}{0.692}
\resdef{E1-FNR-drs-nat-medical}{0.524}
\resdef{E1-FNR-drs-nat-nq}{0.930}
\resdef{E1-FNR-drs-natlm-financial}{0.879}
\resdef{E1-FNR-drs-natlm-legal}{0.876}
\resdef{E1-FNR-drs-natlm-medical}{0.983}
\resdef{E1-FNR-drs-nq-published}{0.854}
\resdef{E1-FNR-drs-released}{0.930}
\resdef{E1-FNR-drs-tpl-financial}{0.833}
\resdef{E1-FNR-drs-tpl-legal}{0.816}
\resdef{E1-FNR-drs-tpl-medical}{0.700}
\resdef{E1-FNR-drs-tpl-nq}{0.854}
\resdef{E1-FNR-dup-lm-financial}{0.829}
\resdef{E1-FNR-dup-lm-legal}{0.830}
\resdef{E1-FNR-dup-lm-medical}{1.000}
\resdef{E1-FNR-dup-lmns-legal}{0.786}
\resdef{E1-FNR-dup-lmns-medical}{0.985}
\resdef{E1-FNR-dup-lmsh-legal}{0.836}
\resdef{E1-FNR-dup-lmsh-medical}{1.000}
\resdef{E1-FNR-dup-nat-financial}{1.000}
\resdef{E1-FNR-dup-nat-legal}{0.823}
\resdef{E1-FNR-dup-nat-medical}{0.991}
\resdef{E1-FNR-dup-nat-nq}{0.989}
\resdef{E1-FNR-dup-natlm-financial}{1.000}
\resdef{E1-FNR-dup-natlm-legal}{0.970}
\resdef{E1-FNR-dup-natlm-medical}{1.000}
\resdef{E1-FNR-dup-nq-published}{0.678}
\resdef{E1-FNR-dup-released}{0.989}
\resdef{E1-FNR-dup-tpl-financial}{0.788}
\resdef{E1-FNR-dup-tpl-legal}{0.343}
\resdef{E1-FNR-dup-tpl-medical}{0.496}
\resdef{E1-FNR-dup-tpl-nq}{0.678}
\resdef{E1-FNR-financial}{0.025 \ensuremath{\pm} 0.004}
\resdef{E1-FNR-general}{0.015 \ensuremath{\pm} 0.000}
\resdef{E1-FNR-legal}{0.000 \ensuremath{\pm} 0.000}
\resdef{E1-FNR-length-matched}{legal 0.036 \ensuremath{\pm} 0.005; medical 0.983 \ensuremath{\pm} 0.011}
\resdef{E1-FNR-lenonly-lm-financial}{1.000}
\resdef{E1-FNR-lenonly-lm-legal}{1.000}
\resdef{E1-FNR-lenonly-lm-medical}{0.999}
\resdef{E1-FNR-lenonly-lmns-legal}{1.000}
\resdef{E1-FNR-lenonly-lmns-medical}{0.999}
\resdef{E1-FNR-lenonly-lmsh-legal}{1.000}
\resdef{E1-FNR-lenonly-lmsh-medical}{0.999}
\resdef{E1-FNR-lenonly-nat-financial}{1.000}
\resdef{E1-FNR-lenonly-nat-legal}{0.708}
\resdef{E1-FNR-lenonly-nat-medical}{0.000}
\resdef{E1-FNR-lenonly-nat-nq}{1.000}
\resdef{E1-FNR-lenonly-natlm-financial}{1.000}
\resdef{E1-FNR-lenonly-natlm-legal}{1.000}
\resdef{E1-FNR-lenonly-natlm-medical}{0.988}
\resdef{E1-FNR-lenonly-nq-published}{1.000}
\resdef{E1-FNR-lenonly-released}{1.000}
\resdef{E1-FNR-lenonly-tpl-financial}{1.000}
\resdef{E1-FNR-lenonly-tpl-legal}{0.992}
\resdef{E1-FNR-lenonly-tpl-medical}{0.000}
\resdef{E1-FNR-lenonly-tpl-nq}{1.000}
\resdef{E1-FNR-medical}{0.000 \ensuremath{\pm} 0.000}
\resdef{E1-FNR-nli-lm-legal}{0.675}
\resdef{E1-FNR-nli-lm-medical}{0.939}
\resdef{E1-FNR-nli-nat-financial}{0.836}
\resdef{E1-FNR-nli-nat-legal}{0.791}
\resdef{E1-FNR-nli-nat-medical}{0.993}
\resdef{E1-FNR-nli-nat-nq}{0.866}
\resdef{E1-FNR-nli-nq-published}{0.935}
\resdef{E1-FNR-nli-released}{0.866}
\resdef{E1-FNR-nli-tpl-financial}{0.961}
\resdef{E1-FNR-nli-tpl-legal}{0.793}
\resdef{E1-FNR-nli-tpl-medical}{1.000}
\resdef{E1-FNR-nli-tpl-nq}{0.935}
\resdef{E1-FNR-normdup-medical}{0.496 \ensuremath{\pm} 0.010}
\resdef{E1-FNR-nq-natural-best}{0.070}
\resdef{E1-FNR-nq-published-best}{0.000}
\resdef{E1-FNR-ppl-lm}{0.361 \ensuremath{\pm} 0.039}
\resdef{E1-FNR-ppl-lm-financial}{1.000}
\resdef{E1-FNR-ppl-lm-legal}{0.775}
\resdef{E1-FNR-ppl-lm-medical}{0.361}
\resdef{E1-FNR-ppl-lmns}{0.000}
\resdef{E1-FNR-ppl-lmns-legal}{0.280}
\resdef{E1-FNR-ppl-lmns-medical}{0.000}
\resdef{E1-FNR-ppl-lmsh}{0.000}
\resdef{E1-FNR-ppl-lmsh-legal}{0.247}
\resdef{E1-FNR-ppl-lmsh-medical}{0.000}
\resdef{E1-FNR-ppl-nat}{0.000}
\resdef{E1-FNR-ppl-nat-financial}{1.000}
\resdef{E1-FNR-ppl-nat-legal}{0.223}
\resdef{E1-FNR-ppl-nat-medical}{0.000}
\resdef{E1-FNR-ppl-nat-nq}{1.000}
\resdef{E1-FNR-ppl-natlm}{0.319}
\resdef{E1-FNR-ppl-natlm-financial}{1.000}
\resdef{E1-FNR-ppl-natlm-legal}{0.644}
\resdef{E1-FNR-ppl-natlm-medical}{0.319}
\resdef{E1-FNR-ppl-nq-published}{1.000 \ensuremath{\pm} 0.000}
\resdef{E1-FNR-ppl-released}{1.000 \ensuremath{\pm} 0.000}
\resdef{E1-FNR-ppl-tpl}{0.062 \ensuremath{\pm} 0.003}
\resdef{E1-FNR-ppl-tpl-financial}{1.000}
\resdef{E1-FNR-ppl-tpl-legal}{0.551}
\resdef{E1-FNR-ppl-tpl-medical}{0.062}
\resdef{E1-FNR-ppl-tpl-nq}{1.000}
\resdef{E1-FNR-ragdefender-lm-financial}{0.087}
\resdef{E1-FNR-ragdefender-lm-legal}{0.382}
\resdef{E1-FNR-ragdefender-lm-medical}{0.405}
\resdef{E1-FNR-ragdefender-lmns-legal}{0.226}
\resdef{E1-FNR-ragdefender-lmns-medical}{0.005}
\resdef{E1-FNR-ragdefender-lmsh-legal}{0.321}
\resdef{E1-FNR-ragdefender-lmsh-medical}{0.151}
\resdef{E1-FNR-ragdefender-nat-financial}{0.198}
\resdef{E1-FNR-ragdefender-nat-legal}{0.286}
\resdef{E1-FNR-ragdefender-nat-medical}{0.029}
\resdef{E1-FNR-ragdefender-nat-nq}{0.172}
\resdef{E1-FNR-ragdefender-natlm-financial}{0.318}
\resdef{E1-FNR-ragdefender-natlm-legal}{0.570}
\resdef{E1-FNR-ragdefender-natlm-medical}{0.762}
\resdef{E1-FNR-ragdefender-nq-published}{0.045}
\resdef{E1-FNR-ragdefender-released}{0.172}
\resdef{E1-FNR-ragdefender-tpl-financial}{0.081}
\resdef{E1-FNR-ragdefender-tpl-legal}{0.339}
\resdef{E1-FNR-ragdefender-tpl-medical}{0.000}
\resdef{E1-FNR-ragdefender-tpl-nq}{0.045}
\resdef{E1-FNR-raguard-lm}{0.519}
\resdef{E1-FNR-raguard-lm-financial}{0.155}
\resdef{E1-FNR-raguard-lm-legal}{0.279}
\resdef{E1-FNR-raguard-lm-medical}{0.519}
\resdef{E1-FNR-raguard-lmns}{0.000}
\resdef{E1-FNR-raguard-lmns-legal}{0.087}
\resdef{E1-FNR-raguard-lmns-medical}{0.000}
\resdef{E1-FNR-raguard-lmsh}{0.000}
\resdef{E1-FNR-raguard-lmsh-legal}{0.071}
\resdef{E1-FNR-raguard-lmsh-medical}{0.000}
\resdef{E1-FNR-raguard-nat}{0.000}
\resdef{E1-FNR-raguard-nat-financial}{0.873}
\resdef{E1-FNR-raguard-nat-legal}{0.085}
\resdef{E1-FNR-raguard-nat-medical}{0.000}
\resdef{E1-FNR-raguard-nat-nq}{0.763}
\resdef{E1-FNR-raguard-natlm}{0.513}
\resdef{E1-FNR-raguard-natlm-financial}{0.900}
\resdef{E1-FNR-raguard-natlm-legal}{0.442}
\resdef{E1-FNR-raguard-natlm-medical}{0.513}
\resdef{E1-FNR-raguard-nq-published}{0.191}
\resdef{E1-FNR-raguard-released}{0.763}
\resdef{E1-FNR-raguard-tpl}{0.000}
\resdef{E1-FNR-raguard-tpl-financial}{0.146}
\resdef{E1-FNR-raguard-tpl-legal}{0.078}
\resdef{E1-FNR-raguard-tpl-medical}{0.000}
\resdef{E1-FNR-raguard-tpl-nq}{0.191}
\resdef{E1-FNR-released-best}{0.070}
\resdef{E1-FNR-released-vs-template}{+0.110}
\resdef{E1-FNR-trust-lm-financial}{0.039}
\resdef{E1-FNR-trust-lm-legal}{0.019 \ensuremath{\pm} 0.004}
\resdef{E1-FNR-trust-lm-medical}{0.979 \ensuremath{\pm} 0.011}
\resdef{E1-FNR-trust-lmns-legal}{0.017}
\resdef{E1-FNR-trust-lmns-medical}{0.955}
\resdef{E1-FNR-trust-lmsh-legal}{0.005}
\resdef{E1-FNR-trust-lmsh-medical}{0.982}
\resdef{E1-FNR-trust-nat-financial}{0.444}
\resdef{E1-FNR-trust-nat-legal}{0.001 \ensuremath{\pm} 0.002}
\resdef{E1-FNR-trust-nat-medical}{0.009 \ensuremath{\pm} 0.002}
\resdef{E1-FNR-trust-nat-nq}{0.086}
\resdef{E1-FNR-trust-natlm-financial}{0.602}
\resdef{E1-FNR-trust-natlm-legal}{0.299}
\resdef{E1-FNR-trust-natlm-medical}{1.000}
\resdef{E1-FNR-trust-nq-published}{0.012}
\resdef{E1-FNR-trust-released}{0.086}
\resdef{E1-FNR-trust-tpl-financial}{0.025}
\resdef{E1-FNR-trust-tpl-legal}{0.000}
\resdef{E1-FNR-trust-tpl-medical}{0.000}
\resdef{E1-FNR-trust-tpl-nq}{0.012}
\resdef{E1-FNR-trustrag-lm-financial}{0.009}
\resdef{E1-FNR-trustrag-lm-legal}{0.044}
\resdef{E1-FNR-trustrag-lm-medical}{0.781}
\resdef{E1-FNR-trustrag-lmns-legal}{0.142}
\resdef{E1-FNR-trustrag-lmns-medical}{0.843}
\resdef{E1-FNR-trustrag-lmsh-legal}{0.038}
\resdef{E1-FNR-trustrag-lmsh-medical}{0.827}
\resdef{E1-FNR-trustrag-nat-financial}{0.036}
\resdef{E1-FNR-trustrag-nat-legal}{0.013}
\resdef{E1-FNR-trustrag-nat-medical}{0.014}
\resdef{E1-FNR-trustrag-nat-nq}{0.070}
\resdef{E1-FNR-trustrag-natlm-financial}{0.164}
\resdef{E1-FNR-trustrag-natlm-legal}{0.240}
\resdef{E1-FNR-trustrag-natlm-medical}{1.000}
\resdef{E1-FNR-trustrag-nq-published}{0.000}
\resdef{E1-FNR-trustrag-released}{0.070}
\resdef{E1-FNR-trustrag-tpl-financial}{0.009}
\resdef{E1-FNR-trustrag-tpl-legal}{0.004}
\resdef{E1-FNR-trustrag-tpl-medical}{0.000}
\resdef{E1-FNR-trustrag-tpl-nq}{0.000}
\resdef{E1-FNR-uniform-lm-financial}{0.039}
\resdef{E1-FNR-uniform-lm-legal}{0.063 \ensuremath{\pm} 0.005}
\resdef{E1-FNR-uniform-lm-medical}{0.985 \ensuremath{\pm} 0.009}
\resdef{E1-FNR-uniform-lmns-legal}{0.033}
\resdef{E1-FNR-uniform-lmns-medical}{0.974}
\resdef{E1-FNR-uniform-lmsh-legal}{0.035}
\resdef{E1-FNR-uniform-lmsh-medical}{0.986}
\resdef{E1-FNR-uniform-nat-financial}{0.448}
\resdef{E1-FNR-uniform-nat-legal}{0.016 \ensuremath{\pm} 0.006}
\resdef{E1-FNR-uniform-nat-medical}{0.017 \ensuremath{\pm} 0.002}
\resdef{E1-FNR-uniform-nat-nq}{0.118}
\resdef{E1-FNR-uniform-natlm-financial}{0.603}
\resdef{E1-FNR-uniform-natlm-legal}{0.610}
\resdef{E1-FNR-uniform-natlm-medical}{1.000}
\resdef{E1-FNR-uniform-nq-published}{0.015}
\resdef{E1-FNR-uniform-released}{0.118}
\resdef{E1-FNR-uniform-tpl-financial}{0.025}
\resdef{E1-FNR-uniform-tpl-legal}{0.002}
\resdef{E1-FNR-uniform-tpl-medical}{0.000}
\resdef{E1-FNR-uniform-tpl-nq}{0.015}
\resdef{E1-FNR-valor-lm-financial}{0.039}
\resdef{E1-FNR-valor-lm-legal}{0.036 \ensuremath{\pm} 0.005}
\resdef{E1-FNR-valor-lm-medical}{0.983 \ensuremath{\pm} 0.011}
\resdef{E1-FNR-valor-lmns-legal}{0.027}
\resdef{E1-FNR-valor-lmns-medical}{0.968}
\resdef{E1-FNR-valor-lmsh-legal}{0.017}
\resdef{E1-FNR-valor-lmsh-medical}{0.986}
\resdef{E1-FNR-valor-nat-financial}{0.446}
\resdef{E1-FNR-valor-nat-legal}{0.002 \ensuremath{\pm} 0.004}
\resdef{E1-FNR-valor-nat-medical}{0.009 \ensuremath{\pm} 0.002}
\resdef{E1-FNR-valor-nat-nq}{0.118}
\resdef{E1-FNR-valor-natlm-financial}{0.602}
\resdef{E1-FNR-valor-natlm-legal}{0.405}
\resdef{E1-FNR-valor-natlm-medical}{1.000}
\resdef{E1-FNR-valor-nq-published}{0.015}
\resdef{E1-FNR-valor-released}{0.118}
\resdef{E1-FNR-valor-tpl-financial}{0.025}
\resdef{E1-FNR-valor-tpl-legal}{0.000}
\resdef{E1-FNR-valor-tpl-medical}{0.000}
\resdef{E1-FNR-valor-tpl-nq}{0.015}
\resdef{E1-FNR5-containonly-lm-financial}{0.022}
\resdef{E1-FNR5-containonly-lm-legal}{0.022}
\resdef{E1-FNR5-containonly-lm-medical}{0.001}
\resdef{E1-FNR5-containonly-lmns-legal}{0.009}
\resdef{E1-FNR5-containonly-lmns-medical}{0.000}
\resdef{E1-FNR5-containonly-lmsh-legal}{0.016}
\resdef{E1-FNR5-containonly-lmsh-medical}{0.000}
\resdef{E1-FNR5-containonly-nat-financial}{0.863}
\resdef{E1-FNR5-containonly-nat-legal}{0.307}
\resdef{E1-FNR5-containonly-nat-medical}{1.000}
\resdef{E1-FNR5-containonly-nat-nq}{0.839}
\resdef{E1-FNR5-containonly-natlm-financial}{0.826}
\resdef{E1-FNR5-containonly-natlm-legal}{0.130}
\resdef{E1-FNR5-containonly-natlm-medical}{0.671}
\resdef{E1-FNR5-containonly-nq-published}{0.000}
\resdef{E1-FNR5-containonly-released}{0.839}
\resdef{E1-FNR5-containonly-tpl-financial}{0.022}
\resdef{E1-FNR5-containonly-tpl-legal}{0.039}
\resdef{E1-FNR5-containonly-tpl-medical}{0.000}
\resdef{E1-FNR5-containonly-tpl-nq}{0.000}
\resdef{E1-FNR5-drs-lm-financial}{0.908}
\resdef{E1-FNR5-drs-lm-legal}{0.933}
\resdef{E1-FNR5-drs-lm-medical}{0.977}
\resdef{E1-FNR5-drs-lmns-legal}{0.643}
\resdef{E1-FNR5-drs-lmns-medical}{0.497}
\resdef{E1-FNR5-drs-lmsh-legal}{0.961}
\resdef{E1-FNR5-drs-lmsh-medical}{0.982}
\resdef{E1-FNR5-drs-nat-financial}{0.925}
\resdef{E1-FNR5-drs-nat-legal}{0.788}
\resdef{E1-FNR5-drs-nat-medical}{0.424}
\resdef{E1-FNR5-drs-nat-nq}{0.995}
\resdef{E1-FNR5-drs-natlm-financial}{0.918}
\resdef{E1-FNR5-drs-natlm-legal}{0.938}
\resdef{E1-FNR5-drs-natlm-medical}{0.983}
\resdef{E1-FNR5-drs-nq-published}{0.970}
\resdef{E1-FNR5-drs-released}{0.995}
\resdef{E1-FNR5-drs-tpl-financial}{0.904}
\resdef{E1-FNR5-drs-tpl-legal}{0.897}
\resdef{E1-FNR5-drs-tpl-medical}{0.635}
\resdef{E1-FNR5-drs-tpl-nq}{0.970}
\resdef{E1-FNR5-dup-lm-financial}{N/A}
\resdef{E1-FNR5-dup-lm-legal}{N/A}
\resdef{E1-FNR5-dup-lm-medical}{N/A}
\resdef{E1-FNR5-dup-lmns-legal}{N/A}
\resdef{E1-FNR5-dup-lmns-medical}{N/A}
\resdef{E1-FNR5-dup-lmsh-legal}{N/A}
\resdef{E1-FNR5-dup-lmsh-medical}{N/A}
\resdef{E1-FNR5-dup-nat-financial}{N/A}
\resdef{E1-FNR5-dup-nat-legal}{N/A}
\resdef{E1-FNR5-dup-nat-medical}{N/A}
\resdef{E1-FNR5-dup-nat-nq}{N/A}
\resdef{E1-FNR5-dup-natlm-financial}{N/A}
\resdef{E1-FNR5-dup-natlm-legal}{N/A}
\resdef{E1-FNR5-dup-natlm-medical}{N/A}
\resdef{E1-FNR5-dup-nq-published}{N/A}
\resdef{E1-FNR5-dup-released}{N/A}
\resdef{E1-FNR5-dup-tpl-financial}{N/A}
\resdef{E1-FNR5-dup-tpl-legal}{N/A}
\resdef{E1-FNR5-dup-tpl-medical}{N/A}
\resdef{E1-FNR5-dup-tpl-nq}{N/A}
\resdef{E1-FNR5-lenonly-lm-financial}{1.000}
\resdef{E1-FNR5-lenonly-lm-legal}{0.992}
\resdef{E1-FNR5-lenonly-lm-medical}{0.952}
\resdef{E1-FNR5-lenonly-lmns-legal}{0.991}
\resdef{E1-FNR5-lenonly-lmns-medical}{0.959}
\resdef{E1-FNR5-lenonly-lmsh-legal}{0.988}
\resdef{E1-FNR5-lenonly-lmsh-medical}{0.969}
\resdef{E1-FNR5-lenonly-nat-financial}{1.000}
\resdef{E1-FNR5-lenonly-nat-legal}{0.095}
\resdef{E1-FNR5-lenonly-nat-medical}{0.000}
\resdef{E1-FNR5-lenonly-nat-nq}{1.000}
\resdef{E1-FNR5-lenonly-natlm-financial}{1.000}
\resdef{E1-FNR5-lenonly-natlm-legal}{0.981}
\resdef{E1-FNR5-lenonly-natlm-medical}{0.856}
\resdef{E1-FNR5-lenonly-nq-published}{1.000}
\resdef{E1-FNR5-lenonly-released}{1.000}
\resdef{E1-FNR5-lenonly-tpl-financial}{1.000}
\resdef{E1-FNR5-lenonly-tpl-legal}{0.183}
\resdef{E1-FNR5-lenonly-tpl-medical}{0.000}
\resdef{E1-FNR5-lenonly-tpl-nq}{1.000}
\resdef{E1-FNR5-nli-lm-legal}{—}
\resdef{E1-FNR5-nli-lm-medical}{—}
\resdef{E1-FNR5-nli-nat-financial}{—}
\resdef{E1-FNR5-nli-nat-legal}{—}
\resdef{E1-FNR5-nli-nat-medical}{—}
\resdef{E1-FNR5-nli-nat-nq}{—}
\resdef{E1-FNR5-nli-nq-published}{—}
\resdef{E1-FNR5-nli-released}{—}
\resdef{E1-FNR5-nli-tpl-financial}{—}
\resdef{E1-FNR5-nli-tpl-legal}{—}
\resdef{E1-FNR5-nli-tpl-medical}{—}
\resdef{E1-FNR5-nli-tpl-nq}{—}
\resdef{E1-FNR5-ppl-lm}{0.092}
\resdef{E1-FNR5-ppl-lm-financial}{1.000}
\resdef{E1-FNR5-ppl-lm-legal}{0.605}
\resdef{E1-FNR5-ppl-lm-medical}{0.092}
\resdef{E1-FNR5-ppl-lmns}{0.000}
\resdef{E1-FNR5-ppl-lmns-legal}{0.259}
\resdef{E1-FNR5-ppl-lmns-medical}{0.000}
\resdef{E1-FNR5-ppl-lmsh}{0.000}
\resdef{E1-FNR5-ppl-lmsh-legal}{0.230}
\resdef{E1-FNR5-ppl-lmsh-medical}{0.000}
\resdef{E1-FNR5-ppl-nat}{0.000}
\resdef{E1-FNR5-ppl-nat-financial}{0.990}
\resdef{E1-FNR5-ppl-nat-legal}{0.180}
\resdef{E1-FNR5-ppl-nat-medical}{0.000}
\resdef{E1-FNR5-ppl-nat-nq}{0.866}
\resdef{E1-FNR5-ppl-natlm}{0.139}
\resdef{E1-FNR5-ppl-natlm-financial}{0.989}
\resdef{E1-FNR5-ppl-natlm-legal}{0.428}
\resdef{E1-FNR5-ppl-natlm-medical}{0.139}
\resdef{E1-FNR5-ppl-nq-published}{0.809}
\resdef{E1-FNR5-ppl-released}{0.866}
\resdef{E1-FNR5-ppl-tpl}{0.046}
\resdef{E1-FNR5-ppl-tpl-financial}{1.000}
\resdef{E1-FNR5-ppl-tpl-legal}{0.392}
\resdef{E1-FNR5-ppl-tpl-medical}{0.046}
\resdef{E1-FNR5-ppl-tpl-nq}{0.809}
\resdef{E1-FNR5-ragdefender-lm-financial}{N/A}
\resdef{E1-FNR5-ragdefender-lm-legal}{N/A}
\resdef{E1-FNR5-ragdefender-lm-medical}{N/A}
\resdef{E1-FNR5-ragdefender-lmns-legal}{N/A}
\resdef{E1-FNR5-ragdefender-lmns-medical}{N/A}
\resdef{E1-FNR5-ragdefender-lmsh-legal}{N/A}
\resdef{E1-FNR5-ragdefender-lmsh-medical}{N/A}
\resdef{E1-FNR5-ragdefender-nat-financial}{N/A}
\resdef{E1-FNR5-ragdefender-nat-legal}{N/A}
\resdef{E1-FNR5-ragdefender-nat-medical}{N/A}
\resdef{E1-FNR5-ragdefender-nat-nq}{N/A}
\resdef{E1-FNR5-ragdefender-natlm-financial}{N/A}
\resdef{E1-FNR5-ragdefender-natlm-legal}{N/A}
\resdef{E1-FNR5-ragdefender-natlm-medical}{N/A}
\resdef{E1-FNR5-ragdefender-nq-published}{N/A}
\resdef{E1-FNR5-ragdefender-released}{N/A}
\resdef{E1-FNR5-ragdefender-tpl-financial}{N/A}
\resdef{E1-FNR5-ragdefender-tpl-legal}{N/A}
\resdef{E1-FNR5-ragdefender-tpl-medical}{N/A}
\resdef{E1-FNR5-ragdefender-tpl-nq}{N/A}
\resdef{E1-FNR5-raguard-lm}{0.694}
\resdef{E1-FNR5-raguard-lm-financial}{0.322}
\resdef{E1-FNR5-raguard-lm-legal}{0.321}
\resdef{E1-FNR5-raguard-lm-medical}{0.694}
\resdef{E1-FNR5-raguard-lmns}{0.000}
\resdef{E1-FNR5-raguard-lmns-legal}{0.087}
\resdef{E1-FNR5-raguard-lmns-medical}{0.000}
\resdef{E1-FNR5-raguard-lmsh}{0.000}
\resdef{E1-FNR5-raguard-lmsh-legal}{0.071}
\resdef{E1-FNR5-raguard-lmsh-medical}{0.000}
\resdef{E1-FNR5-raguard-nat}{0.000}
\resdef{E1-FNR5-raguard-nat-financial}{0.929}
\resdef{E1-FNR5-raguard-nat-legal}{0.101}
\resdef{E1-FNR5-raguard-nat-medical}{0.000}
\resdef{E1-FNR5-raguard-nat-nq}{0.796}
\resdef{E1-FNR5-raguard-natlm}{0.621}
\resdef{E1-FNR5-raguard-natlm-financial}{0.948}
\resdef{E1-FNR5-raguard-natlm-legal}{0.546}
\resdef{E1-FNR5-raguard-natlm-medical}{0.621}
\resdef{E1-FNR5-raguard-nq-published}{0.216}
\resdef{E1-FNR5-raguard-released}{0.796}
\resdef{E1-FNR5-raguard-tpl}{0.000}
\resdef{E1-FNR5-raguard-tpl-financial}{0.309}
\resdef{E1-FNR5-raguard-tpl-legal}{0.087}
\resdef{E1-FNR5-raguard-tpl-medical}{0.000}
\resdef{E1-FNR5-raguard-tpl-nq}{0.216}
\resdef{E1-FNR5-trust-lm-financial}{0.025}
\resdef{E1-FNR5-trust-lm-legal}{0.032}
\resdef{E1-FNR5-trust-lm-medical}{0.044}
\resdef{E1-FNR5-trust-lmns-legal}{0.022}
\resdef{E1-FNR5-trust-lmns-medical}{0.000}
\resdef{E1-FNR5-trust-lmsh-legal}{0.015}
\resdef{E1-FNR5-trust-lmsh-medical}{0.014}
\resdef{E1-FNR5-trust-nat-financial}{0.342}
\resdef{E1-FNR5-trust-nat-legal}{0.001}
\resdef{E1-FNR5-trust-nat-medical}{0.000}
\resdef{E1-FNR5-trust-nat-nq}{0.009}
\resdef{E1-FNR5-trust-natlm-financial}{0.523}
\resdef{E1-FNR5-trust-natlm-legal}{0.351}
\resdef{E1-FNR5-trust-natlm-medical}{0.339}
\resdef{E1-FNR5-trust-nq-published}{0.000}
\resdef{E1-FNR5-trust-released}{0.009}
\resdef{E1-FNR5-trust-tpl-financial}{0.017}
\resdef{E1-FNR5-trust-tpl-legal}{0.003}
\resdef{E1-FNR5-trust-tpl-medical}{0.000}
\resdef{E1-FNR5-trust-tpl-nq}{0.000}
\resdef{E1-FNR5-trustrag-lm-financial}{N/A}
\resdef{E1-FNR5-trustrag-lm-legal}{N/A}
\resdef{E1-FNR5-trustrag-lm-medical}{N/A}
\resdef{E1-FNR5-trustrag-lmns-legal}{N/A}
\resdef{E1-FNR5-trustrag-lmns-medical}{N/A}
\resdef{E1-FNR5-trustrag-lmsh-legal}{N/A}
\resdef{E1-FNR5-trustrag-lmsh-medical}{N/A}
\resdef{E1-FNR5-trustrag-nat-financial}{N/A}
\resdef{E1-FNR5-trustrag-nat-legal}{N/A}
\resdef{E1-FNR5-trustrag-nat-medical}{N/A}
\resdef{E1-FNR5-trustrag-nat-nq}{N/A}
\resdef{E1-FNR5-trustrag-natlm-financial}{N/A}
\resdef{E1-FNR5-trustrag-natlm-legal}{N/A}
\resdef{E1-FNR5-trustrag-natlm-medical}{N/A}
\resdef{E1-FNR5-trustrag-nq-published}{N/A}
\resdef{E1-FNR5-trustrag-released}{N/A}
\resdef{E1-FNR5-trustrag-tpl-financial}{N/A}
\resdef{E1-FNR5-trustrag-tpl-legal}{N/A}
\resdef{E1-FNR5-trustrag-tpl-medical}{N/A}
\resdef{E1-FNR5-trustrag-tpl-nq}{N/A}
\resdef{E1-FNR5-uniform-lm-financial}{0.025}
\resdef{E1-FNR5-uniform-lm-legal}{0.053}
\resdef{E1-FNR5-uniform-lm-medical}{0.570}
\resdef{E1-FNR5-uniform-lmns-legal}{0.030}
\resdef{E1-FNR5-uniform-lmns-medical}{0.000}
\resdef{E1-FNR5-uniform-lmsh-legal}{0.025}
\resdef{E1-FNR5-uniform-lmsh-medical}{0.321}
\resdef{E1-FNR5-uniform-nat-financial}{0.343}
\resdef{E1-FNR5-uniform-nat-legal}{0.002}
\resdef{E1-FNR5-uniform-nat-medical}{0.000}
\resdef{E1-FNR5-uniform-nat-nq}{0.043}
\resdef{E1-FNR5-uniform-natlm-financial}{0.523}
\resdef{E1-FNR5-uniform-natlm-legal}{0.445}
\resdef{E1-FNR5-uniform-natlm-medical}{0.949}
\resdef{E1-FNR5-uniform-nq-published}{0.000}
\resdef{E1-FNR5-uniform-released}{0.043}
\resdef{E1-FNR5-uniform-tpl-financial}{0.017}
\resdef{E1-FNR5-uniform-tpl-legal}{0.005}
\resdef{E1-FNR5-uniform-tpl-medical}{0.000}
\resdef{E1-FNR5-uniform-tpl-nq}{0.000}
\resdef{E1-FNR5-valor-lm-financial}{0.025}
\resdef{E1-FNR5-valor-lm-legal}{0.053}
\resdef{E1-FNR5-valor-lm-medical}{0.570}
\resdef{E1-FNR5-valor-lmns-legal}{0.030}
\resdef{E1-FNR5-valor-lmns-medical}{0.000}
\resdef{E1-FNR5-valor-lmsh-legal}{0.025}
\resdef{E1-FNR5-valor-lmsh-medical}{0.321}
\resdef{E1-FNR5-valor-nat-financial}{0.343}
\resdef{E1-FNR5-valor-nat-legal}{0.002}
\resdef{E1-FNR5-valor-nat-medical}{0.000}
\resdef{E1-FNR5-valor-nat-nq}{0.043}
\resdef{E1-FNR5-valor-natlm-financial}{0.523}
\resdef{E1-FNR5-valor-natlm-legal}{0.445}
\resdef{E1-FNR5-valor-natlm-medical}{0.949}
\resdef{E1-FNR5-valor-nq-published}{0.000}
\resdef{E1-FNR5-valor-released}{0.043}
\resdef{E1-FNR5-valor-tpl-financial}{0.017}
\resdef{E1-FNR5-valor-tpl-legal}{0.005}
\resdef{E1-FNR5-valor-tpl-medical}{0.000}
\resdef{E1-FNR5-valor-tpl-nq}{0.000}
\resdef{E1-FPR-containonly-lm-financial}{0.297}
\resdef{E1-FPR-containonly-lm-legal}{0.025}
\resdef{E1-FPR-containonly-lm-medical}{0.004}
\resdef{E1-FPR-containonly-lmns-legal}{0.023}
\resdef{E1-FPR-containonly-lmns-medical}{0.005}
\resdef{E1-FPR-containonly-lmsh-legal}{0.022}
\resdef{E1-FPR-containonly-lmsh-medical}{0.005}
\resdef{E1-FPR-containonly-nat-financial}{0.335}
\resdef{E1-FPR-containonly-nat-legal}{0.018}
\resdef{E1-FPR-containonly-nat-medical}{0.005}
\resdef{E1-FPR-containonly-nat-nq}{0.023}
\resdef{E1-FPR-containonly-natlm-financial}{0.330}
\resdef{E1-FPR-containonly-natlm-legal}{0.019}
\resdef{E1-FPR-containonly-natlm-medical}{0.003}
\resdef{E1-FPR-containonly-nq-published}{0.025}
\resdef{E1-FPR-containonly-released}{0.023}
\resdef{E1-FPR-containonly-tpl-financial}{0.298}
\resdef{E1-FPR-containonly-tpl-legal}{0.029}
\resdef{E1-FPR-containonly-tpl-medical}{0.005}
\resdef{E1-FPR-containonly-tpl-nq}{0.025}
\resdef{E1-FPR-drs-lm-financial}{0.094}
\resdef{E1-FPR-drs-lm-legal}{0.100}
\resdef{E1-FPR-drs-lm-medical}{0.035}
\resdef{E1-FPR-drs-lmns-legal}{0.097}
\resdef{E1-FPR-drs-lmns-medical}{0.034}
\resdef{E1-FPR-drs-lmsh-legal}{0.101}
\resdef{E1-FPR-drs-lmsh-medical}{0.038}
\resdef{E1-FPR-drs-medical}{0.035}
\resdef{E1-FPR-drs-nat-financial}{0.080}
\resdef{E1-FPR-drs-nat-legal}{0.090}
\resdef{E1-FPR-drs-nat-medical}{0.033}
\resdef{E1-FPR-drs-nat-nq}{0.164}
\resdef{E1-FPR-drs-natlm-financial}{0.084}
\resdef{E1-FPR-drs-natlm-legal}{0.085}
\resdef{E1-FPR-drs-natlm-medical}{0.030}
\resdef{E1-FPR-drs-nq-published}{0.154}
\resdef{E1-FPR-drs-released}{0.164}
\resdef{E1-FPR-drs-tpl-financial}{0.094}
\resdef{E1-FPR-drs-tpl-legal}{0.128}
\resdef{E1-FPR-drs-tpl-medical}{0.035}
\resdef{E1-FPR-drs-tpl-nq}{0.154}
\resdef{E1-FPR-dup-lm-financial}{0.151}
\resdef{E1-FPR-dup-lm-legal}{0.000}
\resdef{E1-FPR-dup-lm-medical}{0.000}
\resdef{E1-FPR-dup-lmns-legal}{0.000}
\resdef{E1-FPR-dup-lmns-medical}{0.000}
\resdef{E1-FPR-dup-lmsh-legal}{0.000}
\resdef{E1-FPR-dup-lmsh-medical}{0.000}
\resdef{E1-FPR-dup-nat-financial}{0.164}
\resdef{E1-FPR-dup-nat-legal}{0.000}
\resdef{E1-FPR-dup-nat-medical}{0.000}
\resdef{E1-FPR-dup-nat-nq}{0.009}
\resdef{E1-FPR-dup-natlm-financial}{0.162}
\resdef{E1-FPR-dup-natlm-legal}{0.003}
\resdef{E1-FPR-dup-natlm-medical}{0.000}
\resdef{E1-FPR-dup-nq-published}{0.005}
\resdef{E1-FPR-dup-released}{0.009}
\resdef{E1-FPR-dup-tpl-financial}{0.150}
\resdef{E1-FPR-dup-tpl-legal}{0.000}
\resdef{E1-FPR-dup-tpl-medical}{0.000}
\resdef{E1-FPR-dup-tpl-nq}{0.005}
\resdef{E1-FPR-financial}{0.017 \ensuremath{\pm} 0.006}
\resdef{E1-FPR-general}{0.010 \ensuremath{\pm} 0.000}
\resdef{E1-FPR-legal}{0.079 \ensuremath{\pm} 0.001}
\resdef{E1-FPR-lenonly-lm-financial}{0.000}
\resdef{E1-FPR-lenonly-lm-legal}{0.000}
\resdef{E1-FPR-lenonly-lm-medical}{0.000}
\resdef{E1-FPR-lenonly-lmns-legal}{0.000}
\resdef{E1-FPR-lenonly-lmns-medical}{0.000}
\resdef{E1-FPR-lenonly-lmsh-legal}{0.000}
\resdef{E1-FPR-lenonly-lmsh-medical}{0.000}
\resdef{E1-FPR-lenonly-nat-financial}{0.000}
\resdef{E1-FPR-lenonly-nat-legal}{0.000}
\resdef{E1-FPR-lenonly-nat-medical}{0.000}
\resdef{E1-FPR-lenonly-nat-nq}{0.000}
\resdef{E1-FPR-lenonly-natlm-financial}{0.000}
\resdef{E1-FPR-lenonly-natlm-legal}{0.003}
\resdef{E1-FPR-lenonly-natlm-medical}{0.000}
\resdef{E1-FPR-lenonly-nq-published}{0.000}
\resdef{E1-FPR-lenonly-released}{0.000}
\resdef{E1-FPR-lenonly-tpl-financial}{0.000}
\resdef{E1-FPR-lenonly-tpl-legal}{0.000}
\resdef{E1-FPR-lenonly-tpl-medical}{0.000}
\resdef{E1-FPR-lenonly-tpl-nq}{0.000}
\resdef{E1-FPR-medical}{0.000 \ensuremath{\pm} 0.000}
\resdef{E1-FPR-nli-lm-legal}{0.513}
\resdef{E1-FPR-nli-lm-medical}{0.183}
\resdef{E1-FPR-nli-nat-financial}{0.519}
\resdef{E1-FPR-nli-nat-legal}{0.539}
\resdef{E1-FPR-nli-nat-medical}{0.212}
\resdef{E1-FPR-nli-nat-nq}{0.388}
\resdef{E1-FPR-nli-nq-published}{0.458}
\resdef{E1-FPR-nli-released}{0.388}
\resdef{E1-FPR-nli-tpl-financial}{0.532}
\resdef{E1-FPR-nli-tpl-legal}{0.507}
\resdef{E1-FPR-nli-tpl-medical}{0.185}
\resdef{E1-FPR-nli-tpl-nq}{0.458}
\resdef{E1-FPR-nq-natural-best}{0.153}
\resdef{E1-FPR-nq-published-best}{0.149}
\resdef{E1-FPR-ppl-lm}{0.003}
\resdef{E1-FPR-ppl-lm-financial}{0.005}
\resdef{E1-FPR-ppl-lm-legal}{0.000}
\resdef{E1-FPR-ppl-lm-medical}{0.003}
\resdef{E1-FPR-ppl-lmns}{0.000}
\resdef{E1-FPR-ppl-lmns-legal}{0.006}
\resdef{E1-FPR-ppl-lmns-medical}{0.000}
\resdef{E1-FPR-ppl-lmsh}{0.000}
\resdef{E1-FPR-ppl-lmsh-legal}{0.000}
\resdef{E1-FPR-ppl-lmsh-medical}{0.000}
\resdef{E1-FPR-ppl-nat}{0.000}
\resdef{E1-FPR-ppl-nat-financial}{0.004}
\resdef{E1-FPR-ppl-nat-legal}{0.006}
\resdef{E1-FPR-ppl-nat-medical}{0.000}
\resdef{E1-FPR-ppl-nat-nq}{0.005}
\resdef{E1-FPR-ppl-natlm}{0.013}
\resdef{E1-FPR-ppl-natlm-financial}{0.004}
\resdef{E1-FPR-ppl-natlm-legal}{0.004}
\resdef{E1-FPR-ppl-natlm-medical}{0.013}
\resdef{E1-FPR-ppl-nq-published}{0.005}
\resdef{E1-FPR-ppl-released}{0.005}
\resdef{E1-FPR-ppl-tpl}{0.000}
\resdef{E1-FPR-ppl-tpl-financial}{0.005}
\resdef{E1-FPR-ppl-tpl-legal}{0.000}
\resdef{E1-FPR-ppl-tpl-medical}{0.000}
\resdef{E1-FPR-ppl-tpl-nq}{0.005}
\resdef{E1-FPR-ragdefender-lm-financial}{0.145}
\resdef{E1-FPR-ragdefender-lm-legal}{0.306}
\resdef{E1-FPR-ragdefender-lm-medical}{0.599}
\resdef{E1-FPR-ragdefender-lmns-legal}{0.112}
\resdef{E1-FPR-ragdefender-lmns-medical}{0.069}
\resdef{E1-FPR-ragdefender-lmsh-legal}{0.158}
\resdef{E1-FPR-ragdefender-lmsh-medical}{0.364}
\resdef{E1-FPR-ragdefender-nat-financial}{0.410}
\resdef{E1-FPR-ragdefender-nat-legal}{0.136}
\resdef{E1-FPR-ragdefender-nat-medical}{0.133}
\resdef{E1-FPR-ragdefender-nat-nq}{0.117}
\resdef{E1-FPR-ragdefender-natlm-financial}{0.484}
\resdef{E1-FPR-ragdefender-natlm-legal}{0.418}
\resdef{E1-FPR-ragdefender-natlm-medical}{0.559}
\resdef{E1-FPR-ragdefender-nq-published}{0.025}
\resdef{E1-FPR-ragdefender-released}{0.117}
\resdef{E1-FPR-ragdefender-tpl-financial}{0.119}
\resdef{E1-FPR-ragdefender-tpl-legal}{0.029}
\resdef{E1-FPR-ragdefender-tpl-medical}{0.055}
\resdef{E1-FPR-ragdefender-tpl-nq}{0.025}
\resdef{E1-FPR-raguard-lm}{0.103}
\resdef{E1-FPR-raguard-lm-financial}{0.104}
\resdef{E1-FPR-raguard-lm-legal}{0.116}
\resdef{E1-FPR-raguard-lm-medical}{0.103}
\resdef{E1-FPR-raguard-lmns}{0.112}
\resdef{E1-FPR-raguard-lmns-legal}{0.119}
\resdef{E1-FPR-raguard-lmns-medical}{0.112}
\resdef{E1-FPR-raguard-lmsh}{0.110}
\resdef{E1-FPR-raguard-lmsh-legal}{0.115}
\resdef{E1-FPR-raguard-lmsh-medical}{0.110}
\resdef{E1-FPR-raguard-nat}{0.110}
\resdef{E1-FPR-raguard-nat-financial}{0.094}
\resdef{E1-FPR-raguard-nat-legal}{0.111}
\resdef{E1-FPR-raguard-nat-medical}{0.110}
\resdef{E1-FPR-raguard-nat-nq}{0.061}
\resdef{E1-FPR-raguard-natlm}{0.092}
\resdef{E1-FPR-raguard-natlm-financial}{0.094}
\resdef{E1-FPR-raguard-natlm-legal}{0.102}
\resdef{E1-FPR-raguard-natlm-medical}{0.092}
\resdef{E1-FPR-raguard-nq-published}{0.065}
\resdef{E1-FPR-raguard-released}{0.061}
\resdef{E1-FPR-raguard-tpl}{0.115}
\resdef{E1-FPR-raguard-tpl-financial}{0.104}
\resdef{E1-FPR-raguard-tpl-legal}{0.108}
\resdef{E1-FPR-raguard-tpl-medical}{0.115}
\resdef{E1-FPR-raguard-tpl-nq}{0.065}
\resdef{E1-FPR-realised-drs}{0.103}
\resdef{E1-FPR-realised-nli}{0.420}
\resdef{E1-FPR-realised-normdup}{0.039}
\resdef{E1-FPR-realised-perplexity}{0.003}
\resdef{E1-FPR-realised-ragdefender}{0.057}
\resdef{E1-FPR-realised-raguard}{0.098}
\resdef{E1-FPR-realised-trustrag}{0.244}
\resdef{E1-FPR-realised-uniform}{0.021}
\resdef{E1-FPR-realised-valor-notrust}{0.026}
\resdef{E1-FPR-realised-valor-trust}{0.026}
\resdef{E1-FPR-released-best}{0.153}
\resdef{E1-FPR-trust-lm-financial}{0.017}
\resdef{E1-FPR-trust-lm-legal}{0.081}
\resdef{E1-FPR-trust-lm-medical}{0.000}
\resdef{E1-FPR-trust-lmns-legal}{0.062}
\resdef{E1-FPR-trust-lmns-medical}{0.000}
\resdef{E1-FPR-trust-lmsh-legal}{0.072}
\resdef{E1-FPR-trust-lmsh-medical}{0.000}
\resdef{E1-FPR-trust-nat-financial}{0.020}
\resdef{E1-FPR-trust-nat-legal}{0.059}
\resdef{E1-FPR-trust-nat-medical}{0.000}
\resdef{E1-FPR-trust-nat-nq}{0.013}
\resdef{E1-FPR-trust-natlm-financial}{0.019}
\resdef{E1-FPR-trust-natlm-legal}{0.064}
\resdef{E1-FPR-trust-natlm-medical}{0.000}
\resdef{E1-FPR-trust-nq-published}{0.008}
\resdef{E1-FPR-trust-released}{0.013}
\resdef{E1-FPR-trust-tpl-financial}{0.017}
\resdef{E1-FPR-trust-tpl-legal}{0.079}
\resdef{E1-FPR-trust-tpl-medical}{0.000}
\resdef{E1-FPR-trust-tpl-nq}{0.008}
\resdef{E1-FPR-trustrag-lm-financial}{0.664}
\resdef{E1-FPR-trustrag-lm-legal}{0.206}
\resdef{E1-FPR-trustrag-lm-medical}{0.017}
\resdef{E1-FPR-trustrag-lmns-legal}{0.199}
\resdef{E1-FPR-trustrag-lmns-medical}{0.019}
\resdef{E1-FPR-trustrag-lmsh-legal}{0.185}
\resdef{E1-FPR-trustrag-lmsh-medical}{0.013}
\resdef{E1-FPR-trustrag-nat-financial}{0.694}
\resdef{E1-FPR-trustrag-nat-legal}{0.191}
\resdef{E1-FPR-trustrag-nat-medical}{0.010}
\resdef{E1-FPR-trustrag-nat-nq}{0.153}
\resdef{E1-FPR-trustrag-natlm-financial}{0.696}
\resdef{E1-FPR-trustrag-natlm-legal}{0.190}
\resdef{E1-FPR-trustrag-natlm-medical}{0.011}
\resdef{E1-FPR-trustrag-nq-published}{0.149}
\resdef{E1-FPR-trustrag-released}{0.153}
\resdef{E1-FPR-trustrag-tpl-financial}{0.665}
\resdef{E1-FPR-trustrag-tpl-legal}{0.147}
\resdef{E1-FPR-trustrag-tpl-medical}{0.014}
\resdef{E1-FPR-trustrag-tpl-nq}{0.149}
\resdef{E1-FPR-uniform-lm-financial}{0.017}
\resdef{E1-FPR-uniform-lm-legal}{0.041}
\resdef{E1-FPR-uniform-lm-medical}{0.000}
\resdef{E1-FPR-uniform-lmns-legal}{0.037}
\resdef{E1-FPR-uniform-lmns-medical}{0.000}
\resdef{E1-FPR-uniform-lmsh-legal}{0.042}
\resdef{E1-FPR-uniform-lmsh-medical}{0.000}
\resdef{E1-FPR-uniform-nat-financial}{0.020}
\resdef{E1-FPR-uniform-nat-legal}{0.036}
\resdef{E1-FPR-uniform-nat-medical}{0.000}
\resdef{E1-FPR-uniform-nat-nq}{0.014}
\resdef{E1-FPR-uniform-natlm-financial}{0.019}
\resdef{E1-FPR-uniform-natlm-legal}{0.029}
\resdef{E1-FPR-uniform-natlm-medical}{0.000}
\resdef{E1-FPR-uniform-nq-published}{0.010}
\resdef{E1-FPR-uniform-released}{0.014}
\resdef{E1-FPR-uniform-tpl-financial}{0.017}
\resdef{E1-FPR-uniform-tpl-legal}{0.059}
\resdef{E1-FPR-uniform-tpl-medical}{0.000}
\resdef{E1-FPR-uniform-tpl-nq}{0.010}
\resdef{E1-FPR-valor-lm-financial}{0.017}
\resdef{E1-FPR-valor-lm-legal}{0.081}
\resdef{E1-FPR-valor-lm-medical}{0.000}
\resdef{E1-FPR-valor-lmns-legal}{0.062}
\resdef{E1-FPR-valor-lmns-medical}{0.000}
\resdef{E1-FPR-valor-lmsh-legal}{0.072}
\resdef{E1-FPR-valor-lmsh-medical}{0.000}
\resdef{E1-FPR-valor-nat-financial}{0.020}
\resdef{E1-FPR-valor-nat-legal}{0.059}
\resdef{E1-FPR-valor-nat-medical}{0.000}
\resdef{E1-FPR-valor-nat-nq}{0.014}
\resdef{E1-FPR-valor-natlm-financial}{0.019}
\resdef{E1-FPR-valor-natlm-legal}{0.066}
\resdef{E1-FPR-valor-natlm-medical}{0.000}
\resdef{E1-FPR-valor-nq-published}{0.010}
\resdef{E1-FPR-valor-released}{0.014}
\resdef{E1-FPR-valor-tpl-financial}{0.017}
\resdef{E1-FPR-valor-tpl-legal}{0.079}
\resdef{E1-FPR-valor-tpl-medical}{0.000}
\resdef{E1-FPR-valor-tpl-nq}{0.010}
\resdef{E1-H-trust-lm-legal}{1535 \ensuremath{\pm} 239}
\resdef{E1-H-trust-lm-medical}{82500 \ensuremath{\pm} 2264}
\resdef{E1-H-trust-nat-legal}{295 \ensuremath{\pm} 115}
\resdef{E1-H-trust-nat-medical}{900 \ensuremath{\pm} 224}
\resdef{E1-H-uniform-lm-legal}{4195 \ensuremath{\pm} 317}
\resdef{E1-H-uniform-lm-medical}{83000 \ensuremath{\pm} 2151}
\resdef{E1-H-uniform-nat-legal}{1100 \ensuremath{\pm} 371}
\resdef{E1-H-uniform-nat-medical}{1600 \ensuremath{\pm} 224}
\resdef{E1-H-valor-lm-legal}{2585 \ensuremath{\pm} 311}
\resdef{E1-H-valor-lm-medical}{82900 \ensuremath{\pm} 2219}
\resdef{E1-H-valor-nat-legal}{345 \ensuremath{\pm} 227}
\resdef{E1-H-valor-nat-medical}{900 \ensuremath{\pm} 224}
\resdef{E1-NAUROC-medical}{3}
\resdef{E1-NIDENT}{12}
\resdef{E1-NIDENT-rho1}{5}
\resdef{E1-NIDENT-total}{26}
\resdef{E1-NSAT-medical}{6}
\resdef{E1-NSATSYS-medical}{4}
\resdef{E1-NSYS-medical}{8}
\resdef{E1-Nflip-normdup-CCR-legal}{0.616 \ensuremath{\pm} 0.011}
\resdef{E1-Nflip-normdup-FPR-legal}{0.000 \ensuremath{\pm} 0.000}
\resdef{E1-Nflip-normdup-N1}{1.000}
\resdef{E1-Nflip-normdup-N5-legal}{0.343 \ensuremath{\pm} 0.010}
\resdef{E1-Nflip-trustrag-CCR-legal}{0.014 \ensuremath{\pm} 0.005}
\resdef{E1-Nflip-trustrag-FPR-legal}{0.147 \ensuremath{\pm} 0.002}
\resdef{E1-Nflip-trustrag-N1}{0.259--1.000}
\resdef{E1-Nflip-trustrag-N5-legal}{0.004 \ensuremath{\pm} 0.002}
\resdef{E1-Nflip-valor-CCR-legal}{0.000 \ensuremath{\pm} 0.000}
\resdef{E1-Nflip-valor-FPR-legal}{0.079 \ensuremath{\pm} 0.001}
\resdef{E1-Nflip-valor-N1}{0.000--0.141}
\resdef{E1-Nflip-valor-N5-legal}{0.000 \ensuremath{\pm} 0.000}
\resdef{E1-OOD-HARM}{-26.6\%}
\resdef{E1-OOD-dFPR}{+2.6}
\resdef{E1-adjudicated-poison}{12--79}
\resdef{E1-adjudication-overlap}{113--120}
\resdef{E1-agreement-table}{\ref{tab:e1-agreement}}
\resdef{E1-band-financial}{0 (0.00\%)}
\resdef{E1-band-legal}{13 (0.65\%)}
\resdef{E1-band-medical}{0 (0.00\%)}
\resdef{E1-band-nq}{0 (0.00\%)}
\resdef{E1-cleanwidth-financial}{5.60}
\resdef{E1-cleanwidth-legal}{1.00}
\resdef{E1-cleanwidth-medical}{2.56}
\resdef{E1-cleanwidth-nq}{24.38}
\resdef{E1-containment-AUROC-financial}{0.995 \ensuremath{\pm} 0.001}
\resdef{E1-containment-AUROC-general}{1.000 \ensuremath{\pm} 0.000}
\resdef{E1-containment-AUROC-legal}{0.979 \ensuremath{\pm} 0.001}
\resdef{E1-containment-AUROC-medical}{1.000 \ensuremath{\pm} 0.000}
\resdef{E1-containment-financial}{1.00}
\resdef{E1-containment-legal}{1.00}
\resdef{E1-containment-medical}{1.00}
\resdef{E1-containment-nq}{1.00}
\resdef{E1-containment-nq-natural}{0.00}
\resdef{E1-containment-nq-published}{1.00}
\resdef{E1-containment-nq-regen}{1.00}
\resdef{E1-delta-FNR-financial}{+0.000 [+0.000, +0.000]}
\resdef{E1-delta-FNR-general}{+0.000 [+0.000, +0.000]}
\resdef{E1-delta-FNR-legal}{-0.002 [-0.006, +0.000]}
\resdef{E1-delta-FNR-medical}{+0.000 [+0.000, +0.000]}
\resdef{E1-length-AUROC-financial}{0.098 \ensuremath{\pm} 0.000}
\resdef{E1-length-AUROC-general}{0.806 \ensuremath{\pm} 0.000}
\resdef{E1-length-AUROC-legal}{0.817 \ensuremath{\pm} 0.001}
\resdef{E1-length-AUROC-medical}{1.000 \ensuremath{\pm} 0.000}
\resdef{E1-lenoverlap-financial}{0.91}
\resdef{E1-lenoverlap-legal}{0.00}
\resdef{E1-lenoverlap-medical}{0.00}
\resdef{E1-lenoverlap-nq}{1.00}
\resdef{E1-lenratio-financial}{5.60}
\resdef{E1-lenratio-legal}{1.00}
\resdef{E1-lenratio-medical}{2.56}
\resdef{E1-lenratio-nq}{24.38}
\resdef{E1-lm-dFNR-financial}{+0.014}
\resdef{E1-lm-dFNR-legal}{+0.036}
\resdef{E1-lm-dFNR-medical}{+0.983}
\resdef{E1-mixed-sources}{0}
\resdef{E1-nat-dFNR-financial}{+0.421}
\resdef{E1-nat-dFNR-legal}{+0.002}
\resdef{E1-nat-dFNR-medical}{+0.009}
\resdef{E1-nat-dFNR-nq}{+0.103}
\resdef{E1-ndistinct-financial}{196}
\resdef{E1-ndistinct-legal}{400}
\resdef{E1-ndistinct-medical}{397}
\resdef{E1-ndistinct-nq}{400}
\resdef{E1-ndistinct-pre}{2}
\resdef{E1-overlap-financial}{0.91}
\resdef{E1-overlap-legal}{0.00}
\resdef{E1-overlap-medical}{0.00}
\resdef{E1-overlap-nq}{1.00}
\resdef{E1-overlap-nq-natural}{1.00}
\resdef{E1-placebo-table}{\ref{tab:placebo}}
\resdef{E1-prevalence-financial}{0.521 \ensuremath{\pm} 0.001}
\resdef{E1-prevalence-legal}{0.746 \ensuremath{\pm} 0.003}
\resdef{E1-prevalence-medical}{0.500 \ensuremath{\pm} 0.000}
\resdef{E1-prevalence-nq}{0.497 \ensuremath{\pm} 0.000}
\resdef{E1-profile-fail}{40}
\resdef{E1-profile-n}{90}
\resdef{E1-profile-table}{\ref{tab:e1-profile}}
\resdef{E1-source-id-leakage-AUROC}{1.000}
\resdef{E1-source-id-leakage-AUROC-corpus}{1.000}
\resdef{E1-source-threat-model}{dedicated Sybil sources, adjudication drawn from the evaluated stream (oracle upper bound)}
\resdef{E1-success-lm-financial}{0.04}
\resdef{E1-success-lm-legal}{0.03}
\resdef{E1-success-lm-medical}{0.83}
\resdef{E1-success-lmns-legal}{0.02}
\resdef{E1-success-lmns-medical}{0.94}
\resdef{E1-success-lmsh-legal}{0.02}
\resdef{E1-success-lmsh-medical}{0.94}
\resdef{E1-success-nat-financial}{0.33}
\resdef{E1-success-nat-legal}{0.00}
\resdef{E1-success-nat-medical}{0.01}
\resdef{E1-success-nat-nq}{0.11}
\resdef{E1-success-natlm-financial}{0.41}
\resdef{E1-success-natlm-legal}{0.20}
\resdef{E1-success-natlm-medical}{0.18}
\resdef{E1-success-tpl-financial}{0.02}
\resdef{E1-success-tpl-legal}{0.00}
\resdef{E1-success-tpl-medical}{0.00}
\resdef{E1-success-tpl-nq}{0.01}
\resdef{E1-survival-financial-lm}{1.00}
\resdef{E1-survival-financial-nat}{0.74}
\resdef{E1-survival-financial-natlm}{0.69}
\resdef{E1-survival-financial-tpl}{1.00}
\resdef{E1-survival-legal-lm}{0.96}
\resdef{E1-survival-legal-lmns}{0.91}
\resdef{E1-survival-legal-lmsh}{0.97}
\resdef{E1-survival-legal-nat}{0.88}
\resdef{E1-survival-legal-natlm}{0.50}
\resdef{E1-survival-legal-tpl}{1.00}
\resdef{E1-survival-lm-financial}{1.00}
\resdef{E1-survival-lm-legal}{0.96}
\resdef{E1-survival-lm-medical}{0.84}
\resdef{E1-survival-lmns-legal}{0.91}
\resdef{E1-survival-lmns-medical}{0.97}
\resdef{E1-survival-lmsh-legal}{0.97}
\resdef{E1-survival-lmsh-medical}{0.96}
\resdef{E1-survival-medical-lm}{0.84}
\resdef{E1-survival-medical-lmns}{0.97}
\resdef{E1-survival-medical-lmsh}{0.96}
\resdef{E1-survival-medical-nat}{0.95}
\resdef{E1-survival-medical-natlm}{0.18}
\resdef{E1-survival-medical-tpl}{1.00}
\resdef{E1-survival-nat-financial}{0.74}
\resdef{E1-survival-nat-legal}{0.88}
\resdef{E1-survival-nat-medical}{0.95}
\resdef{E1-survival-nat-nq}{0.93}
\resdef{E1-survival-natlm-financial}{0.69}
\resdef{E1-survival-natlm-legal}{0.50}
\resdef{E1-survival-natlm-medical}{0.18}
\resdef{E1-survival-nq-nat}{0.93}
\resdef{E1-survival-nq-tpl}{0.99}
\resdef{E1-survival-tpl-financial}{1.00}
\resdef{E1-survival-tpl-legal}{1.00}
\resdef{E1-survival-tpl-medical}{1.00}
\resdef{E1-survival-tpl-nq}{0.99}
\resdef{E1-theircode-allpoison-i}{14}
\resdef{E1-theircode-allpoison-qi}{86}
\resdef{E1-theircode-clean-ragdefender-i}{0.070}
\resdef{E1-theircode-clean-ragdefender-i-ci}{0.070 [0.036,0.106]}
\resdef{E1-theircode-clean-ragdefender-i-top5}{0.246}
\resdef{E1-theircode-clean-ragdefender-i-top5-ci}{0.246 [0.197,0.299]}
\resdef{E1-theircode-clean-ragdefender-qi}{0.052}
\resdef{E1-theircode-clean-ragdefender-qi-ci}{0.052 [0.020,0.090]}
\resdef{E1-theircode-clean-ragdefender-qi-top5}{0.238}
\resdef{E1-theircode-clean-ragdefender-qi-top5-ci}{0.238 [0.059,0.429]}
\resdef{E1-theircode-clean-trustrag-i}{0.122}
\resdef{E1-theircode-clean-trustrag-i-ci}{0.122 [0.078,0.170]}
\resdef{E1-theircode-clean-trustrag-i-top5}{0.077}
\resdef{E1-theircode-clean-trustrag-i-top5-ci}{0.077 [0.033,0.123]}
\resdef{E1-theircode-clean-trustrag-qi}{0.122}
\resdef{E1-theircode-clean-trustrag-qi-ci}{0.122 [0.078,0.172]}
\resdef{E1-theircode-clean-trustrag-qi-top5}{0.095}
\resdef{E1-theircode-clean-trustrag-qi-top5-ci}{0.095 [0.000,0.286]}
\resdef{E1-theircode-commit-poisonedrag}{\texttt{f660d72}}
\resdef{E1-theircode-commit-ragdefender}{\texttt{ba2a17e}}
\resdef{E1-theircode-commit-trustrag}{\texttt{11dcea0}}
\resdef{E1-theircode-empty-ragdefender-i}{0}
\resdef{E1-theircode-empty-ragdefender-i-top5}{0}
\resdef{E1-theircode-empty-ragdefender-qi}{0}
\resdef{E1-theircode-empty-ragdefender-qi-top5}{0}
\resdef{E1-theircode-empty-trustrag-i}{1}
\resdef{E1-theircode-empty-trustrag-i-top5}{15}
\resdef{E1-theircode-empty-trustrag-qi}{2}
\resdef{E1-theircode-empty-trustrag-qi-top5}{84}
\resdef{E1-theircode-left-ragdefender-i}{0.85}
\resdef{E1-theircode-left-ragdefender-i-top5}{0.97}
\resdef{E1-theircode-left-ragdefender-qi}{0.58}
\resdef{E1-theircode-left-ragdefender-qi-top5}{1.37}
\resdef{E1-theircode-left-trustrag-i}{0.10}
\resdef{E1-theircode-left-trustrag-i-top5}{0.36}
\resdef{E1-theircode-left-trustrag-qi}{0.02}
\resdef{E1-theircode-left-trustrag-qi-top5}{0.05}
\resdef{E1-theircode-nq}{100}
\resdef{E1-theircode-poison-ragdefender-i}{0.830}
\resdef{E1-theircode-poison-ragdefender-i-ci}{0.830 [0.776,0.880]}
\resdef{E1-theircode-poison-ragdefender-i-top5}{0.596}
\resdef{E1-theircode-poison-ragdefender-i-top5-ci}{0.596 [0.521,0.674]}
\resdef{E1-theircode-poison-ragdefender-qi}{0.884}
\resdef{E1-theircode-poison-ragdefender-qi-ci}{0.884 [0.832,0.932]}
\resdef{E1-theircode-poison-ragdefender-qi-top5}{0.714}
\resdef{E1-theircode-poison-ragdefender-qi-top5-ci}{0.714 [0.688,0.740]}
\resdef{E1-theircode-poison-trustrag-i}{0.980}
\resdef{E1-theircode-poison-trustrag-i-ci}{0.980 [0.966,0.992]}
\resdef{E1-theircode-poison-trustrag-i-top5}{0.850}
\resdef{E1-theircode-poison-trustrag-i-top5-ci}{0.850 [0.791,0.901]}
\resdef{E1-theircode-poison-trustrag-qi}{0.996}
\resdef{E1-theircode-poison-trustrag-qi-ci}{0.996 [0.990,1.000]}
\resdef{E1-theircode-poison-trustrag-qi-top5}{0.990}
\resdef{E1-theircode-poison-trustrag-qi-top5-ci}{0.990 [0.981,0.998]}
\resdef{E1-theircode-survival-i}{2.40}
\resdef{E1-theircode-survival-qi}{4.79}
\resdef{E1-theircode-table}{\ref{tab:their-code}}
\resdef{E1-theircode-variant-maxdiff}{0.028}
\resdef{E2-HARM-GAIN}{44\%}
\resdef{E2-HPR-financial}{1.311 \ensuremath{\pm} 0.145}
\resdef{E2-HPR-legal}{1.133 \ensuremath{\pm} 0.183}
\resdef{E2-HPR-medical}{N/A}
\resdef{E2-escalation-volume}{medical 0.500; financial 0.000; legal 0.109}
\resdef{E2-floor-FNR-lm-medical}{0.750 \ensuremath{\pm} 0.037}
\resdef{E2-floor-dFNR-natlm-medical}{-0.029 \ensuremath{\pm} 0.023}
\resdef{E2-floor-dFNR-nq}{-0.005 \ensuremath{\pm} 0.005}
\resdef{E2-floor-dFPR-lm-medical}{+1.47 \ensuremath{\pm} 0.23}
\resdef{E2-pi1-harm}{medical 0$\to$0; financial 730$\to$730; legal 200$\to$325}
\resdef{E2-regret-financial}{170.0 \ensuremath{\pm} 69.4}
\resdef{E2-regret-legal}{20.0 \ensuremath{\pm} 27.4}
\resdef{E2-regret-medical}{0.0 \ensuremath{\pm} 0.0}
\resdef{E2-rho-swap}{5 matched pairs (released artifact)}
\resdef{E2-wrapper-ppl}{50100 $\to$ 3725}
\resdef{E2-wrapper-raguard}{18460 $\to$ 3915}
\resdef{E3-B1-nstar}{19}
\resdef{E3-B1-nstar-pi005}{3}
\resdef{E3-B1-poisons-passed}{$\rho{=}1$: 0.00, $\rho{=}5$: 0.00, $\rho{=}20$: 0.00}
\resdef{E3-B1-time-to-demotion}{1 \ensuremath{\pm} 0}
\resdef{E3-B3-FNR-vs-S}{\$S\{=\}1\$: 0.000, \$S\{=\}5\$: 1.000, \$S\{=\}20\$: 1.000, \$S\{=\}50\$: 2.000}
\resdef{E3-B4-recovery}{1 \ensuremath{\pm} 0}
\resdef{E3-B5-peer-drift}{0.000 (propagation off), 0.160 (on)}
\resdef{E3-B6-false-demotion}{0.000}
\resdef{E3-B6-merged-loop}{1.000}
\resdef{E3-B6-recovery}{0}
\resdef{E3-B7-FNR-insider}{1.000}
\resdef{E3-B7-FNR-new}{0.000}
\resdef{E3-B7-benign-retention}{0.980--1.000}
\resdef{E3-eps-grid}{$\{0.02, 0.05, 0.1, 0.2, 0.5\}$}
\resdef{E3-farming-budget}{0.115}
\resdef{E3-p}{0.115}
\resdef{E3-p-pi005}{0.013}
\resdef{E3-quiet-source-drift}{+0.712}
\resdef{E4-A1-evasion}{1.000}
\resdef{E4-A2-delta-FNR}{0.000}
\resdef{E4-A3-AUROC}{0.990$\to$0.963$\to$0.518}
\resdef{E4-A3-AUROC-financial}{0.987$\to$0.740$\to$0.511}
\resdef{E4-A3-AUROC-legal}{0.999$\to$0.976$\to$0.858}
\resdef{E4-A3-AUROC-length-matched}{0.518 \ensuremath{\pm} 0.027}
\resdef{E4-A3-AUROC-length-matched-financial}{0.511 \ensuremath{\pm} 0.050}
\resdef{E4-A3-AUROC-length-matched-legal}{0.858 \ensuremath{\pm} 0.019}
\resdef{E4-A3-AUROC-length-matched-medical}{0.851 \ensuremath{\pm} 0.005}
\resdef{E4-A3-AUROC-length-matched-nq}{0.518 \ensuremath{\pm} 0.027}
\resdef{E4-A3-AUROC-medical}{1.000$\to$1.000$\to$0.851}
\resdef{E4-A3-AUROC-mimicry}{0.963 \ensuremath{\pm} 0.003}
\resdef{E4-A3-AUROC-mimicry-financial}{0.740 \ensuremath{\pm} 0.023}
\resdef{E4-A3-AUROC-mimicry-legal}{0.976 \ensuremath{\pm} 0.003}
\resdef{E4-A3-AUROC-mimicry-medical}{1.000 \ensuremath{\pm} 0.001}
\resdef{E4-A3-AUROC-mimicry-nq}{0.963 \ensuremath{\pm} 0.003}
\resdef{E4-A3-AUROC-natural}{0.993 \ensuremath{\pm} 0.001}
\resdef{E4-A3-AUROC-natural-financial}{0.893 \ensuremath{\pm} 0.005}
\resdef{E4-A3-AUROC-natural-legal}{0.996 \ensuremath{\pm} 0.000}
\resdef{E4-A3-AUROC-natural-medical}{1.000 \ensuremath{\pm} 0.000}
\resdef{E4-A3-AUROC-natural-nq}{0.993 \ensuremath{\pm} 0.001}
\resdef{E4-A3-AUROC-nq}{0.990$\to$0.963$\to$0.518}
\resdef{E4-A3-AUROC-template}{0.990 \ensuremath{\pm} 0.000}
\resdef{E4-A3-AUROC-template-financial}{0.987 \ensuremath{\pm} 0.001}
\resdef{E4-A3-AUROC-template-generic}{1.000 \ensuremath{\pm} 0.000}
\resdef{E4-A3-AUROC-template-generic-nq}{1.000 \ensuremath{\pm} 0.000}
\resdef{E4-A3-AUROC-template-legal}{0.999 \ensuremath{\pm} 0.000}
\resdef{E4-A3-AUROC-template-medical}{1.000 \ensuremath{\pm} 0.000}
\resdef{E4-A3-AUROC-template-nq}{0.990 \ensuremath{\pm} 0.000}
\resdef{E4-A3-FNR-length-matched}{0.948 \ensuremath{\pm} 0.018}
\resdef{E4-A3-FNR-length-matched-financial}{0.841 \ensuremath{\pm} 0.027}
\resdef{E4-A3-FNR-length-matched-legal}{0.497 \ensuremath{\pm} 0.055}
\resdef{E4-A3-FNR-length-matched-medical}{1.000 \ensuremath{\pm} 0.000}
\resdef{E4-A3-FNR-length-matched-nq}{0.948 \ensuremath{\pm} 0.018}
\resdef{E4-A3-FNR-mimicry}{0.523 \ensuremath{\pm} 0.038}
\resdef{E4-A3-FNR-mimicry-financial}{0.735 \ensuremath{\pm} 0.033}
\resdef{E4-A3-FNR-mimicry-legal}{0.095 \ensuremath{\pm} 0.006}
\resdef{E4-A3-FNR-mimicry-medical}{0.164 \ensuremath{\pm} 0.015}
\resdef{E4-A3-FNR-mimicry-nq}{0.523 \ensuremath{\pm} 0.038}
\resdef{E4-A3-FNR-natural}{0.153 \ensuremath{\pm} 0.011}
\resdef{E4-A3-FNR-natural-financial}{0.453 \ensuremath{\pm} 0.009}
\resdef{E4-A3-FNR-natural-legal}{0.002 \ensuremath{\pm} 0.002}
\resdef{E4-A3-FNR-natural-medical}{0.009 \ensuremath{\pm} 0.002}
\resdef{E4-A3-FNR-natural-nq}{0.153 \ensuremath{\pm} 0.011}
\resdef{E4-A3-FNR-template}{0.118 \ensuremath{\pm} 0.000}
\resdef{E4-A3-FNR-template-financial}{0.022 \ensuremath{\pm} 0.003}
\resdef{E4-A3-FNR-template-generic}{0.000 \ensuremath{\pm} 0.000}
\resdef{E4-A3-FNR-template-generic-nq}{0.000 \ensuremath{\pm} 0.000}
\resdef{E4-A3-FNR-template-legal}{0.000 \ensuremath{\pm} 0.000}
\resdef{E4-A3-FNR-template-medical}{0.000 \ensuremath{\pm} 0.000}
\resdef{E4-A3-FNR-template-nq}{0.118 \ensuremath{\pm} 0.000}
\resdef{E4-A3-retrieval-dropout}{0.457 \ensuremath{\pm} 0.021}
\resdef{E4-A3-retrieved-length-matched-financial}{0.494 \ensuremath{\pm} 0.020}
\resdef{E4-A3-retrieved-length-matched-legal}{0.757 \ensuremath{\pm} 0.015}
\resdef{E4-A3-retrieved-length-matched-medical}{0.958 \ensuremath{\pm} 0.011}
\resdef{E4-A3-retrieved-length-matched-nq}{0.862 \ensuremath{\pm} 0.030}
\resdef{E4-A3-retrieved-mimicry-financial}{0.356 \ensuremath{\pm} 0.014}
\resdef{E4-A3-retrieved-mimicry-legal}{0.451 \ensuremath{\pm} 0.012}
\resdef{E4-A3-retrieved-mimicry-medical}{0.682 \ensuremath{\pm} 0.023}
\resdef{E4-A3-retrieved-mimicry-nq}{0.543 \ensuremath{\pm} 0.021}
\resdef{E4-A3-retrieved-natural-financial}{0.730 \ensuremath{\pm} 0.008}
\resdef{E4-A3-retrieved-natural-legal}{0.881 \ensuremath{\pm} 0.004}
\resdef{E4-A3-retrieved-natural-medical}{0.948 \ensuremath{\pm} 0.003}
\resdef{E4-A3-retrieved-natural-nq}{0.778 \ensuremath{\pm} 0.006}
\resdef{E4-A3-retrieved-template-financial}{1.000 \ensuremath{\pm} 0.000}
\resdef{E4-A3-retrieved-template-generic-nq}{0.961 \ensuremath{\pm} 0.002}
\resdef{E4-A3-retrieved-template-legal}{1.000 \ensuremath{\pm} 0.000}
\resdef{E4-A3-retrieved-template-medical}{1.000 \ensuremath{\pm} 0.000}
\resdef{E4-A3-retrieved-template-nq}{0.930 \ensuremath{\pm} 0.000}
\resdef{E4-A4-ASR}{0.275}
\resdef{E4-A4-queries}{47}
\resdef{E4-A4-random}{0.200}
\resdef{E4-A5-ASR}{HotFlip 0.050, feature adversary 0.400}
\resdef{E4-A5-PPL}{45.7$\to$30.2}
\resdef{E4-A6-FNR-vs-m}{\$m\{=\}0\$: 0.118, \$m\{=\}2\$: 0.118, \$m\{=\}4\$: 0.122, \$m\{=\}6\$: 0.122}
\resdef{E4-A7-queries}{8}
\resdef{E4-A7-requery}{0.125}
\resdef{E4-C1-delta-FNR}{0.000}
\resdef{E4-D1-FNR-baselines-range}{0.069--1.000}
\resdef{E4-D1-FNR-valor}{0.977}
\resdef{E4-D1-FNR-valor-static}{0.030}
\resdef{E4-D1-FNR-valor-trust-medical}{1.000}
\resdef{E4-D1-table}{\ref{tab:e4-d1}}
\resdef{E4-budget-iters}{60}
\resdef{E4-budget-queries}{500}
\resdef{E7-AUROC-full}{0.992 \ensuremath{\pm} 0.003}
\resdef{E7-AUROC-nolength}{0.998 \ensuremath{\pm} 0.000}
\resdef{E7-AUROC-uncontrolled-all}{1.000}
\resdef{E7-ECE-full}{0.416 \ensuremath{\pm} 0.013}
\resdef{E7-ECE-nolength}{0.111 \ensuremath{\pm} 0.007}
\resdef{E7-FNR-full}{0.319 \ensuremath{\pm} 0.073}
\resdef{E7-FNR-nolength}{0.035 \ensuremath{\pm} 0.005}
\resdef{E7-FNR-uncontrolled-all}{0.000}
\resdef{E7-GBT-AUROC}{1.000 \ensuremath{\pm} 0.000}
\resdef{E7-GBT-FNR}{0.000 \ensuremath{\pm} 0.000}
\resdef{E7-ablation-table}{\ref{tab:e7-ablation}}
\resdef{E7-esc-vol}{0.284 \ensuremath{\pm} 0.038}
\resdef{E7-heldout-skeleton-AUROC}{financial 1.000; legal 0.999; medical 1.000}
\resdef{E7-noesc-FNR}{0.980 \ensuremath{\pm} 0.009}
\resdef{E7-nolength-delta-FNR}{medical/poisonedrag\_bb\_lengthmatch -0.284; medical/poisonedrag\_bb +0.000; nq/poisonedrag\_bb -0.027}
\resdef{E7NT-AUROC-full-lm}{0.885}
\resdef{E7NT-AUROC-full-lmsh}{0.962}
\resdef{E7NT-AUROC-full-nat}{1.000}
\resdef{E7NT-AUROC-full-natlm}{0.483}
\resdef{E7NT-AUROC-full-tpl}{1.000}
\resdef{E7NT-AUROC-gbt-lm}{1.000}
\resdef{E7NT-AUROC-gbt-lmsh}{1.000}
\resdef{E7NT-AUROC-gbt-nat}{1.000}
\resdef{E7NT-AUROC-gbt-natlm}{0.571}
\resdef{E7NT-AUROC-gbt-tpl}{1.000}
\resdef{E7NT-AUROC-nocontain-lm}{0.680}
\resdef{E7NT-AUROC-nocontain-lmsh}{0.817}
\resdef{E7NT-AUROC-nocontain-nat}{1.000}
\resdef{E7NT-AUROC-nocontain-natlm}{0.398}
\resdef{E7NT-AUROC-nocontain-tpl}{1.000}
\resdef{E7NT-AUROC-nolength-lm}{0.992}
\resdef{E7NT-AUROC-nolength-lmsh}{0.999}
\resdef{E7NT-AUROC-nolength-nat}{0.996}
\resdef{E7NT-AUROC-nolength-natlm}{0.671}
\resdef{E7NT-AUROC-nolength-tpl}{1.000}
\resdef{E7NT-AUROC-nolex-lm}{0.627}
\resdef{E7NT-AUROC-nolex-lmsh}{0.841}
\resdef{E7NT-AUROC-nolex-nat}{0.986}
\resdef{E7NT-AUROC-nolex-natlm}{0.486}
\resdef{E7NT-AUROC-nolex-tpl}{1.000}
\resdef{E7NT-AUROC-noset-lm}{0.957}
\resdef{E7NT-AUROC-noset-lmsh}{0.966}
\resdef{E7NT-AUROC-noset-nat}{1.000}
\resdef{E7NT-AUROC-noset-natlm}{0.785}
\resdef{E7NT-AUROC-noset-tpl}{1.000}
\resdef{E7NT-AUROC-uniform-lm}{0.885}
\resdef{E7NT-AUROC-uniform-lmsh}{0.962}
\resdef{E7NT-AUROC-uniform-nat}{1.000}
\resdef{E7NT-AUROC-uniform-natlm}{0.483}
\resdef{E7NT-AUROC-uniform-tpl}{1.000}
\resdef{E7NT-ECE-full-lm}{0.416}
\resdef{E7NT-ECE-full-lmsh}{0.473}
\resdef{E7NT-ECE-full-nat}{0.010}
\resdef{E7NT-ECE-full-natlm}{0.093}
\resdef{E7NT-ECE-full-tpl}{0.004}
\resdef{E7NT-ECE-gbt-lm}{0.000}
\resdef{E7NT-ECE-gbt-lmsh}{0.000}
\resdef{E7NT-ECE-gbt-nat}{0.005}
\resdef{E7NT-ECE-gbt-natlm}{0.097}
\resdef{E7NT-ECE-gbt-tpl}{0.000}
\resdef{E7NT-ECE-nocontain-lm}{0.418}
\resdef{E7NT-ECE-nocontain-lmsh}{0.476}
\resdef{E7NT-ECE-nocontain-nat}{0.026}
\resdef{E7NT-ECE-nocontain-natlm}{0.091}
\resdef{E7NT-ECE-nocontain-tpl}{0.006}
\resdef{E7NT-ECE-nolength-lm}{0.111}
\resdef{E7NT-ECE-nolength-lmsh}{0.017}
\resdef{E7NT-ECE-nolength-nat}{0.070}
\resdef{E7NT-ECE-nolength-natlm}{0.094}
\resdef{E7NT-ECE-nolength-tpl}{0.005}
\resdef{E7NT-ECE-nolex-lm}{0.397}
\resdef{E7NT-ECE-nolex-lmsh}{0.387}
\resdef{E7NT-ECE-nolex-nat}{0.083}
\resdef{E7NT-ECE-nolex-natlm}{0.096}
\resdef{E7NT-ECE-nolex-tpl}{0.002}
\resdef{E7NT-ECE-noset-lm}{0.420}
\resdef{E7NT-ECE-noset-lmsh}{0.477}
\resdef{E7NT-ECE-noset-nat}{0.037}
\resdef{E7NT-ECE-noset-natlm}{0.096}
\resdef{E7NT-ECE-noset-tpl}{0.002}
\resdef{E7NT-ECE-uniform-lm}{0.416}
\resdef{E7NT-ECE-uniform-lmsh}{0.473}
\resdef{E7NT-ECE-uniform-nat}{0.010}
\resdef{E7NT-ECE-uniform-natlm}{0.093}
\resdef{E7NT-ECE-uniform-tpl}{0.004}
\resdef{E7NT-FNR-full-lm}{0.986}
\resdef{E7NT-FNR-full-lmsh}{0.988}
\resdef{E7NT-FNR-full-nat}{0.010}
\resdef{E7NT-FNR-full-natlm}{1.000}
\resdef{E7NT-FNR-full-tpl}{0.000}
\resdef{E7NT-FNR-gbt-lm}{0.000}
\resdef{E7NT-FNR-gbt-lmsh}{0.000}
\resdef{E7NT-FNR-gbt-nat}{0.000}
\resdef{E7NT-FNR-gbt-natlm}{1.000}
\resdef{E7NT-FNR-gbt-tpl}{0.000}
\resdef{E7NT-FNR-nocontain-lm}{0.992}
\resdef{E7NT-FNR-nocontain-lmsh}{0.998}
\resdef{E7NT-FNR-nocontain-nat}{0.037}
\resdef{E7NT-FNR-nocontain-natlm}{0.991}
\resdef{E7NT-FNR-nocontain-tpl}{0.000}
\resdef{E7NT-FNR-nolength-lm}{0.204}
\resdef{E7NT-FNR-nolength-lmsh}{0.026}
\resdef{E7NT-FNR-nolength-nat}{0.126}
\resdef{E7NT-FNR-nolength-natlm}{1.000}
\resdef{E7NT-FNR-nolength-tpl}{0.000}
\resdef{E7NT-FNR-nolex-lm}{0.897}
\resdef{E7NT-FNR-nolex-lmsh}{0.713}
\resdef{E7NT-FNR-nolex-nat}{0.114}
\resdef{E7NT-FNR-nolex-natlm}{0.990}
\resdef{E7NT-FNR-nolex-tpl}{0.000}
\resdef{E7NT-FNR-noset-lm}{0.990}
\resdef{E7NT-FNR-noset-lmsh}{0.991}
\resdef{E7NT-FNR-noset-nat}{0.068}
\resdef{E7NT-FNR-noset-natlm}{1.000}
\resdef{E7NT-FNR-noset-tpl}{0.000}
\resdef{E7NT-FNR-uniform-lm}{0.986}
\resdef{E7NT-FNR-uniform-lmsh}{0.988}
\resdef{E7NT-FNR-uniform-nat}{0.017}
\resdef{E7NT-FNR-uniform-natlm}{1.000}
\resdef{E7NT-FNR-uniform-tpl}{0.000}
\resdef{E7NT-FNR5-full-lm}{0.550}
\resdef{E7NT-FNR5-full-lmsh}{0.280}
\resdef{E7NT-FNR5-full-nat}{0.000}
\resdef{E7NT-FNR5-full-natlm}{0.957}
\resdef{E7NT-FNR5-full-tpl}{0.000}
\resdef{E7NT-FNR5-gbt-lm}{0.000}
\resdef{E7NT-FNR5-gbt-lmsh}{0.000}
\resdef{E7NT-FNR5-gbt-nat}{0.000}
\resdef{E7NT-FNR5-gbt-natlm}{0.930}
\resdef{E7NT-FNR5-gbt-tpl}{0.000}
\resdef{E7NT-FNR5-nocontain-lm}{0.927}
\resdef{E7NT-FNR5-nocontain-lmsh}{0.824}
\resdef{E7NT-FNR5-nocontain-nat}{0.000}
\resdef{E7NT-FNR5-nocontain-natlm}{0.982}
\resdef{E7NT-FNR5-nocontain-tpl}{0.000}
\resdef{E7NT-FNR5-nolength-lm}{0.028}
\resdef{E7NT-FNR5-nolength-lmsh}{0.002}
\resdef{E7NT-FNR5-nolength-nat}{0.014}
\resdef{E7NT-FNR5-nolength-natlm}{0.923}
\resdef{E7NT-FNR5-nolength-tpl}{0.000}
\resdef{E7NT-FNR5-nolex-lm}{0.852}
\resdef{E7NT-FNR5-nolex-lmsh}{0.561}
\resdef{E7NT-FNR5-nolex-nat}{0.052}
\resdef{E7NT-FNR5-nolex-natlm}{0.981}
\resdef{E7NT-FNR5-nolex-tpl}{0.000}
\resdef{E7NT-FNR5-noset-lm}{0.201}
\resdef{E7NT-FNR5-noset-lmsh}{0.133}
\resdef{E7NT-FNR5-noset-nat}{0.000}
\resdef{E7NT-FNR5-noset-natlm}{0.758}
\resdef{E7NT-FNR5-noset-tpl}{0.000}
\resdef{E7NT-FNR5-uniform-lm}{0.550}
\resdef{E7NT-FNR5-uniform-lmsh}{0.280}
\resdef{E7NT-FNR5-uniform-nat}{0.000}
\resdef{E7NT-FNR5-uniform-natlm}{0.957}
\resdef{E7NT-FNR5-uniform-tpl}{0.000}
\resdef{E7NT-FPR-full-lm}{0.000}
\resdef{E7NT-FPR-full-lmsh}{0.000}
\resdef{E7NT-FPR-full-nat}{0.000}
\resdef{E7NT-FPR-full-natlm}{0.000}
\resdef{E7NT-FPR-full-tpl}{0.000}
\resdef{E7NT-FPR-gbt-lm}{0.000}
\resdef{E7NT-FPR-gbt-lmsh}{0.000}
\resdef{E7NT-FPR-gbt-nat}{0.000}
\resdef{E7NT-FPR-gbt-natlm}{0.000}
\resdef{E7NT-FPR-gbt-tpl}{0.000}
\resdef{E7NT-FPR-nocontain-lm}{0.000}
\resdef{E7NT-FPR-nocontain-lmsh}{0.000}
\resdef{E7NT-FPR-nocontain-nat}{0.000}
\resdef{E7NT-FPR-nocontain-natlm}{0.000}
\resdef{E7NT-FPR-nocontain-tpl}{0.000}
\resdef{E7NT-FPR-nolength-lm}{0.007}
\resdef{E7NT-FPR-nolength-lmsh}{0.006}
\resdef{E7NT-FPR-nolength-nat}{0.000}
\resdef{E7NT-FPR-nolength-natlm}{0.004}
\resdef{E7NT-FPR-nolength-tpl}{0.000}
\resdef{E7NT-FPR-nolex-lm}{0.022}
\resdef{E7NT-FPR-nolex-lmsh}{0.021}
\resdef{E7NT-FPR-nolex-nat}{0.019}
\resdef{E7NT-FPR-nolex-natlm}{0.014}
\resdef{E7NT-FPR-nolex-tpl}{0.005}
\resdef{E7NT-FPR-noset-lm}{0.000}
\resdef{E7NT-FPR-noset-lmsh}{0.000}
\resdef{E7NT-FPR-noset-nat}{0.000}
\resdef{E7NT-FPR-noset-natlm}{0.000}
\resdef{E7NT-FPR-noset-tpl}{0.000}
\resdef{E7NT-FPR-uniform-lm}{0.000}
\resdef{E7NT-FPR-uniform-lmsh}{0.000}
\resdef{E7NT-FPR-uniform-nat}{0.000}
\resdef{E7NT-FPR-uniform-natlm}{0.000}
\resdef{E7NT-FPR-uniform-tpl}{0.000}
\resdef{E7NT-GBT-AUROC-lm}{1.000}
\resdef{E7NT-GBT-AUROC-lmsh}{1.000}
\resdef{E7NT-GBT-AUROC-nat}{1.000}
\resdef{E7NT-GBT-AUROC-natlm}{0.571}
\resdef{E7NT-GBT-AUROC-tpl}{1.000}
\resdef{E7NT-GBT-ECE-lm}{0.000}
\resdef{E7NT-GBT-ECE-lmsh}{0.000}
\resdef{E7NT-GBT-ECE-nat}{0.005}
\resdef{E7NT-GBT-ECE-natlm}{0.097}
\resdef{E7NT-GBT-ECE-tpl}{0.000}
\resdef{E7NT-GBT-FNR-lm}{0.000}
\resdef{E7NT-GBT-FNR-lmsh}{0.000}
\resdef{E7NT-GBT-FNR-nat}{0.000}
\resdef{E7NT-GBT-FNR-natlm}{1.000}
\resdef{E7NT-GBT-FNR-tpl}{0.000}
\resdef{E7NT-GBT-FNR5-lm}{0.000}
\resdef{E7NT-GBT-FNR5-lmsh}{0.000}
\resdef{E7NT-GBT-FNR5-nat}{0.000}
\resdef{E7NT-GBT-FNR5-natlm}{0.930}
\resdef{E7NT-GBT-FNR5-tpl}{0.000}
\resdef{E7NT-GBT-FPR-lm}{0.000}
\resdef{E7NT-GBT-FPR-lmsh}{0.000}
\resdef{E7NT-GBT-FPR-nat}{0.000}
\resdef{E7NT-GBT-FPR-natlm}{0.000}
\resdef{E7NT-GBT-FPR-tpl}{0.000}
\resdef{E7NT-table}{\ref{tab:e7-ablation}}
\resdef{E8-hardware}{AMD Ryzen 7 5700G (16 threads), 15\,GiB RAM, PyTorch 2.14.0, CPU only}
\resdef{E8-latency-drs}{0.369}
\resdef{E8-latency-nli}{7258.4}
\resdef{E8-latency-nodefense}{0.014}
\resdef{E8-latency-normdup}{0.421}
\resdef{E8-latency-perplexity}{503.1}
\resdef{E8-latency-population}{4400--7600 queries per row, pooled over 199 record shards of every E1 lane}
\resdef{E8-latency-ragdefender}{19.07}
\resdef{E8-latency-raguard}{487.0}
\resdef{E8-latency-table}{\ref{tab:e8-latency}}
\resdef{E8-latency-trustrag}{125.0}
\resdef{E8-latency-uniform}{6.25}
\resdef{E8-latency-valor-notrust}{11.91}
\resdef{E8-latency-valor-trust}{12.22}
\resdef{E8-total-cpu-hours}{35.8}
\resdef{E9-ASR-table}{\ref{tab:e9-asr}}
\resdef{E9-ASR-valor}{0.150}
\resdef{N}{5}
\resdef{S}{5}
\resdef{SV-auroc}{1}
\resdef{SV-avfilter-containment}{0.002}
\resdef{SV-code}{3}
\resdef{SV-control}{3}
\resdef{SV-dscan-clean-words}{100}
\resdef{SV-dscan-containment}{0.001--0.005}
\resdef{SV-dscan-length-AUROC}{0.999}
\resdef{SV-dscan-length-AUROC-sample}{1.000}
\resdef{SV-dscan-poison-words}{119--162}
\resdef{SV-fpr}{9}
\resdef{SV-lengthcue}{5}
\resdef{SV-meas}{5}
\resdef{SV-n}{21}
\resdef{SV-nofpr}{12}
\resdef{SV-notprefixed}{7}
\resdef{SV-othercue}{2}
\resdef{SV-prefixed}{8}
\resdef{SV-radar-containment}{0.002}
\resdef{SV-ragorigin-clean-contained}{0}
\resdef{SV-ragorigin-clean-n}{9{,}500}
\resdef{SV-ragorigin-contained}{500}
\resdef{SV-ragorigin-n}{500}
\resdef{SV-robustrag-containment}{0.002}
\resdef{SV-undet}{6}
\resdef{X5-artifacts-table}{\ref{tab:artifacts}}
\resdef{X5-hijackrag-containment}{1.00}
\resdef{X5-ragorigin-contained}{500}
\resdef{X5-ragorigin-n}{500}
\resdef{X5-ragorigin-overlap}{1.00}
\resdef{X5-trojanrag-overlap}{0.00}
\resdef{rho-financial}{5}
\resdef{rho-general}{1}
\resdef{rho-legal}{10}
\resdef{rho-medical}{20}

\makeatletter
\def\res@bare@split#1\ensuremath#2\@nil{#1\unskip}
\newcommand{\resbare}[1]{%
  \ifcsname res@val@\detokenize{#1}\endcsname
    \expandafter\expandafter\expandafter\res@bare@split
      \csname res@val@\detokenize{#1}\endcsname\ensuremath\@nil
  \else
    \res{#1}%
  \fi}
\makeatother
\newcommand{\zkip}{RAGuard-ZKIP}      
\newcommand{\etal}{et al.}

\newcommand{\phat}{\ensuremath{\hat{p}}}                 
\newcommand{\ptilde}{\ensuremath{\tilde{p}}}             
\newcommand{\ks}{\ensuremath{\kappa_{s}}}                
\newcommand{\kmin}{\ensuremath{\kappa_{\min}}}
\newcommand{\kmax}{\ensuremath{\kappa_{\max}}}
\newcommand{\rhod}{\ensuremath{\rho_{d}}}                
\newcommand{\taud}{\ensuremath{\tau_{d}}}                
\newcommand{\tauds}{\ensuremath{\tau_{d,s}}}             
\newcommand{\CFN}{\ensuremath{C_{\mathrm{FN}}}}
\newcommand{\CFP}{\ensuremath{C_{\mathrm{FP}}}}
\newcommand{\CA}{\ensuremath{C_{\mathrm{A}}}}            
\newcommand{\pizero}{\ensuremath{\pi_{0}}}               
\newcommand{\pione}{\ensuremath{\pi_{1}}}                
\newcommand{\FNR}{\ensuremath{\mathrm{FNR}}}
\newcommand{\FPR}{\ensuremath{\mathrm{FPR}}}

\newcommand{\HPR}{\ensuremath{\mathrm{HPR}}}
\newcommand{\ECE}{\ensuremath{\mathrm{ECE}}}
\newcommand{\AUROC}{\ensuremath{\mathrm{AUROC}}}
\newcommand{\Harm}{H}
\newcommand{\Rq}{\ensuremath{\mathcal{R}(q)}}            
\newcommand{\PASS}{\textsc{pass}}
\newcommand{\BLOCK}{\textsc{block}}
\newcommand{\ESCALATE}{\textsc{escalate}}

\newcommand{\Prob}{\mathbb{P}}
\newcommand{\ind}[1]{\mathbf{1}\{#1\}}

\newcommand{\inputsection}[1]{%
  \IfFileExists{sections/#1.tex}{\input{sections/#1}}{\par\res{missing sections/#1.tex}\par}}

\begin{document}

\title{Ranking Is Not Deciding: Benchmark Shortcuts and\\
       Operating-Point Collapse in Query-Time RAG Poisoning Defenses}

\author{\IEEEauthorblockN{Pushkal Kumar}
        \IEEEauthorblockA{North Creek High School\\
        pushkalku@gmail.com}
        \and
        \IEEEauthorblockN{Arvin Sahaym}
        \IEEEauthorblockA{Washington State University\\
        arvin@wsu.edu}}

\maketitle

\begin{abstract}
Query-time filters that screen each retrieved document before it reaches the
generator are the dominant published defense against retrieval-augmented
generation poisoning, and they report near-perfect detection on domain
benchmarks. We show that this performance is predicted by how the poison set
was built rather than by what the filter reads: removing a single
construction cue destroys the deployed decision before it destroys the
ranking. Two statistics computed from a poison set and its clean
corpus alone, with no defense, scores or labels, profile the cue:
containment, how often the target question appears verbatim in the poison,
and length overlap. Containment captures a construction choice used in
PoisonedRAG: its released code prepends the target question to every inserted
passage. Of the \resbare{SV-n} published defense evaluations we survey,
\resbare{SV-prefixed} verifiably run poison built this way and
\resbare{SV-lengthcue} substitute a length cue; separately, \resbare{SV-nofpr}
report no false-positive rate at the operating point they deploy. Deleting the
prefix from the attack authors' own passages, and nothing else, leaves
\resbare{E1-survival-nat-nq} of them still retrieved for their target and
takes RAGuard, a published content filter we re-implement, from \AUROC{}
\resbare{E1-AUROC-raguard-nq-published} to \resbare{E1-AUROC-raguard-released}
and its miss rate from \resbare{E1-FNR-raguard-nq-published} to
\resbare{E1-FNR-raguard-released}. Across four corpora, four attack families
and ten per-document defenses on one shared candidate set, the profile
predicts where detection saturates. Padding medical poison to clean length
with word-shuffled filler that carries no content takes TrustRAG's set-level
stage from a miss rate of \resbare{E1-FNR-trustrag-tpl-medical} to
\resbare{E1-FNR-trustrag-lmsh-medical}, and one learned scorer of ours from
\resbare{E1-FNR-valor-tpl-medical} to \resbare{E1-FNR-valor-lmsh-medical}
(\resbare{E1-FNR5-valor-lmsh-medical} at a matched 5\% false-positive rate)
while its \AUROC{} holds at \resbare{E1-AUROC-valor-lmsh-medical} and the
share of injected poison reaching the context rises from
\resbare{E1-success-tpl-medical} to \resbare{E1-success-lmsh-medical}. The
ranking survives; the decision does not. \AUROC{} alone can conceal this
collapse, and a defense-aware attacker defeats our own scorer, so we draw no
defense claim from it. \textbf{A detection result is evidence only when it is
reported with its poison set's construction profile and the false-positive
rate at the threshold actually deployed}; we release the script that computes
the profile.
\end{abstract}



\section{Introduction}
\label{sec:intro}

\begin{figure}[t]
\centering
\input{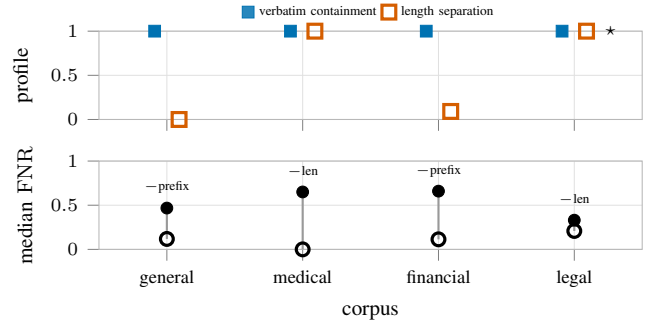}
\begin{tikzpicture}
\begin{groupplot}[group style={group size=1 by 2, vertical sep=0.55cm,
                                xticklabels at=edge bottom},
    width=0.99\columnwidth, height=2.75cm, x3base, x3rate,
    xmin=-0.5, xmax=3.5, xtick={0, 1, 2, 3}, xticklabels={general, medical, financial, legal},
    ytick={0,0.5,1}, xticklabel style={font=\scriptsize},
    legend style={at={(0.5,1.04)}, anchor=south, legend columns=-1, draw=none,
                  fill=none, font=\tiny}]
\nextgroupplot[ylabel={profile}]
\addplot[x3b, only marks, mark=square*, mark size=2.2pt]
  table[x=xc, y=containment] {figures/f5_profile.dat};
\addlegendentry{verbatim containment}
\addplot[x3c, only marks, mark=square, mark size=2.6pt, very thick]
  table[x=xl, y=separation] {figures/f5_profile.dat};
\addlegendentry{length separation}
\node[font=\scriptsize, x3a, anchor=west, xshift=3.5pt] at (axis cs:3.090,1.000) {$\star$};
\nextgroupplot[ylabel={median \FNR{}}, xlabel={corpus}]
\addplot[x3grey, mark=none, thick, forget plot] table[x=x, y=y] {figures/f5_link.dat};
\addplot[x3a, only marks, mark=o, mark size=2.4pt, very thick, forget plot]
  table[x=i, y=tpl] {figures/f5_consequence.dat};
\addplot[x3a, only marks, mark=*, mark size=2.2pt, forget plot]
  table[x=i, y=ctl] {figures/f5_consequence.dat};
\node[font=\tiny, x3a, anchor=south] at (axis cs:0,0.528) {$-$prefix};
\node[font=\tiny, x3a, anchor=south] at (axis cs:1,0.710) {$-$len};
\node[font=\tiny, x3a, anchor=south] at (axis cs:2,0.719) {$-$prefix};
\node[font=\tiny, x3a, anchor=south] at (axis cs:3,0.391) {$-$len};
\end{groupplot}
\end{tikzpicture}
\caption{\textbf{Two numbers, computable with no defense, no scores and no labels, say whether a published detection number means anything.} \emph{Upper:} each poison set's construction profile --- how often the target question occurs verbatim in the poison, and how far the poison's word lengths fall outside the clean band ($\star$: that corpus's \emph{clean} band has no width at all, so its separation is degenerate). \emph{Lower:} median miss rate at the deployed point over the six defenses of \cref{fig:ladder}, on the template rung ($\circ$) and under the one control that moves that corpus's median most ($\bullet$, named above each pair). Per-defense values and the signed one-feature \AUROC{}s are in the released data.}
\label{fig:headline}
\end{figure}

Retrieval-augmented generation (RAG) grounds a language model in documents
retrieved from an external corpus~\cite{lewis2020rag}, so the corpus is an
attack surface: an adversary who can insert text can steer the output for
chosen queries without touching model weights. Five passages per target
question suffice for over 90\% attack success~\cite{zou2025poisonedrag}, later
work needs one~\cite{zhang2026corruptrag,chang2025authchain,
shafran2025jamming} or hides the payload behind
triggers~\cite{chaudhari2024phantom,xue2024badrag}, and content-level
detectors have ``negligible impact'' on optimized attacks~\cite{zhang2025rsb}.
The response has been query-time filters that decide, before generation, which
retrieved documents may enter the context~\cite{cheng2025raguard,
kim2025ragdefender,zhou2025trustrag,shen2025reliabilityrag,xian2025drs,
chen2026radar} (\cref{sec:background}).

\noindent\textbf{Two construction numbers predict the result.}
A poison set can be profiled before any defense runs, from the poison and the
clean corpus alone, by two numbers: how often the target question occurs
verbatim as a contiguous substring of the poison document, and how far the
poison and clean word-length distributions separate. The first is not an
artifact of our regeneration. PoisonedRAG's released implementation constructs each
inserted passage as the target question followed by the generated
text~\cite{zou2025poisonedrag},\footnote{\texttt{src/attack.py}:
\texttt{adv\_text\_a = question + "."}, then \texttt{adv\_text\_a + i for i
in adv\_texts\_b}; the white-box path initialises \texttt{adv\_a = query}.}
and its released archive stores the generated half
alone. Across our four corpora the pair runs,
as containment~/~length overlap,
\res{E1-containment-nq-published}~/~\res{E1-overlap-nq} on Natural Questions
(NQ), \res{E1-containment-medical}~/~\res{E1-overlap-medical} on medical,
\res{E1-containment-financial}~/~\res{E1-overlap-financial} on financial, and
\res{E1-containment-legal}~/~\res{E1-overlap-legal} on legal, whose clean
corpus has no length variance at all
($p_{95}/p_{5} = \res{E1-cleanwidth-legal}$). Containment is thus
\res{E1-containment-medical} in every template cell of \cref{tab:e1-profile}
and on the passages PoisonedRAG's code inserts: it is the literature's
condition, not a variable of ours, and the across-corpus prediction rests on
the length number and on the one released arm where we can set containment to
zero. Deleting the prefix from the attack authors' own passages, changing
nothing else, is the leading control of this paper (\cref{fig:headline}): it
moves the best published content filter we re-implement, RAGuard, from
\AUROC{} \res{E1-AUROC-raguard-nq-published} to
\res{E1-AUROC-raguard-released}. Every saturating cell has at least one
extreme number; where both are, as on medical,
\res{E1-NSATSYS-medical} of the \res{E1-NSYS-medical} distinct score systems
miss nothing (ten defenses, eleven table rows, eight score systems: three
rows are our one learned scorer at three thresholds, \cref{sec:stack:stack}).
Every miss rate here
is reported against the false-positive rate it was bought at.

\noindent\textbf{Ranking is not deciding.}
On medical poison padded to the clean length with word-shuffled filler ---
identical length and vocabulary, no coherent content --- our learned scorer
keeps \AUROC{} \res{E1-AUROC-valor-lmsh-medical} and misses
\res{E1-FNR5-valor-lmsh-medical} at a matched 5\% false-positive rate, while
at the threshold it deploys it misses \res{E1-FNR-valor-lmsh-medical};
TrustRAG's set-level stage moves from \res{E1-FNR-trustrag-tpl-medical} to
\res{E1-FNR-trustrag-lmsh-medical} under the same edit. The ranking is
degraded; the operating point is gone; and \AUROC{}-only reporting cannot tell
the two apart. Padding with real clean prose also collapses a GPT-2
perplexity threshold~\cite{radford2019gpt2,jain2023baselinedefenses} and
RAGuard's chunk score, but by content dilution, not length: a mean-token
likelihood is set by its padding, and shuffled filler returns their miss rates
to \res{E1-FNR-ppl-lmsh} and \res{E1-FNR-raguard-lmsh}
(\cref{sec:eval:ladder}); we lead with the detectors whose collapse survives
the placebo. Once a cue has
driven score mass to the extremes every threshold rule selects the same
documents, so benchmark validity and operating-point evaluability are one
quantity measured twice.

\noindent\textbf{What this does not show.}
One of the ten defenses is our own learned scorer, the one system whose
features can be ablated. Every negative result below indicts it first and the constructed
benchmarks second: a defense-aware attacker passes \res{E4-D1-FNR-valor} of
its poisoned documents past it (\cref{sec:eval:ladder}), and we propose no
defense. The profile is validated at four corpora and no more, and at single
(corpus, attack, defense) points its direction is wrong for
\res{E1-profile-fail} of \res{E1-profile-n} (\cref{fig:profile}). Per-source
trust is a mechanism reported
beside a provenance leakage of \res{E1-source-id-leakage-AUROC}; it is not a
detection result, and nothing here is a guarantee.

\noindent\textbf{Contributions.} \textbf{(I)} A two-number construction profile
of a poison set, computable before any defense runs, that
predicts which published benchmarks saturate --- measured on regenerated
templates, on the attack authors' own released passages, on the poison sets of
\res{SV-n} published defense evaluations and on every question-targeted attack
that releases poison as text (\cref{sec:background:gap,sec:profile}).
\textbf{(II)} The dissociation the profile exposes and
its two failure modes: \emph{calibration collapse}, which leaves the ranking
usable and the deployed point dead and is partly repairable at a stated
false-positive cost, and \emph{signal collapse}, which nothing in the decision
layer repairs; with a five-rung ladder of payload-preserving edits scored under
a conjunctive criterion (\cref{sec:profile}). \textbf{(III)} The consequence
for evaluation: on the cells the profile flags, no operating-point claim ---
ours included --- is falsifiable, verified by building the best available
cost-derived rule and finding that it changes no decision in
\res{E1-NIDENT} of \res{E1-NIDENT-total} cells (\cref{sec:eval}).


\section{Background, Related Work, and Lineage}
\label{sec:background}

\subsection{Poisoning attacks on RAG}
\label{sec:background:attacks}

Attacks differ in objective (answer
manipulation~\cite{zou2025poisonedrag,zhang2026corruptrag,
zhang2024hijackrag,chang2025authchain}, jamming~\cite{shafran2025jamming},
backdoors~\cite{xue2024badrag,chaudhari2024phantom,cheng2024trojanrag},
universal corruption~\cite{zhong2023corpuspoisoning}, stance
flipping~\cite{gong2025topicfliprag}, utility
degradation~\cite{cho2024garag}, prompt injection~\cite{greshake2023ipi,
abdelnabi2025taskdrift}), in knowledge (black-box, gray-box with retriever
gradients~\cite{zhong2023corpuspoisoning,xue2024badrag,ebrahimi2018hotflip},
supply chain~\cite{cheng2024trojanrag}), and in budget, from $N$ documents per
target query~\cite{zou2025poisonedrag} down to
one~\cite{zhang2026corruptrag,chang2025authchain,shafran2025jamming,
chaudhari2024phantom,zhang2025rsb}.

Two properties matter here. First, black-box poison is
fluent text, so the standing reading is that perplexity separates only
gradient-crafted strings and jamming
blockers~\cite{zou2025poisonedrag,shafran2025jamming,chang2025authchain},
and that adversarial decoding evades it
outright~\cite{zhang2024advdecoding}. On PoisonedRAG's own NQ passages that
reading holds: a GPT-2 perplexity threshold misses
\resbare{E1-FNR-ppl-nq-published} of them with the question prefix its code
prepends and \resbare{E1-FNR-ppl-released} without it, at \AUROC{}
\resbare{E1-AUROC-ppl-nq-published} and \resbare{E1-AUROC-ppl-released}. On
poison regenerated from that attack's template for a medical corpus the same
threshold misses \resbare{E1-FNR-ppl-tpl} at \AUROC{}
\resbare{E1-AUROC-ppl-tpl}: our skeletons are perplexity-detectable and the
attack authors' text is not, which is a statement about corpora and
generators, not about the prefix. Second, attacks frame the injection as a
\emph{correction}: CorruptRAG asserts that ``the latest data confirms'' the
false answer~\cite{zhang2026corruptrag}, AuthChain adds authority cues and
timestamps~\cite{chang2025authchain}, and the Purpose attack writes poison as a
non-contradicting update~\cite{wang2026purpose} --- the surface form of a
legitimate restatement, which no construction cue marks.

\subsection{Defenses by stage and granularity}
\label{sec:background:defenses}

Defenses differ in where they act and what they decide on
(\cref{tab:defenses}). \emph{Offline}: duplicate
hashing~\cite{zou2025poisonedrag}, DRS's covariance outlier
score~\cite{xian2025drs}, provenance attestation~\cite{patil2026ragshield},
offline per-source reliability~\cite{hwang2025rarag}. \emph{Retrieval-time}:
chunk-wise perplexity and query-similarity outlier
removal~\cite{cheng2025raguard}, source-aware~\cite{koganti2026sourcererank}
and graph~\cite{zheng2025grada} re-ranking. \emph{Pre-generation}, over the
retrieved set: perplexity~\cite{jain2023baselinedefenses}, gradient-masked
token probability~\cite{kim2025gmtp}, $k$-means removal (TrustRAG
stage~1~\cite{zhou2025trustrag}), hierarchical
clustering~\cite{kim2025ragdefender}, a natural-language-inference (NLI)
contradiction graph~\cite{shen2025reliabilityrag}, min-cut
selection~\cite{chen2026radar}, conflict-free assembly~\cite{si2025seconrag},
supervised injection filters~\cite{edemacu2025filterrag,wang2026datafilter}.
\emph{Generation-time}: isolate-then-aggregate with a certified
bound~\cite{xiang2024robustrag}, rationale generation and source-aware
consolidation~\cite{wei2025instructrag,wang2025astuterag}, attention-based
filtering~\cite{ren2026dscan,choudhary2026avfilter}, capability-based
isolation~\cite{debenedetti2026camel}. \emph{Post-generation}: answer
activations~\cite{tan2025revprag} and forensic
attribution~\cite{zhang2025ragforensics,zhang2026ragorigin}.

Three published works share the name RAGuard: a misleading-retrieval
benchmark~\cite{zeng2025raguardbench} and two defenses. Here ``RAGuard'' is
Cheng \etal's query-time chunk filter~\cite{cheng2025raguard}; the other,
\zkip{}~\cite{kumar2026raguardzkip}, decides from generator outputs and exposes
no per-document decision or operating point, so it is a capability row in
\cref{tab:capability}, not a baseline.

\begin{table*}[tp]
\centering
\caption{Query-time RAG poisoning defenses, column by column.
NLL is negative log-likelihood. Stage: Off offline, Ret retrieval-time, PreG pre-generation.
Gran.: D document, RS retrieved set.
\emph{Fit?} is Yes when anything is estimated from defender data before deployment; in
rows 9--11 that is the scorer, never the threshold. \emph{FPR}: the rate nominally targeted
against the rate realised here (four corpora, template rung, PoisonedRAG, five seeds);
\emph{none}: none is specified; RAGuard's is the union bound over its three tail tests (one
two-sided, two one-sided). Every fitted threshold is a quantile of the calibration split's clean
top-$k$: $\le 200$ documents for perplexity and RAGuard, the top-50 held out in halves for DRS (\cref{sec:stack:stack}). Latency: mean per query on \res{E8-hardware} over
\res{E8-latency-population}. Rows 9--11 are one score at three thresholds.
\Cref{tab:capability} lists the systems that expose no per-document decision.}
\label{tab:defenses}
\scriptsize
\setlength{\tabcolsep}{3pt}
\begin{tabular}{@{}p{1.20in} l l p{1.28in} p{1.20in} c p{0.86in} p{0.56in}@{}}
\toprule
Defense & Stage & Gran. & Signal & Operating-point rule & Fit? & FPR nom.\,$\to$\,real. & ms/q \\
\midrule
No defense (control) & --- & --- & none & admits every candidate & No & --- $\to$ 0.000 & \res{E8-latency-nodefense} \\
Perplexity~\cite{jain2023baselinedefenses,zou2025poisonedrag} & PreG & D & GPT-2 mean token NLL & $p99$ of clean top-$k$ NLL & Yes & 1\% $\to$ \res{E1-FPR-realised-perplexity} & \res{E8-latency-perplexity} \\
RAGuard (Cheng \etal)~\cite{cheng2025raguard} & Ret/PreG & D & half-chunk NLL gap; query similarity & 2.5\% clean tail, OR of three & Yes & 10\% $\to$ \res{E1-FPR-realised-raguard} & \res{E8-latency-raguard} \\
DRS~\cite{xian2025drs} & Off & D & shift along low-variance clean eigendirections & $p99$ of held-out clean top-50 & Yes & 1\% $\to$ \res{E1-FPR-realised-drs} & \res{E8-latency-drs} \\
TrustRAG stage~1~\cite{zhou2025trustrag} & PreG & D, RS & $k$-means(2) cohesion; ROUGE-L overlap & fixed 0.88 and 0.25 & No & none $\to$ \res{E1-FPR-realised-trustrag} & \res{E8-latency-trustrag} \\
RAGDefender~\cite{kim2025ragdefender} & PreG & D, RS & Ward cluster size; TF--IDF terms; concentration & drop the $N_{\mathrm{adv}}$ closest & No & none $\to$ \res{E1-FPR-realised-ragdefender} & \res{E8-latency-ragdefender} \\
NLI consistency~\cite{shen2025reliabilityrag} & PreG & RS & cross-encoder contradiction; max independent set & fixed edge at $\beta = 0.5$ & No & none $\to$ \res{E1-FPR-realised-nli} & \res{E8-latency-nli} \\
Dup-only~\cite{zou2025poisonedrag,zhang2025rsb} & Off/PreG & D & exact-hash and cosine near-duplicate; norm test inert on unit-norm vectors & fixed cosine 0.95 & No & none $\to$ \res{E1-FPR-realised-normdup} & \res{E8-latency-normdup} \\
Same scorer at $\tau = 0.5$ & PreG & D & learned calibrated posterior \phat & fixed $\tau = 0.5$, harm-blind & Yes & none $\to$ \res{E1-FPR-realised-uniform} & \res{E8-latency-uniform} \\
This paper, $-$trust & PreG & D & the same \phat & \taud${}=1/(1{+}\rhod)$, from a stated harm ratio & Yes & none $\to$ \res{E1-FPR-realised-valor-notrust} & \res{E8-latency-valor-notrust} \\
This paper, $+$trust & PreG & D & \phat{} shifted by a Beta source posterior & \tauds${}=1/(1{+}\rhod\ks)$, \ks{} clipped & Yes & none $\to$ \res{E1-FPR-realised-valor-trust} & \res{E8-latency-valor-trust} \\
\bottomrule
\end{tabular}
\end{table*}

\subsection{Lineage of the decision rule}
\label{sec:background:lineage}

Every ingredient of the decision rule of \cref{sec:stack:decision} is old and
claimed by no one here~\cite{elkan2001costsensitive,zadrozny2001costs,
guo2017calibration,naeini2015ece,chow1970reject,drummond2006costcurves,
androutsopoulos2000costsensitivespam,axelsson2000baserate,provost2001robust,
josang2002beta,sun2006trustattacks}. We say \emph{query-time admission gate}
because observing the query is the escape Pathak and Sharma's coverage bound
names~\cite{pathak2026coverage}.

\subsection{Evaluating a security classifier}
\label{sec:background:methodology}

The failure modes measured here are catalogued by this venue's own literature:
experimental bias~\cite{pendlebury2019tesseract,chow2026beyondtesseract}, Arp
\etal's sampling bias, spurious correlations and inappropriate
baselines~\cite{arp2022dosdonts}, base-rate and closed-world
problems~\cite{sommer2010closedworld}, shortcut features a benchmark rewards
and a deployment does not~\cite{geirhos2020shortcut,risse2024limits}, and
robustness that is an artifact of how the attack was
constructed~\cite{athalye2018obfuscated,carlini2019evaluating,tramer2020adaptive}.
That a ranking metric can hide what a detector does at a deployable
false-positive rate is this community's own
lesson~\cite{carlini2022lira,jacob2025promptshield,hand2009measuring}.
The control ladder of \cref{sec:profile} follows the adaptive-evaluation recipe.

\subsection{The gap}
\label{sec:background:gap}

Over \cref{tab:defenses} and the capability-only systems of
\cref{tab:capability}---whose rules are certified budgets, graph cuts, fixed
sieves in series, or static provenance and trust weights, none a
harm-conditioned threshold---no surveyed defense conditions its
operating point on deployment harm---thresholds are clean-score
quantiles~\cite{cheng2025raguard,xian2025drs}, fixed hyperparameters, or
consistency rules with no threshold semantics~\cite{zhou2025trustrag,
kim2025ragdefender,shen2025reliabilityrag,chen2026radar}---and none maintains
per-source trust with a defined feedback
channel~\cite{hwang2025rarag,patil2026ragshield,koganti2026sourcererank}. The
third gap is measured rather than asserted (\cref{tab:survey}). Of \res{SV-n} published defense evaluations,
\res{SV-prefixed} verifiably run PoisonedRAG poison with the target question
inside it --- \res{SV-code} ship the attack's \texttt{attack.py} line unchanged,
and one releases poison in which we measure the question in
\res{SV-ragorigin-contained} of \res{SV-ragorigin-n} passages ---
\res{SV-lengthcue} substitute a length or repetition cue (a fragment repeated
2--10 times, or poison of \res{SV-dscan-poison-words} words against clean
chunks fixed at \res{SV-dscan-clean-words}, where word count alone reaches
\AUROC{} \res{SV-dscan-length-AUROC}), \res{SV-othercue} evaluate a
construction with no question in it (gradient-optimised tokens, a rewritten
gold passage), and \res{SV-undet} cannot be determined from what was
published. No false-positive rate at the deployed operating point is reported
by \res{SV-nofpr} of them; only \res{SV-auroc} reports a ranking metric at all; and
the \res{SV-control} that remove the
prefix~\cite{zhou2025trustrag,munir2026tris,mohanty2026trishieldrag} report
attack success at one fixed threshold with no false-positive side --- two
observe the filter collapse, and the third's text describes the prefix-removed
setting as its default while its released attack code has no such path. One
concurrent per-document detector reports an \AUROC{} beside a 5\%
clean-removal budget~\cite{xu2026ragsieve}, on query-inclusion poison. The
attack literature has meanwhile named query inclusion as a filterable artifact
in order to build poison without
it~\cite{jung2026camodocs,wang2026refinerag,zhang2026inceptionrag}; CamoDocs'
fixed-threshold query detector is the containment-only floor of
\cref{sec:profile}. None measures what a published filter's ranking and
deployed point each lose when the artifact is removed from the attack authors'
own text, whether the loss is calibration or signal, or the length cue. That
is this paper.

\begin{table*}[tp]
\centering
\caption{Twenty-one published RAG-poisoning defense evaluations, by the construction of the poison they report against.
\emph{Prefixed}: the target question appears verbatim in the evaluated poison (Code: verified in the released repository; Text: stated in the paper; Meas.: measured on the released artifact; ---: undetermined from what was published).
\emph{Cue instead}: the construction that stands in for the prefix where there is none.
\emph{FPR}: a false-positive rate at the deployed operating point is reported.
\emph{Control}: any evaluation with the prefix removed, and what it reports.
$n$: target queries per dataset. Every row is sourced in the artifact.}
\label{tab:survey}
\scriptsize
\setlength{\tabcolsep}{3pt}
\begin{tabular}{@{}l >{\raggedright\arraybackslash}p{1.02in} >{\raggedright\arraybackslash}p{1.62in} >{\raggedright\arraybackslash}p{0.80in} >{\raggedright\arraybackslash}p{0.95in} >{\raggedright\arraybackslash}p{0.78in} >{\raggedright\arraybackslash}p{0.70in} r@{}}
\toprule
\# & Evaluation & Poison evaluated & Prefixed & Cue instead & FPR & Control & $n$ \\
\midrule
1 & RAGuard~\cite{cheng2025raguard} & PoisonedRAG, PIA, jamming; 2 adaptive & --- (no repo) & --- & Yes, Table~I & No (paraphrase only) & 100 \\ 
2 & TrustRAG~\cite{zhou2025trustrag} & PoisonedRAG release via its \texttt{attack.py}; PIA, AdvDecoding, jamming & Yes, Code (\texttt{attack.py}:87); text says removed & --- & No (CRR, F1 only) & Yes, Table~4, ASR only & 100 \\ 
3 & RAGDefender~\cite{kim2025ragdefender} & PoisonedRAG (\texttt{LM\_targeted}), GARAG; own GPT-4 poison & Yes, Code (\texttt{attack.py}:92; \texttt{main.py}:225) & --- & Partial (NQ mis-partition rate) & No & 100 \\ 
4 & ReliabilityRAG~\cite{shen2025reliabilityrag} & own GPT-4o context, repeated $10\times$ (QA), $3\times$ (Bio); PIA with query & No, Text (App.~E.1.1) & repetition $\times$3--10 & No & No & 100--500 \\ 
5 & RobustRAG~\cite{xiang2024robustrag} & own GPT-4 poison ``following PoisonedRAG'', repeated 5--10$\times$ & No, Code+Meas. (\texttt{attack.py}:49; cont.\ \res{SV-robustrag-containment}) & repetition $\times$5--10 & N/A (not a detector) & No & 100 \\ 
6 & RevPRAG~\cite{tan2025revprag} & ``following PoisonedRAG''; GARAG; PRCAP & --- (no repo) & --- & Yes, TPR/FPR tables & No & 1{,}500 \\ 
7 & D-SCAN~\cite{ren2026dscan} & PoisonedRAG-generated, 2 of 5 docs replaced & No, Meas.\ (cont.\ \res{SV-dscan-containment}) & length: poison \res{SV-dscan-poison-words} words vs clean fixed at 100; word count alone AUROC \res{SV-dscan-length-AUROC} & No & No & 2{,}412--12{,}576 \\ 
8 & AV-Filter~\cite{choudhary2026avfilter} & PoisonedRAG / RobustRAG releases, fragment $\times 2$; PIA with question & No, Code+Meas.\ (\texttt{attack.py}:30; cont.\ \res{SV-avfilter-containment}) & repetition $\times$2 & Yes, App.~D.4 (0.05--0.45) & No & 100 \\ 
9 & RADAR~\cite{chen2026radar} & RobustRAG's poison repeated $5\times$; DeepSeek-generated dynamic poison & No, Code+Meas.\ (\texttt{attack.py}:110; cont.\ \res{SV-radar-containment}) & repetition $\times$5 & No & No & 500 \\ 
10 & TRIS~\cite{munir2026tris} & PoisonedRAG black-box, HotFlip, own ``Poison Factory'' & Yes, Text (``triggers repeat the query verbatim'') & --- & No (stated as future work) & Yes, Table~7, ASR only & 100 \\ 
11 & SecureCollaRAG~\cite{wang2026securecollarag} & SAA, PoisonedRAG, own ATA & --- (SAA: Text yes; PoisonedRAG row: no code path) & --- & Yes, Table~II & No & 50 \\ 
12 & TriShieldRAG~\cite{mohanty2026trishieldrag} & PoisonedRAG release, full 2.68M NQ corpus & Yes, Text (``target question verbatim'') & --- & No (poison block counts only) & Yes, adaptive arm, ASR only & 100 \\ 
13 & RAGForensics~\cite{zhang2025ragforensics} & PoisonedRAG-B/W, InstruInject & Yes, Text (``only the targeted query as subtext'') & --- & Yes, Table~1 & No & 50 \\ 
14 & RAGOrigin~\cite{zhang2026ragorigin} & 15 attacks incl.\ PoisonedRAG-B & Yes, Meas.\ (\res{SV-ragorigin-contained} of \res{SV-ragorigin-n} poison texts; \res{SV-ragorigin-clean-contained} of \res{SV-ragorigin-clean-n} clean) & --- & Yes, Table~III & No & 100 \\ 
15 & GMTP~\cite{kim2025gmtp} & PoisonedRAG white-box (HotFlip), Phantom, AdvDecoding & No, Text (30 masked tokens, no verbatim question) & gradient-optimised tokens & Yes, App.~H & No & 200 \\ 
16 & FilterRAG~\cite{edemacu2025filterrag} & PoisonedRAG release + GPT-4o extras & --- (no repo) & --- & No & No & 100 \\ 
17 & RAGuard-ZKIP~\cite{kumar2026raguardzkip} & own LLM rewrites of gold documents; no PoisonedRAG & No, Text & LLM rewrite of the gold passage & No & No & 33--190 \\ 
18 & PRA-RAG~\cite{tan2026prarag} & ``according to the PoisonedRAG method''; AdvDec, DoS, CorruptRAG & --- (no repo) & --- & No & No & --- \\ 
19 & SeCon-RAG~\cite{si2025seconrag} & ``following PoisonedRAG''; PIA & Yes, Code (\texttt{attack.py}:92, PoisonedRAG's line) & --- & No & No & 100 \\ 
20 & DRS~\cite{xian2025drs} & PoisonedRAG on medical QA; agent attacks & --- (no code; PDF only) & --- & Nominal only (p99 in-sample) & No & --- \\ 
21 & RSB benchmark~\cite{zhang2025rsb} & 13 attacks $\times$ 7 defenses & Yes, Text (``contains the exact targeted query'') & --- & Yes, Tables~23--37 & No & 100 \\ 
\bottomrule
\end{tabular}
\end{table*}

\section{Threat Model and the Semantics of a Control}
\label{sec:threat}

\subsection{System and trust boundary}
\label{sec:threat:system}

In a deployment $d$, a dense retriever over an index $I$ returns a candidate
set $\Rq$ for a query $q$. A \emph{query-time admission gate} --- the
defense class measured here --- scores each $x_i\in\Rq$ and emits an action
$a_i\in\{\PASS,\ESCALATE,\BLOCK\}$; only passed documents enter the generator
context. Ingestion binds each document to a source identifier $s(x_i)$. The
defender controls everything except document content; the adversary controls
the content of documents in sources it owns or has compromised. The
deployment label $d$ is a property of the query path, not of a document, so no
document field can select the threshold applied to it.

\subsection{Adversary and defender}
\label{sec:threat:adversary}

\begin{description}[leftmargin=1.6em,itemsep=1pt,parsep=0pt]
\item[G1--G3 (integrity, availability, stealth).] Have a chosen $x^\ast$ passed
  so its payload reaches the
  generator~\cite{zou2025poisonedrag,zhang2026corruptrag,zhang2024hijackrag};
  have benign documents blocked or honest sources
  down-trusted~\cite{shafran2025jamming,hoffman2009reputation,
  sun2006trustattacks}; do the first with no \ESCALATE{} or \BLOCK{}
  attributable to $x^\ast$ or $s(x^\ast)$.
\item[C1.] Insert or edit $N$ documents per target query across $S$ controlled
  identities.
\item[C2.] Choose all document text and every self-declared field; no such
  field is a scorer input.
\item[C3.] Query the system as a user and learn whether a document reached the
  context: a hard-label oracle for decision-based
  evasion~\cite{tramer2020adaptive}.
\item[C4.] Corrupt a fraction $\varepsilon$ of adjudication records
  (simulation only, \cref{app:derivations}; every pipeline result is at
  $\varepsilon=0$).
\item[C5.] Cannot alter model weights, the ledger except through C4, or $I$
  other than by insertion.
\item[C6.] Per Kerckhoffs the adversary knows the algorithm, $\rhod$, the
  thresholds and the trust update rule. Every attack reported here is black-box
  (C3 plus a surrogate encoder~\cite{reimers2019sbert,wang2020minilm}); white-box
  results are labelled where they appear~\cite{nasr2025attacker,carlini2019evaluating}.
\item[D1--D4.] Identities are bound by the identity provider, so Sybil
  resistance is inherited, not provided~\cite{douceur2002sybil}; adjudication
  is independent of the gate; and the calibration split carries no adversarial
  document.
\end{description}
Out of scope: retriever-weight and index tampering beyond
insertion~\cite{cheng2024trojanrag,dai2026disarmrag}, generator-side effects of
instructions that reach the context~\cite{greshake2023ipi}, and the veracity of
a fluent falsehood with legitimate provenance.

\subsection{What makes a rung an attack}
\label{sec:threat:control}

A \emph{control} is a poison set that removes one profiled construction cue
(\cref{sec:profile}) while holding the payload and the attack family fixed.
Each rung is a payload-preserving edit available under C1 and C2 alone: no
gradients, no surrogate, no queries to the defender. A rung that lowers
detection is therefore an attack, not dataset hygiene. A rung counts as a
\emph{successful} attack only under a conjunction: the document is retrieved
after the edited corpus is re-embedded and re-ranked, its payload is intact
against the pre-edit text, and it passes the gate. A length or lexical edit
moves the embedding, and a document that leaves $\Rq$ has evaded nothing, so
retrieval survival is re-measured at every rung and printed beside the miss
rate (\cref{sec:eval}). The published attacks we reproduce are scored under the
same conjunction.

\noindent\textbf{The defense-aware attacker.} One query-based attacker is run,
on NQ, against our own scorer: the mimicry rung (question and cue phrases
removed, length matched) followed by hard-label word substitution under C3
against the scorer's admission decision, at most \res{E4-budget-queries}
oracle queries and \res{E4-budget-iters} iterations per document, with a
22M-parameter surrogate encoder for retrieval survival. The resulting documents
are then scored by every other defense unchanged, so for those defenses the
number is a \emph{transfer} attack and a lower bound on what an attacker tuned
to each of them would achieve~\cite{tramer2020adaptive}.

\section{The Measured Stack, and the Decision Layer as Instrument}
\label{sec:stack}

Every number here comes from one harness. This section fixes it, then states
the decision layer \cref{sec:eval} uses as an instrument, not a contribution.


\subsection{Corpora, poison, and defenses}
\label{sec:stack:stack}

\noindent\textbf{Corpora.}
Four: general (NQ targets from PoisonedRAG~\cite{zou2025poisonedrag} over a
BEIR sub-corpus~\cite{thakur2021beir}, 24{,}295 documents), medical
(PubMedQA~\cite{jin2019pubmedqa}, 21{,}000), financial
(FinQA~\cite{chen2021finqa} and TAT-QA~\cite{zhu2021tatqa}, 24{,}991) and legal
(CUAD~\cite{hendrycks2021cuad} contracts chunked to a fixed 150 words, 33{,}530,
which makes the clean legal lengths constant). Each corpus carries 100 target
questions split 40/20/40 by query into scorer-train, calibration and evaluation;
every cell is evaluated on 40 queries, NQ included. Retrieval is
all-MiniLM-L6-v2~\cite{reimers2019sbert,wang2020minilm}, no reranker, $k{=}10$
candidates. The scale is small and fixed (400 candidates per cell and seed,
one 384-dimensional retriever, every clean-quantile threshold of
\cref{tab:defenses} fitted on the 20 calibration queries' clean top-$k$);
\cref{sec:discussion} states what that costs. Realised corpus poison rate \res{E0-poison-rates};
the scorer is calibrated at its training bags' candidate-level prevalence
$\pizero{=}\res{E0-pi0}$ and deployed at \res{E0-pi1-range}.

\noindent\textbf{Poison.}
Four published families: PoisonedRAG ($N{=}5$)~\cite{zou2025poisonedrag},
CorruptRAG-AS ($N{=}1$)~\cite{zhang2026corruptrag}, HijackRAG-style instruction
injection~\cite{zhang2024hijackrag} and a jamming
blocker~\cite{shafran2025jamming}. In fifteen of sixteen corpus--family cells
our generator regenerates the poison from the published template. The
sixteenth, NQ under PoisonedRAG, is the attack authors' own text, assembled the
way their code assembles it (the target question, then the released passage;
\cref{sec:profile}); deleting that prefix and changing nothing else is the
leading control of \cref{sec:eval}. Source identities are ours
--- three of the four attacks specify no source model --- and the identifier
is an exact label indicator here (provenance-only \AUROC{}
\res{E1-source-id-leakage-AUROC}), so no trust number is a detection result.

\noindent\textbf{Defenses.}
The ten defenses of \cref{tab:defenses} score one identical candidate set and
return one keep/remove decision per document; fitted rows use the condition's
own calibration split. Three rows are one learned scorer of ours at three
thresholds, and \cref{tab:e1-main} splits the last at two trust weights, so the
paper counts eleven table rows, ten defenses, and eight distinct score
systems, and every count below says which it uses. DRS is fitted as its
Algorithm~2 specifies, on the clean top-50 candidates of the 20 calibration
queries: one held-out half defines the directions and holds more documents
than the encoder has dimensions, the other supplies the 99th-percentile
threshold, which its published protocol takes in-sample with no held-out
false-positive rate (\cref{tab:defenses} prints the held-out one;
\cref{sec:discussion} the rank-deficient fit an earlier draft ran).

\subsection{Metrics and statistics}
\label{sec:stack:metrics}

Reported separately, no composite index: false-negative (\FNR) and
false-positive (\FPR) rates at each defense's deployed rule and at matched
\FPR; area under the receiver-operating-characteristic (ROC) curve (\AUROC);
expected calibration error (\ECE) over fifteen equal-width bins at the realised
prevalence~\cite{naeini2015ece} (the reliability diagram of \cref{sec:eval}
bins by equal mass instead, and says so); harm per 1{,}000 queries
$\Harm_d$~\cite{drummond2006costcurves}; and, in \cref{app:grids}, context
contamination rate (CCR), the poisoned fraction of the top-$k$.

Five seeds, $\{11,22,33,44,55\}$. Per-seed sd is not a precision claim and never
appears alone: candidate sets are seed-invariant (Jaccard
\res{E0-candidate-jaccard}), which forces it toward zero. Every comparison is a
paired cluster bootstrap on the \emph{difference} against a named comparator:
40 question clusters, each carrying all five of its seeds, 1000 resamples.
Hardware \res{E8-hardware}, \res{E8-total-cpu-hours} CPU-hours for the
main-table lanes.

\subsection{Cost-sensitive admission}
\label{sec:stack:decision}

Each candidate $x_i\in\Rq$ has an authenticated source $s(x_i)$, a latent label
$y_i$ ($1$ adversarial) and a score $\phat_i=f(x_i,q)$ estimating
$\Prob(y_i{=}1)$. Deployment $d$ prices a miss, a false block and an
adjudication at $\CFN,\CFP,\CA>0$~\cite{elkan2001costsensitive}; only
$\rhod{:=}\CFN/\CFP$ and $c_A{:=}\CA/\CFP$ enter, and \rhod{} is a governance
input, illustrative at $\{1,5,10,20\}$ and swept in \cref{app:grids}. The
objective is expected harm over the retrieval stream~\cite{zadrozny2001costs},
\begin{equation}
\begin{split}
\Harm=\sum_i\big[&\CFN\,\Prob(a_i{=}\PASS,\,y_i{=}1)\\
 &+\CFP\,\Prob(a_i{=}\BLOCK,\,y_i{=}0)\\
 &+\CA\,\Prob(a_i{=}\ESCALATE)\big],
\end{split}
\label{eq:harm}
\end{equation}
whose third term prevents escalating everything. Four assumptions.
\textbf{A1 (calibration):} $\Prob(y{=}1\mid\phat{=}t)=t$ at the prevalence where
the gate runs --- the one that fails, because the scorer is calibrated at
\pizero{} and deployed elsewhere (\cref{sec:eval} measures the \ECE).
\textbf{A2 (independence):} evidence and source identity are conditionally
independent given $y$; $f$ never sees $s(x)$ or a self-declared field (C2).
\textbf{A3 (per-document):} an individually innocuous poison set is outside a
per-document rule. \textbf{A4 (cost stationarity):} costs depend on the
deployment, not the document.

\begin{proposition}[Elkan~\cite{elkan2001costsensitive}]
\label{prop:bayes}
Under A1, A3 and A4, with $p{=}\Prob(y_i{=}1\mid\text{evidence})$ and $c_A$
unset, the decision minimising \eqref{eq:harm} pointwise is
\begin{equation}
a_i=\BLOCK\iff p\;\ge\;\taud:=\frac{1}{1+\rhod}.
\label{eq:tau}
\end{equation}
\end{proposition}

\noindent Trust enters as a prior-odds multiplier \ks{} on \phat{}
(\cref{sec:stack:system}), clipped to $[\kmin,\kmax]{=}[0.25,20]$, so under A2
the decision is again one threshold on the raw score,
\begin{equation}
\phat\;\ge\;\tauds:=\frac{1}{1+\rhod\,\ks},
\label{eq:tauds}
\end{equation}
equal to \taud{} at $\ks{=}1$; \cref{app:derivations} carries the derivation
and the closed forms. Both dials move only documents whose score lies in the
bounded band $[1/(1+\rhod\kmax),\,1/(1+\rhod\kmin)]$, which a construction cue
empties by driving score mass to the extremes (\cref{sec:eval:agreement}).
With $c_A$ set the optimum is Chow's three-way rule~\cite{chow1970reject}; no
main-table cell exercises it.

\subsection{The scored system}
\label{sec:stack:system}

Our scorer's features, the two instrumented groups first: near-verbatim query
containment --- the cue \cref{sec:profile} profiles --- and log document length;
then embeddings, further lexical features and set-context features over the
candidate set, all defender-computed from the two texts, their embeddings and
the retrieval geometry, none from self-declared metadata (C2). A
multilayer-perceptron (MLP) trunk $[512,256,128]$ over $(q,\Rq)$ bags,
temperature scaled~\cite{guo2017calibration}, with a gradient-boosted tree as an
ablation arm; a feature mask drops any group, which is what the ablation of
\cref{sec:eval} uses.

Trust is a per-source Beta posterior over adjudicated evidence, keyed to an
authenticated identity (D1) and starting at $\ks{=}1$; its update and closed
forms are \cref{app:derivations}.

\section{Cue Profiles and the Control Ladder}
\label{sec:profile}

Two quantities profile a poison set before any defense runs; a ladder of
controls removes what they find (\cref{fig:headline} states both halves in
advance). This section defines them; \cref{sec:eval} measures them.

\subsection{Two construction quantities}
\label{sec:profile:cues}

Let $P$ be a poison set built for a target question set $Q$, and $C$ the clean
corpus it is inserted into.

\emph{Verbatim containment.} Normalise a string by lower-casing, stripping
punctuation, and collapsing whitespace. A poison document $x\in P$ built for
target $q\in Q$ is \emph{contained} if the normalised $q$ occurs as a contiguous
substring of the normalised $x$. Containment is the fraction of $P$ that is
contained.

\emph{Length overlap.} Let $[\ell_5,\ell_{95}]$ be the fifth and ninety-fifth
percentiles of the word counts in $C$; length overlap is the fraction of $P$
inside that interval. We print the clean spread $\ell_{95}/\ell_5$ beside it:
with no length variance in $C$ a low overlap describes the corpus rather than
the attack, and CUAD's fixed 150-word chunking gives it a spread of one.

Neither quantity runs a defense, reads a score, or needs a held-out label
(\cref{tab:e1-profile}). Containment is \res{E1-containment-medical} wherever poison is regenerated
from a published template and \res{E1-containment-nq-published} on the one
condition whose poison we did not write, because PoisonedRAG's released
implementation prepends the target question to every inserted
passage~\cite{zou2025poisonedrag}. Nor is the profile specific to our corpora
(\cref{tab:artifacts}).
Every question-targeted attack that releases poison as text inserts it at full
containment by its own construction: PoisonedRAG on three corpora, HijackRAG
at \res{X5-hijackrag-containment} (``we set
$R=q_i$'')~\cite{zhang2024hijackrag}, and CorruptRAG, whose authors state the
convention outright ($p^s_i=q_i$)~\cite{zhang2026corruptrag}; a second group's independent run of
PoisonedRAG carries the question in
\res{X5-ragorigin-contained} of \res{X5-ragorigin-n} passages at length
overlap \res{X5-ragorigin-overlap}~\cite{zhang2026ragorigin}. The one released
artifact with no question cue, TrojanRAG, targets fixed-width Wikipedia chunks
and sits at length overlap \res{X5-trojanrag-overlap}~\cite{cheng2024trojanrag}.
A defense evaluated by running PoisonedRAG sees containment
\res{E1-containment-nq-published}; one evaluated on the fragment its archive
stores sees \res{E1-containment-nq-natural}; \cref{tab:survey} says which each
published evaluation did.

Length overlap cuts the other way: \res{E1-lenoverlap-medical} on medical
against \res{E1-lenoverlap-financial} on financial and \res{E1-overlap-nq} on
NQ, whose clean band is \res{E1-cleanwidth-nq} times as wide at its
ninety-fifth percentile as at its fifth. Overlap is coarse, so we print the \emph{signed} one-feature \AUROC{} of
each cue beside it, never its absolute value: length alone is worth
\resbare{E1-length-AUROC-medical} on medical against
\resbare{E1-length-AUROC-financial} on financial, where the poison is the longer
class and normalising the sign would report no cue at all.

\begin{table}[tbp]
\centering\scriptsize\setlength{\tabcolsep}{1.5pt}
\caption{Poison-set construction profile: two quantities computable from a poison set and its clean corpus alone.}
\label{tab:e1-profile}
\begin{tabular}{lcccccc}
\toprule
Corpus & PoisonedRAG & CorruptRAG & HijackRAG & jam. & words & clean \\
\midrule
nq & 1.00/1.00 & 1.00/1.00 & 1.00/1.00 & 1.00/1.00 & 33/37/43 & 24.38 \\
medical & 1.00/0.00 & 1.00/0.00 & 1.00/0.00 & 1.00/0.00 & 37/42/50 & 2.56 \\
financial & 1.00/0.91 & 1.00/1.00 & 1.00/1.00 & 1.00/1.00 & 38/46/57 & 5.60 \\
legal & 1.00/0.00 & 1.00/0.00 & 1.00/0.00 & 1.00/0.00 & 41/85/183 & 1.00 \\
nq $-$prefix & 0.00/1.00 & N/A & N/A & N/A & 25/28/32 & 24.38 \\
\bottomrule
\end{tabular}
\\[2pt]
{\footnotesize Each cell is containment / length overlap (\cref{sec:profile:cues}); \emph{words} is the realised $p_5/p_{50}/p_{95}$ word count of the headline attack's poison and \emph{clean} the clean corpus's own $p_{95}/p_5$, at 1.00 a corpus with no length variance, where a length cue is free. \texttt{nq} is the one corpus whose poison the attack's authors wrote, assembled as their released code inserts it: the target question, a full stop, then the released text. The \emph{prefix removed} row is those same passages with that prefix deleted and nothing else changed. Length overlap is interpolated from the recorded $\{\min, p_5, p_{50}, p_{95}, \max\}$ ladder in financial (poisonedrag), legal (corruptrag), legal (poisonedrag); elsewhere the poison distribution lies wholly inside or wholly outside the clean band and the value is exact.}
\end{table}

\begin{table}[tbp]
\centering\scriptsize\setlength{\tabcolsep}{3pt}
\caption{Released poison of five attacks under the two profile quantities of \cref{sec:profile}. Every question-targeted artifact is inserted at containment 1.00 by the attack's own construction; the one artifact with no question cue sits at length overlap 0.00.}
\label{tab:artifacts}
\begin{tabular}{lccccc}
\toprule
artifact & $n$ & cont. & overlap & poison words & clean words \\
\midrule
PoisonedRAG NQ, archive ($I$) & 500 & 0.00 & 1.00 & 24/28/32 & 8/81/195 \\
PoisonedRAG NQ, inserted ($S{\oplus}I$) & 500 & 1.00 & 1.00 & 31/36/41 & 8/81/195 \\
PoisonedRAG HotpotQA, inserted & 500 & 1.00 & 1.00 & 34/42/57 & 10/79/182 \\
PoisonedRAG MS~MARCO, inserted & 500 & 1.00 & 0.97 & 28/32/37 & 26/52/105 \\
PoisonedRAG NQ, 3rd party~\cite{zhang2026ragorigin} & 500 & 1.00 & 1.00 & 24/30/40 & 18/70/213 \\
HijackRAG, $R{=}q_i \oplus H$~\cite{zhang2024hijackrag} & 2{,}251 & 1.00 & 1.00 & 19/27/42 & 8/81/195 \\
TrojanRAG, DPR chunks~\cite{cheng2024trojanrag} & 60 & 0.00 & 0.00 & 10/28/36 & 100/100/100 \\
Topic-FlipRAG (topic-level)~\cite{gong2025topicfliprag} & 45 & 0.02 & 1.00 & 40/66/119 & 25/66/209 \\
\bottomrule
\end{tabular}
\\[2pt]
{\footnotesize Containment and overlap as in \cref{tab:e1-profile} (overlap: the ladder interpolation on the word-length p5/p50/p95 ladders, with the exact fraction inside the clean $[p_5,p_{95}]$ band in parentheses where the two differ by more than 0.03); word counts are p5/p50/p95. Clean corpus: the paper's BEIR sub-corpus for PoisonedRAG and HijackRAG, the file's own retrieved clean passages for the third-party run, the file's own DPR negatives for TrojanRAG, and the attack's own database for Topic-FlipRAG. HijackRAG released $H$ only, so its row assembles $R\oplus H$ with PoisonedRAG's NQ questions as stand-ins; Topic-FlipRAG has no target question and shows the profile flags nothing there. The artifact's profiling script recomputes every row.}
\end{table}

The consequence, as a rule that applies to a paper and not only to our data:

\begin{quote}
A defense reporting near-perfect detection on a poison set whose containment
rate is $1.00$, or whose poison length distribution lies outside the clean
$[\ell_5,\ell_{95}]$, has published no evidence that it uses anything else.
\end{quote}

\noindent\textbf{A floor for the profile.} Two reference detectors read one
scalar each: \emph{length-only} scores $-\log(1+|\mathrm{tok}(x)|)$ and
\emph{containment-only} scores $\ind{\mathrm{norm}(q)\subseteq\mathrm{norm}(x)}$
with a smaller token-coverage term breaking ties. Each is one input of our own
scorer's feature vector and fits a threshold as every fitted defense here
does; neither is a defense, and both bound what a detection number is worth.

\subsection{The control ladder}
\label{sec:profile:ladder}

A control removes one profiled cue while holding payload and attack family
fixed. \textbf{Template:} the published generator, unmodified.
\textbf{Prefix removed:} every verbatim occurrence of the target question
deleted and query-token coverage capped, payload untouched; containment falls to
zero. On NQ that rung is applied to the attack authors' own passages, deleting
the prefix their code writes and nothing else, which is why it is the control
\cref{sec:eval} leads with. \textbf{Length matched:} poison padded with benign
filler from the clean corpus until its word count lies inside
$[\ell_5,\ell_{95}]$; two placebo fillers of the same length, the clean
sentences with their words shuffled and pseudo-words, separate length from
content (\cref{sec:eval:ladder}). \textbf{Both.} A fifth rung,
\textbf{mimicry}, edits a real domain passage so that only the target fact
changes~\cite{thornton2026chameleon,zhang2026corruptrag}, and is length-matched
in turn.

Every rung is scored under the conjunctive criterion of
\cref{sec:threat:control}~\cite{carlini2019evaluating,tramer2020adaptive},
with retrieval survival, its first leg, printed under every rung of
\cref{fig:ladder}. For adjacent rungs $r,r'$ and defense $f$, the
detection attributable to the cue $r'$ removes is $\FNR_{r'}(f)-\FNR_{r}(f)$ at
the deployed point, what the deployment loses, and the same difference at
matched $\FPR = 5\%$, what the ranking loses; every table below prints both.

\noindent\textbf{Template leakage and hold-out.} Domain-regime scorers train on
a disjoint half of our skeletons (\cref{sec:stack}) and are evaluated on the
other; the only shared structure is the verbatim prefix and the answer slot,
which the prefix-removed and mimicry rungs delete. The general-regime scorer
trains on HotpotQA and MS~MARCO poison only, so both NQ arms are scored by a
model that has seen neither their corpus nor their generator
(held-out-skeleton \AUROC{} \res{E7-heldout-skeleton-AUROC}, on the
calibration split, so it certifies only a shared register).

\noindent\textbf{What the protocol does not establish.}
\label{sec:profile:limits}
Four corpora do not validate a predictor, so we report the per-point scatter
(\cref{fig:profile}) rather than a correlation coefficient; the profile's
direction is wrong at \res{E1-profile-fail} of \res{E1-profile-n}
(corpus, attack, defense) points there, so it predicts which corpora saturate,
not which defense does on which cell. A clean profile is necessary for an
informative benchmark, not sufficient: unprofiled cues may remain, and
surviving all five rungs certifies nothing beyond those five edits.

\section{Results}
\label{sec:eval}

Five questions, one per subsection, each closing with a one-sentence verdict;
statistics throughout are \cref{sec:stack:metrics}'s.

\subsection{Prediction, and the published construction}
\label{sec:eval:prediction}

The cleanest control here needs no defense-side change and no poison we
wrote: the NQ passages as PoisonedRAG's own code inserts them (containment
\res{E1-containment-nq-published}, what a defense evaluated by running that
attack saw) against the same passages with that prefix deleted
(\res{E1-containment-nq-natural}), every other word, question, retriever and
candidate set held fixed.

\begin{table*}[tp]
\centering\scriptsize\setlength{\tabcolsep}{3pt}
\caption{Eleven rows, ten defenses, eight score systems (\cref{sec:stack:stack}) and two one-feature reference detectors on one shared candidate set, each cell the miss rate, the false-positive rate it was bought at, and the miss rate at a matched 5\% false-positive rate. The general columns are the one condition whose poison the attack's own authors wrote, assembled the way their code assembles it.}
\label{tab:e1-main}
\begin{tabular}{lccccc}
\toprule
Defense & general & general, $-$prefix & medical & financial & legal \\
\midrule
No defense & 1.000 / 0.000 / N/A & 1.000 / 0.000 / N/A & 1.000 / 0.000 / N/A & 1.000 / 0.000 / N/A & 1.000 / 0.000 / N/A \\
Perplexity (GPT-2) & 1.000 / 0.005 / 0.809 & 1.000 / 0.005 / 0.866 & 0.062 / 0.000 / 0.046 & 1.000 / 0.005 / 1.000 & 0.551 / 0.000 / 0.392 \\
RAGuard (Cheng) & 0.191 / 0.065 / 0.216 & 0.763 / 0.061 / 0.796 & 0.000 / 0.115 / 0.000 & 0.146 / 0.104 / 0.309 & 0.078 / 0.108 / 0.087 \\
DRS & 0.854 / 0.154 / 0.970 & 0.930 / 0.164 / 0.995 & 0.700 / 0.035 / 0.635 & 0.833 / 0.094 / 0.904 & 0.816 / 0.128 / 0.897 \\
TrustRAG stage 1 & 0.000 / 0.149 / N/A & 0.070 / 0.153 / N/A & 0.000 / 0.014 / N/A & 0.009 / 0.665 / N/A & 0.004 / 0.147 / N/A \\
RAGDefender & 0.045 / 0.025 / N/A & 0.172 / 0.117 / N/A & 0.000 / 0.055 / N/A & 0.081 / 0.119 / N/A & 0.339 / 0.029 / N/A \\
Dup-only & 0.678 / 0.005 / N/A & 0.989 / 0.009 / N/A & 0.496 / 0.000 / N/A & 0.788 / 0.150 / N/A & 0.343 / 0.000 / N/A \\
Ours, $\tau{=}0.5$ & 0.015 / 0.010 / 0.000 & 0.118 / 0.014 / 0.043 & 0.000 / 0.000 / 0.000 & 0.025 / 0.017 / 0.017 & 0.002 / 0.059 / 0.005 \\
Ours, $\tau_d$ & 0.015 / 0.010 / 0.000 & 0.118 / 0.014 / 0.043 & 0.000 / 0.000 / 0.000 & 0.025 / 0.017 / 0.017 & 0.000 / 0.079 / 0.005 \\
Ours, $\tau_{d,s}$ ($a{=}0.3$) & 0.012 / 0.008 / 0.000 & 0.086 / 0.013 / 0.009 & 0.000 / 0.000 / 0.000 & 0.025 / 0.017 / 0.017 & 0.000 / 0.079 / 0.003 \\
Ours, $\tau_{d,s}$ ($a{=}0.1$) & 0.015 / 0.007 / 0.000 & 0.104 / 0.014 / 0.025 & 0.000 / 0.000 / 0.000 & 0.025 / 0.017 / 0.017 & 0.000 / 0.079 / 0.003 \\
NLI (ReliabilityRAG) & 0.935 / 0.458 / — & 0.866 / 0.388 / — & 1.000 / 0.185 / — & 0.961 / 0.532 / — & 0.793 / 0.507 / — \\
length-only (ref.) & 1.000 / 0.000 / 1.000 & 1.000 / 0.000 / 1.000 & 0.000 / 0.000 / 0.000 & 1.000 / 0.000 / 1.000 & 0.992 / 0.000 / 0.183 \\
containment-only (ref.) & 0.000 / 0.025 / 0.000 & 0.882 / 0.023 / 0.839 & 0.000 / 0.005 / 0.000 & 0.009 / 0.298 / 0.022 & 0.077 / 0.029 / 0.039 \\
\bottomrule
\end{tabular}
\\[2pt]
{\footnotesize Each cell is FNR / FPR at the deployed operating point / FNR at a matched 5\% FPR (the third number is omitted, or N/A, where the defense emits no score), lower is better, point estimates over seeds [11, 22, 33, 44, 55], attack \texttt{poisonedrag\_bb}; per-seed sd is not a precision claim (\cref{sec:stack:metrics}); dispersion and every paired cluster-bootstrap interval are in the appendix grids (\cref{tab:e1-delta}). The two reference rows read one scalar each (word count; verbatim occurrence of the target question) and are the floor a defense has to clear, not defenses; Dup-only appears under both its lane names and the learned score at two trust weights, so the defense rows are the eleven configurations of \cref{tab:defenses}. The DRS row was re-measured after a defect in our port: the earlier fit calibrated on 199 clean documents, fewer than the 384 embedding dimensions, and took its threshold cross-fit on fold directions that differ from the deployed ones; every DRS cell here uses the corrected fit (clean top-50 of the 20 calibration queries, held-out threshold; \cref{sec:discussion}), which moves its clean FPR medical 0.945 $\to$ 0.035, financial 0.669 $\to$ 0.094, legal 0.517 $\to$ 0.128, nq 0.985 $\to$ 0.154. The $-$prefix column is the released NQ poison with the question prefix its code prepends deleted and nothing else changed (\cref{tab:e1-prefix} in \cref{app:grids} adds AUROC).}
\end{table*}

\begin{table}[tbp]
\centering\scriptsize\setlength{\tabcolsep}{3pt}
\caption{Their code on their poison: the released TrustRAG and RAGDefender filters on PoisonedRAG's released NQ passages, with and without the question prefix its code prepends.}
\label{tab:their-code}
\begin{tabular}{lcccc}
\toprule
Defense & P & poison removed & clean removed & left / query \\
\midrule
TrustRAG-S1 (10) & $Q{\oplus}I$ & 0.996 [0.990,1.000] & 0.122 [0.078,0.172] & 0.02 \\
 & $I$ & 0.980 [0.966,0.992] & 0.122 [0.078,0.170] & 0.10 \\
TrustRAG-S1 (top-5) & $Q{\oplus}I$ & 0.990 [0.981,0.998] & 0.095 [0.000,0.286] & 0.05 \\
 & $I$ & 0.850 [0.791,0.901] & 0.077 [0.033,0.123] & 0.36 \\
RAGDefender (10) & $Q{\oplus}I$ & 0.884 [0.832,0.932] & 0.052 [0.020,0.090] & 0.58 \\
 & $I$ & 0.830 [0.776,0.880] & 0.070 [0.036,0.106] & 0.85 \\
RAGDefender (top-5) & $Q{\oplus}I$ & 0.714 [0.688,0.740] & 0.238 [0.059,0.429] & 1.37 \\
 & $I$ & 0.596 [0.521,0.674] & 0.246 [0.197,0.299] & 0.97 \\
\bottomrule
\end{tabular}
\\[2pt]
{\footnotesize Released repositories, imported unchanged and called as their own drivers call them: TrustRAG \texttt{k\_mean\_filtering} at its default (\texttt{kmeans\_ngram} over SimCSE; commit \texttt{11dcea0}) and RAGDefender \texttt{RAGDefender.defend} with its \texttt{artifacts/main.py} embedder (\texttt{ba2a17e}), on PoisonedRAG's released NQ poison (100 targets, five passages each; \texttt{f660d72}) and its released clean Contriever top-5, each passage once as \texttt{src/attack.py} inserts it ($Q\oplus I$) and once as the archive stores it ($I$). \emph{(10)}: the clean top-5 plus the five poison passages, the candidate set of \cref{tab:e1-main}; \emph{(top-5)}: the top five of those by their Contriever score, what their code filters, holding 4.79 of the five poison passages per query under $Q\oplus I$ (all five in 86 of 100 queries) and 2.40 under $I$ (all five in 14). Rates are shares of the released poison / clean passages removed [95\% query bootstrap]. Under $Q\oplus I$ TrustRAG returns an empty context on 84 queries.}
\end{table}

\Cref{tab:e1-main} gives every defense on one shared candidate set; the
paired cluster-bootstrap differences (\cref{tab:e1-delta}) are in
\cref{app:grids}. On templated medical
poison --- containment~\res{E1-containment-medical}, poison length disjoint
from the clean distribution (\cref{tab:e1-profile}) ---
\res{E1-NSATSYS-medical} of the \res{E1-NSYS-medical} distinct score systems
miss nothing (\FNR{} \resbare{E1-FNR-medical}) and \res{E1-NAUROC-medical} score
families reach \AUROC{} \res{E1-AUROC-medical}. Two systems saturate
nowhere. DRS~\cite{xian2025drs}, fitted as
its Algorithm~2 specifies, misses \res{E1-FNR-drs-medical} at \FPR{}
\res{E1-FPR-drs-medical} and peaks at \AUROC{} \resbare{E1-AUROC-drs-medical};
the near-zero miss rate an earlier draft printed for it, bought by blocking
most of the benign corpus, was our port's rank-deficient fit, not DRS
(\cref{tab:e1-main}'s note, \cref{sec:discussion}). The
norm/duplicate filter misses \resbare{E1-FNR-normdup-medical}. Saturation is what
the profile predicts; competence is not.

\begin{figure}[t]
\centering
\input{figures/x3_style}
\begin{tikzpicture}
\begin{axis}[x3base, x3rate, width=0.99\columnwidth, height=5.2cm,
    xmode=log, log basis x=10,
    xmin=0.001, xmax=1.0,
    xlabel={realised \FPR{} at the defense's own deployed rule},
    ylabel={\FNR{} at that same rule},
    legend to name=x3f8, legend cell align=left,
    legend style={legend columns=-1, draw=none, fill=none}]
\addplot[draw=none, fill=x3grey, fill opacity=0.35, forget plot] coordinates {(0.001,0) (0.0018,0) (0.0018,1) (0.001,1)} \closedcycle;
\addplot[draw=none, fill=x3c, fill opacity=0.10, forget plot] coordinates {(0.5,0) (1,0) (1,1) (0.5,1)} \closedcycle;
\node[font=\tiny, x3c, anchor=north east] at (axis cs:0.98,0.985) {blocking, not detecting};
\draw[x3a, densely dotted] (axis cs:0.05,0) -- (axis cs:0.05,1);
\node[font=\tiny, x3a, anchor=south west, inner sep=1.5pt] at (axis cs:0.05,0.86) {matched-\FPR{} ruler, 5\%};
\node[font=\tiny, x3grey, anchor=center, rotate=90] at (axis cs:0.00134164,0.74) {\FPR{} $=0$};
\addplot[x3b, only marks, mark=asterisk, mark size=1.9pt, opacity=0.85]
  table[x=fprplot, y=fnr] {figures/f8_nq.dat};
\addlegendentry{nq}
\addplot[x3b, only marks, mark=square*, mark size=1.9pt, opacity=0.85]
  table[x=fprplot, y=fnr] {figures/f8_medical.dat};
\addlegendentry{medical}
\addplot[x3b, only marks, mark=triangle*, mark size=1.9pt, opacity=0.85]
  table[x=fprplot, y=fnr] {figures/f8_financial.dat};
\addlegendentry{financial}
\addplot[x3b, only marks, mark=diamond*, mark size=1.9pt, opacity=0.85]
  table[x=fprplot, y=fnr] {figures/f8_legal.dat};
\addlegendentry{legal}
\addplot[x3a, only marks, mark=o, mark size=3.2pt, thick]
  table[x=fprplot, y=fnr] {figures/f8_reference.dat};
\addlegendentry{one-feature reference detector}
\draw[x3grey, thin] (axis cs:0.005000,0.000000) -- ([yshift=32.0pt] axis cs:0.005000,0.000000);
\node[font=\tiny, x3a, anchor=south west, inner sep=2pt, yshift=32.0pt] at (axis cs:0.005000,0.000000) {containment-only, medical: 0.000 at 0.005};
\addplot[x3a, only marks, mark=square, mark size=3.6pt, thick, forget plot] coordinates {(0.664572,0.008631)};
\draw[x3grey, thin] (axis cs:0.664572,0.008631) -- ([yshift=24.0pt] axis cs:0.664572,0.008631);
\node[font=\tiny, x3a, anchor=south east, inner sep=2pt, yshift=24.0pt] at (axis cs:0.664572,0.008631) {TrustRAG-1, financial: 0.009 at 0.665};
\end{axis}
\end{tikzpicture}
\par\smallskip
\pgfplotslegendfromname{x3f8}
\caption{\textbf{``Near-perfect detection'' is bought at a false-positive rate nobody reports.} One mark per (corpus, defense) on the rung the literature reports against, each at the operating point that defense deploys for itself; two marks carry their own numbers. The same near-zero miss rate is realised across more than two orders of magnitude of false-positive rate --- at one end by blocking most of the benign corpus, at the other by a single boolean feature. Right-hand shading: \cref{sec:report}'s rule that a zero miss rate above $\FPR{}=0.5$ is a closed door and not a detector; dotted rule, its matched-$\FPR{}$ ruler. Marks in the grey floor band realise \emph{exactly} zero false positives, pinned so a log abscissa can show them.}
\label{fig:operating}
\end{figure}

\Cref{fig:operating} puts every defense at the point it chooses for itself:
near-zero miss rates are bought across more than two orders of magnitude of
\FPR{}, the cheapest of them by a single boolean feature, and where a zero
miss rate costs an \FPR{} above one half we read the cell as blocking, not
detecting (\cref{sec:report}). The profile's four values predict four outcomes
(\cref{fig:headline}): medical, containment with disjoint lengths, is where
every content defense that saturates anywhere saturates; legal's constant
clean width makes a length cue free; financial, containment with overlapping
lengths, is where the prefix alone collapses the deployed point; NQ,
containment \res{E1-containment-nq-published} against overlap
\res{E1-overlap-nq} on the attack authors' own text, is the diagnostic cell.

Delete the prefix and the picture inverts (\cref{tab:e1-main}, $-$prefix
column; \cref{tab:e1-prefix} in \cref{app:grids} adds every \AUROC{}).
RAGuard, the best published content filter we re-implement, falls from \AUROC{}
\res{E1-AUROC-raguard-nq-published} on the passage the code inserts to
\res{E1-AUROC-raguard-released} without the prefix, and its deployed miss rate
from \res{E1-FNR-raguard-nq-published} to \res{E1-FNR-raguard-released}: there
both ranking and decision go, and neither of the other two scored baselines
does better. Our own scorer, trained on HotpotQA and MS~MARCO poison and never
on NQ, holds \AUROC{} \res{E1-AUROC-valor-released} at a miss rate of
\res{E1-FNR-valor-released} --- not because it reads content, but because the
released passages are short in a corpus whose clean band is
\res{E1-cleanwidth-nq} times as wide at its ninety-fifth percentile as at its
fifth (length alone is worth \AUROC{} \resbare{E1-length-AUROC-general} there), and
the mimicry-plus-length rung below takes it to
\resbare{E4-A3-AUROC-length-matched}. Survival of the prefix-less passages is
\res{E1-survival-nat-nq} here; at PoisonedRAG's million-document scale it
would be lower, the attacker's problem and not the filter's evidence.

\noindent\textbf{Their code, their poison.} Fifteen of our sixteen cells
carry poison we regenerated and seven baselines are our ports, so every
separability number on them is an upper bound; \cref{tab:their-code} has
neither: the released code of TrustRAG's stage~1 and of RAGDefender,
unmodified and at their own defaults, on PoisonedRAG's released NQ poison and
clean top-5 under both constructions.
On the ten-candidate set of \cref{tab:e1-main}, TrustRAG removes
\res{E1-theircode-poison-trustrag-qi} of the poison as the attack's code
inserts it and \res{E1-theircode-poison-trustrag-i} as the archive stores it
(clean removal \res{E1-theircode-clean-trustrag-qi}); RAGDefender,
\res{E1-theircode-poison-ragdefender-qi} against
\res{E1-theircode-poison-ragdefender-i}. Our port agrees (\res{E1-FNR-trustrag-nq-published} and
\res{E1-FNR-trustrag-released} missed in \cref{tab:e1-main}), and no
embedding or gate variant moves a rate by more than
\res{E1-theircode-variant-maxdiff}. Neither reads containment: both are
set-level redundancy filters, and five mutually similar passages per target
is PoisonedRAG's construction as much as the prefix is, and survives its
deletion. The prefix acts on them upstream, in their own retrieval: their
Contriever top-5 holds \res{E1-theircode-survival-qi} of the five poison
passages per query with it and \res{E1-theircode-survival-i} without, and
removal among what is retrieved falls to
\res{E1-theircode-poison-trustrag-i-top5} and
\res{E1-theircode-poison-ragdefender-i-top5}, leaving
\res{E1-theircode-left-trustrag-i-top5} and
\res{E1-theircode-left-ragdefender-i-top5} poison passages per query. With
the prefix that top-5 is all poison on \res{E1-theircode-allpoison-qi} of
\res{E1-theircode-nq} queries and TrustRAG returns an empty context on
\res{E1-theircode-empty-trustrag-qi-top5}: on the literature's construction
it blanks the context rather than discriminating within it, which no removal
rate shows. The prefix is the cue of the per-document content filters; the
set-level filters read a second construction cue, redundancy, that a
$-$prefix control leaves in place.

\noindent\textbf{The floor.} The one-feature reference detectors of
\cref{sec:profile} reach \AUROC{} \resbare{E1-length-AUROC-medical} and
\resbare{E1-containment-AUROC-medical} on templated medical poison and
\resbare{E1-length-AUROC-general} and \resbare{E1-containment-AUROC-general} on the
published NQ construction (\cref{fig:dissociation}, \cref{app:grids}); on a
profiled corpus no row of \cref{tab:e1-main} clears that floor.

\textbf{Verdict: yes, at corpus level.} The profile predicts which corpora
saturate, including the one whose poison we did not write, where the cue it
reads is the attack's own construction. It does not predict an individual
defense (wrong at \res{E1-profile-fail} of \res{E1-profile-n} points,
\cref{fig:profile}) and stays a hypothesis with four observations
(\cref{sec:profile:limits}).

\subsection{The control ladder}
\label{sec:eval:ladder}

\begin{figure*}[tp]
\centering
\input{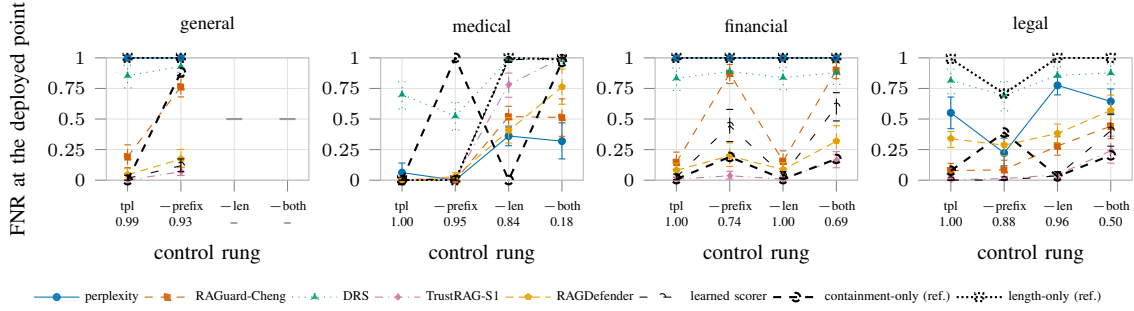}
\begin{tikzpicture}
\begin{groupplot}[group style={group size=4 by 1, horizontal sep=0.95cm},
    width=0.235\textwidth, height=3.2cm, x3base,
    xmin=-0.4, xmax=3.4, xtick={0, 1, 2, 3}, xticklabels={tpl, $-$prefix, $-$len, $-$both},
    xlabel={control rung},
    xticklabel style={font=\tiny}, legend style={legend columns=-1, draw=none, fill=none}]
\nextgroupplot[title={general}, ylabel={FNR at the deployed point}, legend to name=x3f2upper, x3rate, xticklabels={{tpl\\0.99}, {$-$prefix\\0.93}, {$-$len\\--}, {$-$both\\--}}, xticklabel style={font=\tiny, align=center}]
\addplot[x3miss, forget plot] table[x=rung, y=y] {figures/f2_missing_nq.dat};
\addplot[x3s1, error bars/.cd, y dir=both, y explicit]
  table[x=rung, y=ppl,
        y error plus expr=\thisrow{ppl_hi}-\thisrow{ppl},
        y error minus expr=\thisrow{ppl}-\thisrow{ppl_lo}]
  {figures/f2_ladder_nq.dat};
\addlegendentry{perplexity}
\addplot[x3s2, error bars/.cd, y dir=both, y explicit]
  table[x=rung, y=raguard,
        y error plus expr=\thisrow{raguard_hi}-\thisrow{raguard},
        y error minus expr=\thisrow{raguard}-\thisrow{raguard_lo}]
  {figures/f2_ladder_nq.dat};
\addlegendentry{RAGuard-Cheng}
\addplot[x3s3, error bars/.cd, y dir=both, y explicit]
  table[x=rung, y=drs,
        y error plus expr=\thisrow{drs_hi}-\thisrow{drs},
        y error minus expr=\thisrow{drs}-\thisrow{drs_lo}]
  {figures/f2_ladder_nq.dat};
\addlegendentry{DRS}
\addplot[x3s4, error bars/.cd, y dir=both, y explicit]
  table[x=rung, y=trustrag,
        y error plus expr=\thisrow{trustrag_hi}-\thisrow{trustrag},
        y error minus expr=\thisrow{trustrag}-\thisrow{trustrag_lo}]
  {figures/f2_ladder_nq.dat};
\addlegendentry{TrustRAG-S1}
\addplot[x3s5, error bars/.cd, y dir=both, y explicit]
  table[x=rung, y=ragdef,
        y error plus expr=\thisrow{ragdef_hi}-\thisrow{ragdef},
        y error minus expr=\thisrow{ragdef}-\thisrow{ragdef_lo}]
  {figures/f2_ladder_nq.dat};
\addlegendentry{RAGDefender}
\addplot[x3s6, error bars/.cd, y dir=both, y explicit]
  table[x=rung, y=ours,
        y error plus expr=\thisrow{ours_hi}-\thisrow{ours},
        y error minus expr=\thisrow{ours}-\thisrow{ours_lo}]
  {figures/f2_ladder_nq.dat};
\addlegendentry{learned scorer}
\addplot[x3ref1] table[x=rung, y=containonly] {figures/f2_ladder_nq.dat};
\addlegendentry{containment-only (ref.)}
\addplot[x3ref2] table[x=rung, y=lenonly] {figures/f2_ladder_nq.dat};
\addlegendentry{length-only (ref.)}
\nextgroupplot[title={medical},  x3rate, xticklabels={{tpl\\1.00}, {$-$prefix\\0.95}, {$-$len\\0.84}, {$-$both\\0.18}}, xticklabel style={font=\tiny, align=center}]
\addplot[forget plot, x3s1, error bars/.cd, y dir=both, y explicit]
  table[x=rung, y=ppl,
        y error plus expr=\thisrow{ppl_hi}-\thisrow{ppl},
        y error minus expr=\thisrow{ppl}-\thisrow{ppl_lo}]
  {figures/f2_ladder_medical.dat};
\addplot[forget plot, x3s2, error bars/.cd, y dir=both, y explicit]
  table[x=rung, y=raguard,
        y error plus expr=\thisrow{raguard_hi}-\thisrow{raguard},
        y error minus expr=\thisrow{raguard}-\thisrow{raguard_lo}]
  {figures/f2_ladder_medical.dat};
\addplot[forget plot, x3s3, error bars/.cd, y dir=both, y explicit]
  table[x=rung, y=drs,
        y error plus expr=\thisrow{drs_hi}-\thisrow{drs},
        y error minus expr=\thisrow{drs}-\thisrow{drs_lo}]
  {figures/f2_ladder_medical.dat};
\addplot[forget plot, x3s4, error bars/.cd, y dir=both, y explicit]
  table[x=rung, y=trustrag,
        y error plus expr=\thisrow{trustrag_hi}-\thisrow{trustrag},
        y error minus expr=\thisrow{trustrag}-\thisrow{trustrag_lo}]
  {figures/f2_ladder_medical.dat};
\addplot[forget plot, x3s5, error bars/.cd, y dir=both, y explicit]
  table[x=rung, y=ragdef,
        y error plus expr=\thisrow{ragdef_hi}-\thisrow{ragdef},
        y error minus expr=\thisrow{ragdef}-\thisrow{ragdef_lo}]
  {figures/f2_ladder_medical.dat};
\addplot[forget plot, x3s6, error bars/.cd, y dir=both, y explicit]
  table[x=rung, y=ours,
        y error plus expr=\thisrow{ours_hi}-\thisrow{ours},
        y error minus expr=\thisrow{ours}-\thisrow{ours_lo}]
  {figures/f2_ladder_medical.dat};
\addplot[forget plot, x3ref1] table[x=rung, y=containonly] {figures/f2_ladder_medical.dat};
\addplot[forget plot, x3ref2] table[x=rung, y=lenonly] {figures/f2_ladder_medical.dat};
\nextgroupplot[title={financial},  x3rate, xticklabels={{tpl\\1.00}, {$-$prefix\\0.74}, {$-$len\\1.00}, {$-$both\\0.69}}, xticklabel style={font=\tiny, align=center}]
\addplot[forget plot, x3s1, error bars/.cd, y dir=both, y explicit]
  table[x=rung, y=ppl,
        y error plus expr=\thisrow{ppl_hi}-\thisrow{ppl},
        y error minus expr=\thisrow{ppl}-\thisrow{ppl_lo}]
  {figures/f2_ladder_financial.dat};
\addplot[forget plot, x3s2, error bars/.cd, y dir=both, y explicit]
  table[x=rung, y=raguard,
        y error plus expr=\thisrow{raguard_hi}-\thisrow{raguard},
        y error minus expr=\thisrow{raguard}-\thisrow{raguard_lo}]
  {figures/f2_ladder_financial.dat};
\addplot[forget plot, x3s3, error bars/.cd, y dir=both, y explicit]
  table[x=rung, y=drs,
        y error plus expr=\thisrow{drs_hi}-\thisrow{drs},
        y error minus expr=\thisrow{drs}-\thisrow{drs_lo}]
  {figures/f2_ladder_financial.dat};
\addplot[forget plot, x3s4, error bars/.cd, y dir=both, y explicit]
  table[x=rung, y=trustrag,
        y error plus expr=\thisrow{trustrag_hi}-\thisrow{trustrag},
        y error minus expr=\thisrow{trustrag}-\thisrow{trustrag_lo}]
  {figures/f2_ladder_financial.dat};
\addplot[forget plot, x3s5, error bars/.cd, y dir=both, y explicit]
  table[x=rung, y=ragdef,
        y error plus expr=\thisrow{ragdef_hi}-\thisrow{ragdef},
        y error minus expr=\thisrow{ragdef}-\thisrow{ragdef_lo}]
  {figures/f2_ladder_financial.dat};
\addplot[forget plot, x3s6, error bars/.cd, y dir=both, y explicit]
  table[x=rung, y=ours,
        y error plus expr=\thisrow{ours_hi}-\thisrow{ours},
        y error minus expr=\thisrow{ours}-\thisrow{ours_lo}]
  {figures/f2_ladder_financial.dat};
\addplot[forget plot, x3ref1] table[x=rung, y=containonly] {figures/f2_ladder_financial.dat};
\addplot[forget plot, x3ref2] table[x=rung, y=lenonly] {figures/f2_ladder_financial.dat};
\nextgroupplot[title={legal},  x3rate, xticklabels={{tpl\\1.00}, {$-$prefix\\0.88}, {$-$len\\0.96}, {$-$both\\0.50}}, xticklabel style={font=\tiny, align=center}]
\addplot[forget plot, x3s1, error bars/.cd, y dir=both, y explicit]
  table[x=rung, y=ppl,
        y error plus expr=\thisrow{ppl_hi}-\thisrow{ppl},
        y error minus expr=\thisrow{ppl}-\thisrow{ppl_lo}]
  {figures/f2_ladder_legal.dat};
\addplot[forget plot, x3s2, error bars/.cd, y dir=both, y explicit]
  table[x=rung, y=raguard,
        y error plus expr=\thisrow{raguard_hi}-\thisrow{raguard},
        y error minus expr=\thisrow{raguard}-\thisrow{raguard_lo}]
  {figures/f2_ladder_legal.dat};
\addplot[forget plot, x3s3, error bars/.cd, y dir=both, y explicit]
  table[x=rung, y=drs,
        y error plus expr=\thisrow{drs_hi}-\thisrow{drs},
        y error minus expr=\thisrow{drs}-\thisrow{drs_lo}]
  {figures/f2_ladder_legal.dat};
\addplot[forget plot, x3s4, error bars/.cd, y dir=both, y explicit]
  table[x=rung, y=trustrag,
        y error plus expr=\thisrow{trustrag_hi}-\thisrow{trustrag},
        y error minus expr=\thisrow{trustrag}-\thisrow{trustrag_lo}]
  {figures/f2_ladder_legal.dat};
\addplot[forget plot, x3s5, error bars/.cd, y dir=both, y explicit]
  table[x=rung, y=ragdef,
        y error plus expr=\thisrow{ragdef_hi}-\thisrow{ragdef},
        y error minus expr=\thisrow{ragdef}-\thisrow{ragdef_lo}]
  {figures/f2_ladder_legal.dat};
\addplot[forget plot, x3s6, error bars/.cd, y dir=both, y explicit]
  table[x=rung, y=ours,
        y error plus expr=\thisrow{ours_hi}-\thisrow{ours},
        y error minus expr=\thisrow{ours}-\thisrow{ours_lo}]
  {figures/f2_ladder_legal.dat};
\addplot[forget plot, x3ref1] table[x=rung, y=containonly] {figures/f2_ladder_legal.dat};
\addplot[forget plot, x3ref2] table[x=rung, y=lenonly] {figures/f2_ladder_legal.dat};
\end{groupplot}
\end{tikzpicture}
\par\smallskip
\pgfplotslegendfromname{x3f2upper}
\caption{\textbf{No single control reveals the problem.} Miss rate at the deployed operating point under four payload-preserving edits of the same attack (\texttt{tpl} template; $-$prefix, verbatim question removed; $-$len, poison padded to the clean word-length distribution; $-$both), one panel per corpus. Which rung moves a defense is a property of the corpus --- length carries medical, the prefix carries financial --- so either control run alone reverses the conclusion on the other. Black dashes: the two one-feature reference detectors of \cref{sec:profile} (drawn per rung in \cref{fig:dissociation}, \cref{app:grids}). NQ has no length rungs: its released poison already lies inside the clean band (overlap 1.00), so the rung is inert by construction. The number under each rung label is retrieval survival, the first leg of the conjunction of \cref{sec:threat}: the fraction of injected poison retrieved into its own target's candidate set with no defense. Whiskers: the mean over seeds of the per-seed cluster-bootstrap 95\% interval. All ten defenses, each with its false-positive rate, are in \cref{tab:e1-main}.}
\label{fig:ladder}
\end{figure*}

\begin{table*}[tp]
\centering\scriptsize\setlength{\tabcolsep}{3pt}
\caption{Placebo fillers for the length-matched rung on medical. The two negative-log-likelihood detectors collapse under clean filler and recover under the shuffled and pseudo-word fillers; our scorer and TrustRAG's set-level stage collapse at their deployed points under all three.}
\label{tab:placebo}
\begin{tabular}{lccccccccc}
\toprule
 & \multicolumn{3}{c}{clean-prose filler} & \multicolumn{3}{c}{word-shuffled filler} & \multicolumn{3}{c}{pseudo-word filler} \\
\cmidrule(lr){2-4} \cmidrule(lr){5-7} \cmidrule(lr){8-10}
Defense & FNR / FPR & AUROC & FNR@5\% & FNR / FPR{} & AUROC{} & FNR@5\%{} & FNR / FPR{}{} & AUROC{}{} & FNR@5\%{}{} \\
\midrule
Perplexity (GPT-2) & 0.361 / 0.003 & 0.981 & 0.092 & 0.000 / 0.000 & 1.000 & 0.000 & 0.000 / 0.000 & 1.000 & 0.000 \\
RAGuard (Cheng) & 0.519 / 0.103 & 0.838 & 0.694 & 0.000 / 0.110 & 1.000 & 0.000 & 0.000 / 0.112 & 1.000 & 0.000 \\
DRS & 0.988 / 0.035 & 0.536 & 0.977 & 0.990 / 0.038 & 0.526 & 0.982 & 0.605 / 0.034 & 0.907 & 0.497 \\
TrustRAG stage 1 & 0.781 / 0.017 & N/A & N/A & 0.827 / 0.013 & N/A & N/A & 0.843 / 0.019 & N/A & N/A \\
RAGDefender & 0.405 / 0.599 & N/A & N/A & 0.151 / 0.364 & N/A & N/A & 0.005 / 0.069 & N/A & N/A \\
Dup-only & 1.000 / 0.000 & N/A & N/A & 1.000 / 0.000 & N/A & N/A & 0.985 / 0.000 & N/A & N/A \\
Ours, $\tau{=}0.5$ & 0.985 / 0.000 & 0.880 & 0.570 & 0.986 / 0.000 & 0.953 & 0.321 & 0.974 / 0.000 & 0.999 & 0.000 \\
Ours, $\tau_d$ & 0.983 / 0.000 & 0.880 & 0.570 & 0.986 / 0.000 & 0.953 & 0.321 & 0.968 / 0.000 & 0.999 & 0.000 \\
Ours, $\tau_{d,s}$ ($a{=}0.3$) & 0.979 / 0.000 & 0.993 & 0.044 & 0.982 / 0.000 & 0.998 & 0.014 & 0.955 / 0.000 & 1.000 & 0.000 \\
length-only (ref.) & 0.999 / 0.000 & 0.486 & 0.952 & 0.999 / 0.000 & 0.454 & 0.969 & 0.999 / 0.000 & 0.525 & 0.959 \\
containment-only (ref.) & 0.001 / 0.004 & 1.000 & 0.001 & 0.000 / 0.005 & 1.000 & 0.000 & 0.000 / 0.005 & 1.000 & 0.000 \\
retrieval survival & 0.84 &  &  & 0.96 &  &  & 0.97 &  &  \\
\bottomrule
\end{tabular}
\\[2pt]
{\footnotesize Each column block pads the same template poison to the same draw from the clean word-length distribution with a different filler (\cref{sec:eval:ladder}). FNR / FPR at the deployed threshold, AUROC, and the miss rate at a matched 5\% false-positive rate, point estimates over five seeds; N/A where the defense emits no score. The last row is retrieval survival with no defense, the first leg of the conjunction of \cref{sec:threat}.}
\end{table*}

\Cref{fig:ladder} runs the ladder of \cref{sec:profile:ladder}, with retrieval
survival, the first leg of the conjunction, printed under every rung. The two
cues are independent, and each is load-bearing on a different corpus. Removing
the verbatim prefix moves our scorer's deployed miss rate by
\res{E1-nat-dFNR-medical} on medical and \res{E1-nat-dFNR-financial} on
financial; matching poison length moves it by \res{E1-lm-dFNR-medical} on
medical and \res{E1-lm-dFNR-legal} on legal, while staying inert on financial
(\res{E1-lm-dFNR-financial}). \Cref{fig:dissociation} (\cref{app:grids}) shows
the same crossing on the two one-feature detectors: each dies under the
control that removes its own cue and is untouched by the other, so a study
running one control reads the other cue as content. Both readings have been
published: TrustRAG's Table~4 and
TriShieldRAG's adaptive arm reach opposite verdicts from one control
each~\cite{zhou2025trustrag,mohanty2026trishieldrag}.

Length matching pads and never truncates, so on medical the padded
document is mostly clean prose. Two placebo arms hold length fixed and vary
the filler (\cref{tab:placebo}): the same clean
sentences with their words shuffled across the pool (identical length and
vocabulary, no content), and pseudo-words of matched length. Our scorer and
TrustRAG's set-level stage collapse at the deployed point under all three
fillers; the shuffled arm gives \res{E1-FNR-valor-lmsh-medical} at \AUROC{}
\res{E1-AUROC-valor-lmsh-medical} for our scorer and
\res{E1-FNR-trustrag-lmsh-medical} for TrustRAG, so the collapse is not content
dilution. The two negative-log-likelihood
detectors do not survive the placebo: perplexity and RAGuard miss
\resbare{E1-FNR-ppl-lm} and \res{E1-FNR-raguard-lm} under clean filler and
\res{E1-FNR-ppl-lmsh} and \res{E1-FNR-raguard-lmsh} under shuffled filler,
because a mean-token likelihood is set by whatever it is padded with: fluent
prose lowers it, shuffled words raise it. Padding cannot give an NLL
detector a length control at all, so their collapse is withdrawn and both
arms are reported as bounds. The pseudo-word arm is a
vocabulary control rather than a length control: it pushes the document off
the clean vocabulary, our scorer's ranking recovers to \AUROC{}
\res{E1-AUROC-valor-lmns-medical} while its deployed miss rate stays at
\res{E1-FNR-valor-lmns-medical}, and DRS, an embedding-distance rule, recovers
from \res{E1-FNR-drs-lmsh-medical} to \res{E1-FNR-drs-lmns-medical}; the
shuffled arm holds vocabulary fixed and is the placebo we read. Where the
clean corpus is a constant width, as on legal, padding overshoots and installs
an inverted cue instead of removing one (\cref{tab:placebo-legal},
\cref{app:grids}).

Survival is the leg that fails on medical with both cues removed:
\res{E1-survival-natlm-medical} of the poison is retrieved for its target, so
the miss rate of \res{E1-FNR-valor-natlm-medical} drawn there is over the
survivors and the attack's success under the conjunction is
\res{E1-success-natlm-medical}; on financial and legal the both-cues rung
survives at \res{E1-survival-natlm-financial} and
\res{E1-survival-natlm-legal}. Stacking the two rungs and adding style mimicry
gives the strongest attack in this paper: on NQ the miss rate goes from \resbare{E4-A3-FNR-template}
at the first rung (the archive-stored passages, already without the prefix) to
\resbare{E4-A3-FNR-length-matched} at the last, at a survival after
re-embedding and re-ranking of \resbare{E4-A3-retrieved-length-matched-nq};
on medical from \resbare{E4-A3-FNR-template-medical} to
\resbare{E4-A3-FNR-length-matched-medical} at
\resbare{E4-A3-retrieved-length-matched-medical}. Legal is the exception
(\resbare{E4-A3-FNR-length-matched-legal} at survival
\resbare{E4-A3-retrieved-length-matched-legal}): the fixed-width chunking that
makes CUAD a length constant~\cite{hendrycks2021cuad} also makes padded poison
expensive to retrieve.

The query-based attacker of \cref{sec:threat:control}, tuned to our own
scorer on NQ (the grid is \cref{app:grids}'s E4-D1 table), passes
\res{E4-D1-FNR-valor} of its documents past the learned score; its static
column is the archive-stored, prefix-less construction, which is why our
scorer's static miss rate there (\res{E4-D1-FNR-valor-static}) is not
\cref{tab:e1-main}'s. For the baselines (\res{E4-D1-FNR-baselines-range}, the
low end bought by blocking almost every benign candidate) the same documents
are a \emph{transfer} attack, a lower bound (\cref{sec:threat:control}).

\textbf{Verdict: the load-bearing cue is corpus-specific.} One control
understates the problem on some corpora and overstates it on others, so a single
ablation is never evidence.

\subsection{The dissociation, and two collapse modes}
\label{sec:eval:dissociation}

\begin{figure*}[tp]
\centering
\input{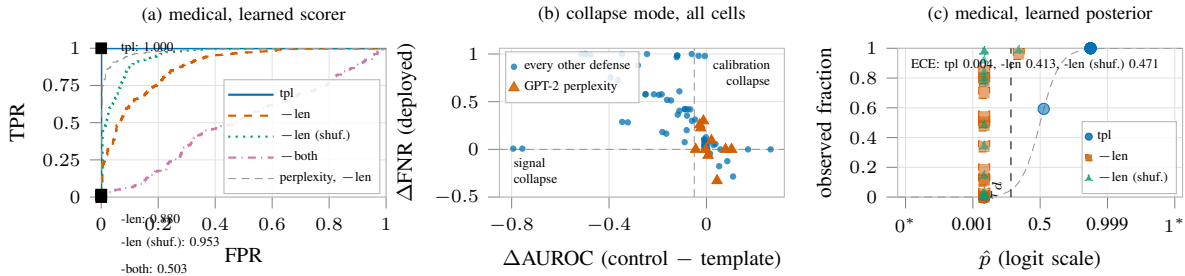}
\begin{tikzpicture}
\begin{groupplot}[group style={group size=3 by 1, horizontal sep=1.5cm},
    width=0.295\textwidth, height=3.55cm, x3base,
    legend style={at={(0.98,0.02)}, anchor=south east, legend columns=1}]
\nextgroupplot[title={(a) medical, learned scorer}, xlabel={FPR}, ylabel={TPR}, x3rate, xmin=0, xmax=1]
\addplot[x3s1, mark=none, thick] table[x=fpr, y=tpr] {figures/f3_roc_medical_template.dat};
\addlegendentry{tpl}
\addplot[x3s2, mark=none, thick] table[x=fpr, y=tpr] {figures/f3_roc_medical_lengthmatch.dat};
\addlegendentry{$-$len}
\addplot[x3s3, mark=none, thick] table[x=fpr, y=tpr] {figures/f3_roc_medical_lmsh.dat};
\addlegendentry{$-$len (shuf.)}
\addplot[x3s4, mark=none, thick] table[x=fpr, y=tpr] {figures/f3_roc_medical_natlm.dat};
\addlegendentry{$-$both}
\addplot[x3grey, mark=none, thin, densely dashed] table[x=fpr, y=tpr] {figures/f3_roc_medical_lengthmatch_ppl.dat};
\addlegendentry{perplexity, $-$len}
\addplot[x3a, only marks, mark=square*, mark size=2.0pt, forget plot] coordinates {(0.000000,1.000000)};
\node[font=\tiny, anchor=west, xshift=4pt, yshift=0pt] at (axis cs:0.000000,1.000000) {tpl: 1.000};
\addplot[x3a, only marks, mark=square*, mark size=2.0pt, forget plot] coordinates {(0.000000,0.016546)};
\node[font=\tiny, anchor=west, xshift=4pt, yshift=-9pt] at (axis cs:0.000000,0.016546) {-len: 0.880};
\addplot[x3a, only marks, mark=square*, mark size=2.0pt, forget plot] coordinates {(0.000000,0.013546)};
\node[font=\tiny, anchor=west, xshift=4pt, yshift=-18pt] at (axis cs:0.000000,0.013546) {-len (shuf.): 0.953};
\addplot[x3a, only marks, mark=square*, mark size=2.0pt, forget plot] coordinates {(0.000000,0.000000)};
\node[font=\tiny, anchor=west, xshift=4pt, yshift=-27pt] at (axis cs:0.000000,0.000000) {-both: 0.503};
\nextgroupplot[title={(b) collapse mode, all cells}, xlabel={$\Delta$AUROC (control $-$ template)},
    ylabel={$\Delta$FNR (deployed)}, x3base,
    xmin=-0.85, xmax=0.32, ymin=-0.5, ymax=1.06,
    xtick={-0.8,-0.4,0}, ytick={-0.5,0,0.5,1}, legend style={at={(0.02,0.98)}, anchor=north west}]
\draw[x3grey, densely dashed] (axis cs:-0.85,0) -- (axis cs:0.32,0);
\draw[x3grey, densely dashed] (axis cs:-0.05,-0.5) -- (axis cs:-0.05,1.06);
\node[font=\tiny, x3a, anchor=north east, align=right] at (axis cs:0.30,1.04) {calibration\\collapse};
\node[font=\tiny, x3a, anchor=south west, align=left] at (axis cs:-0.83,-0.48) {signal\\collapse};
\addplot[x3b, only marks, mark=*, mark size=1.0pt, opacity=0.75] table[x=dauroc, y=dfnr] {figures/f7_modes.dat};
\addlegendentry{every other defense}
\addplot[x3c, only marks, mark=triangle*, mark size=2.2pt] table[x=dauroc, y=dfnr] {figures/f7_modes_ppl.dat};
\addlegendentry{GPT-2 perplexity}
\nextgroupplot[title={(c) medical, learned posterior}, xlabel={$\hat{p}$ (logit scale)}, ylabel={observed fraction}, x3base, ymin=0, ymax=1, xmin=-14.616, xmax=14.616,
    xtick={-13.8155,-6.907,0,6.907,13.8155},
    xticklabels={$0^{*}$,0.001,0.5,0.999,$1^{*}$}]
\addplot[x3grey, densely dashed, mark=none, samples=121, domain=-14.616:14.616, forget plot] {1/(1+exp(-x))};
\addplot[x3s1, only marks, scatter, scatter src=explicit,
          scatter/use mapped color={draw=x3b, fill=x3b, fill opacity=0.55},
          visualization depends on={\thisrow{template_ms} \as \x3msize},
          scatter/@pre marker code/.append style={/tikz/mark size=\x3msize pt}]
  table[x=template_logit, y=template_f, meta=template_n] {figures/f3_reliability.dat};
\addlegendentry{tpl}
\addplot[x3s2, only marks, scatter, scatter src=explicit,
          scatter/use mapped color={draw=x3c, fill=x3c, fill opacity=0.55},
          visualization depends on={\thisrow{lengthmatch_ms} \as \x3msize},
          scatter/@pre marker code/.append style={/tikz/mark size=\x3msize pt}]
  table[x=lengthmatch_logit, y=lengthmatch_f, meta=lengthmatch_n] {figures/f3_reliability.dat};
\addlegendentry{$-$len}
\addplot[x3s3, only marks, scatter, scatter src=explicit,
          scatter/use mapped color={draw=x3d, fill=x3d, fill opacity=0.55},
          visualization depends on={\thisrow{lmsh_ms} \as \x3msize},
          scatter/@pre marker code/.append style={/tikz/mark size=\x3msize pt}]
  table[x=lmsh_logit, y=lmsh_f, meta=lmsh_n] {figures/f3_reliability.dat};
\addlegendentry{$-$len (shuf.)}
\draw[x3a, dashed] (axis cs:-2.9957,0) -- (axis cs:-2.9957,1);
\node[font=\tiny, anchor=south, rotate=90] at (axis cs:-2.9957,0.05) {$\tau_d$};
\node[font=\tiny, anchor=north west] at (axis cs:-14.216,0.99) {ECE: tpl 0.004, -len 0.413, -len (shuf.) 0.471};
\end{groupplot}
\end{tikzpicture}
\caption{\textbf{Ranking survives; the operating point does not, and \AUROC{} alone cannot tell the two apart.} (a)~ROC of our learned scorer, trust off, on medical under the template rung, the length-matched rung with clean-prose filler, the same rung with the words of the filler shuffled (identical length and vocabulary, no content), and both cues removed. The filled square is the \emph{deployed} point, labelled with the rung and that rung's \AUROC{}: under shuffled filler the curve is degraded but usable while the square falls to the floor. The thin grey curve is a GPT-2 perplexity threshold under clean filler: the same shape, but for content dilution --- under shuffled filler perplexity recovers to a miss rate of zero (\cref{tab:placebo}), so it is drawn and not headlined. (b)~Every cell, one mark per (corpus, defense, rung): abscissa the \AUROC{} the rung costs, ordinate the miss rate it adds. High and to the right is \emph{calibration collapse}, which a threshold repairs at a stated false-positive cost; left of the dashed rule the ranking itself is gone. (c)~Fifteen equal-mass reliability bins for the deployed posterior under the three fillers of (a), on a logit abscissa; marker area $\propto$ bin count, grey curve perfect calibration, black rule $\tau_d$, ticks $0^{*}$ and $1^{*}$ bins saturated to double precision.}
\label{fig:roc}
\end{figure*}

\Cref{fig:roc} plots ranking against decision under four cue conditions, with a
reliability diagram at the realised
prevalence~\cite{naeini2015ece,guo2017calibration}; the deployed operating
points slide off the curves, and \AUROC{} cannot see it. The headline row is our scorer under shuffled filler, where nothing but length
and word order changed: \AUROC{} \res{E1-AUROC-valor-lmsh-medical}, miss rate
\res{E1-FNR5-valor-lmsh-medical} at 5\% \FPR{},
\res{E1-FNR-valor-lmsh-medical} at the deployed threshold; perplexity under
clean filler has the same shape for the content-dilution reason of
\cref{sec:eval:ladder}, so it is drawn and not headlined.

Sorting every defense-by-rung pair by the ratio of $\Delta\FNR$ at the deployed
point to $\Delta\AUROC$ separates two modes. \emph{Calibration collapse}: the
ranking is degraded but usable and the deployed point dies --- the headline
row, at \ECE{} \res{E1-ECE-lmsh-medical} at the realised prevalence. \emph{Signal collapse}: the ordering goes to chance
and no threshold rule recovers anything --- medical with both cues removed
reaches \AUROC{} \res{E1-AUROC-valor-natlm-medical}, NQ at the
mimicry-plus-length-matched rung \resbare{E4-A3-AUROC-length-matched}, financial
\resbare{E4-A3-AUROC-length-matched-financial} (the boundary,
$\Delta\AUROC=-0.05$ in \cref{fig:roc}(b), is a stated convention). The
profile predicts which mode a cell is in: where the
removed cue was the only cue the detector had, the signal goes with it; where
content structure remains, only the threshold is wrong.

\textbf{Verdict: they dissociate.} \AUROC{} reported alone cannot separate a
calibration failure, which a stated false-positive cost repairs, from a signal
failure, which nothing in the decision layer repairs --- and on these cells it
reports both of them as success.

\subsection{Mechanism and repair}
\label{sec:eval:mechanism}

\begin{table*}[tp]
\centering\scriptsize\setlength{\tabcolsep}{1.5pt}
\caption{Feature-group ablation of the one scorer whose features we control, trust off and no escalation band (the E1 reading, \cref{sec:stack:stack}); medical, one column group per poison construction, point estimates over three seeds (dispersion in the released \texttt{e7nt\_summary.json}). $-$ query-overlap deletes the two verbatim-question features; GBT is a gradient-boosted tree on the identical features; $\tau{=}0.5$ is the full scorer at the harm-blind threshold.}
\label{tab:e7-ablation}
\begin{tabular}{lcccccccccccccccccccc}
\toprule
 & \multicolumn{5}{c}{template} & \multicolumn{5}{c}{$-$len, clean} & \multicolumn{5}{c}{$-$len, shuffled} & \multicolumn{5}{c}{$-$both} \\
\cmidrule(lr){2-6} \cmidrule(lr){7-11} \cmidrule(lr){12-16} \cmidrule(lr){17-21}
Variant & FNR & FPR & AUROC & FNR@5\% & ECE & FNR{} & FPR{} & AUROC{} & FNR@5\%{} & ECE{} & FNR{}{} & FPR{}{} & AUROC{}{} & FNR@5\%{}{} & ECE{}{} & FNR{}{}{} & FPR{}{}{} & AUROC{}{}{} & FNR@5\%{}{}{} & ECE{}{}{} \\
\midrule
full & 0.000 & 0.000 & 1.000 & 0.000 & 0.004 & 0.986 & 0.000 & 0.885 & 0.550 & 0.416 & 0.988 & 0.000 & 0.962 & 0.280 & 0.473 & 1.000 & 0.000 & 0.483 & 0.957 & 0.093 \\
$-$ doc-length & 0.000 & 0.000 & 1.000 & 0.000 & 0.005 & 0.204 & 0.007 & 0.992 & 0.028 & 0.111 & 0.026 & 0.006 & 0.999 & 0.002 & 0.017 & 1.000 & 0.004 & 0.671 & 0.923 & 0.094 \\
$-$ query-overlap & 0.000 & 0.000 & 1.000 & 0.000 & 0.006 & 0.992 & 0.000 & 0.680 & 0.927 & 0.418 & 0.998 & 0.000 & 0.817 & 0.824 & 0.476 & 0.991 & 0.000 & 0.398 & 0.982 & 0.091 \\
$-$ lexical & 0.000 & 0.005 & 1.000 & 0.000 & 0.002 & 0.897 & 0.022 & 0.627 & 0.852 & 0.397 & 0.713 & 0.021 & 0.841 & 0.561 & 0.387 & 0.990 & 0.014 & 0.486 & 0.981 & 0.096 \\
$-$ set-context & 0.000 & 0.000 & 1.000 & 0.000 & 0.002 & 0.990 & 0.000 & 0.957 & 0.201 & 0.420 & 0.991 & 0.000 & 0.966 & 0.133 & 0.477 & 1.000 & 0.000 & 0.785 & 0.758 & 0.096 \\
GBT scorer & 0.000 & 0.000 & 1.000 & 0.000 & 0.000 & 0.000 & 0.000 & 1.000 & 0.000 & 0.000 & 0.000 & 0.000 & 1.000 & 0.000 & 0.000 & 1.000 & 0.000 & 0.571 & 0.930 & 0.097 \\
$\tau{=}0.5$ & 0.000 & 0.000 & 1.000 & 0.000 & 0.004 & 0.986 & 0.000 & 0.885 & 0.550 & 0.416 & 0.988 & 0.000 & 0.962 & 0.280 & 0.473 & 1.000 & 0.000 & 0.483 & 0.957 & 0.093 \\
\bottomrule
\end{tabular}
\end{table*}

\Cref{tab:e7-ablation} ablates the one scorer whose features we control. On
length-matched medical, deleting the document-length feature moves the
deployed miss rate from \res{E7NT-FNR-full-lm} to \res{E7NT-FNR-nolength-lm}
under clean filler and from \res{E7NT-FNR-full-lmsh} to
\res{E7NT-FNR-nolength-lmsh} under shuffled filler, \ECE{} from
\res{E7NT-ECE-full-lm} to \res{E7NT-ECE-nolength-lm}, while \AUROC{}
\emph{rises} from \res{E7NT-AUROC-full-lm} to \res{E7NT-AUROC-nolength-lm}:
the feature carried the threshold and was hurting the ordering, because a
scorer that learned short documents are poison reads padded poison as an
argument for admission. The repair is a fallback, not a recovery of content:
containment is still \res{E1-containment-medical} on that cell, deleting the
query-overlap features instead gives \res{E7NT-FNR-nocontain-lm} at \AUROC{}
\res{E7NT-AUROC-nocontain-lm}, and with both cues removed nothing repairs
anything --- every row of that column misses at least
\res{E7NT-FNR-nolex-natlm} and the full scorer's \AUROC{} is
\res{E7NT-AUROC-full-natlm}. A gradient-boosted tree on the identical features
reaches \FNR{} \res{E7NT-GBT-FNR-lm} at \AUROC{} \res{E7NT-GBT-AUROC-lm} on the
same cell, so the collapse is a property of this MLP's calibration, which
indicts our scorer first; the tree dies with everything else at
\res{E7NT-GBT-AUROC-natlm} once both cues are gone. Source identity is an exact label indicator here
(\cref{sec:stack:stack}), so an earlier draft's trust-on version of this table,
like every trust-weighted \AUROC{} in \cref{tab:placebo}, re-ranked on a
leaked label.

Repair follows the taxonomy. An empirical clean-quantile floor --- refuse to
admit below the $q$-th percentile of clean scores, cross-fitted on held-out
clean queries --- returns the length-matched medical miss rate from
\resbare{E1-FNR-valor-lm-medical} to \resbare{E2-floor-FNR-lm-medical} for
\resbare{E2-floor-dFPR-lm-medical} points of \FPR{}; where the signal has collapsed
no ordering is left for a quantile to exploit, and on medical with both cues
removed it moves the miss rate by \resbare{E2-floor-dFNR-natlm-medical}.

\textbf{Verdict: the profiled cue that remains carries the calibration.}
Remove one and the threshold moves onto the other; remove both and the ranking
goes with them.

\subsection{Decision agreement and inertness}
\label{sec:eval:agreement}

\Cref{tab:e1-agreement} (\cref{app:grids}) counts what the decision layer
changes. Across the \res{E1-NIDENT-total} corpus-by-attack cells the
harm-blind $\tau{=}0.5$ classifier, the cost-derived \taud{} and the
trust-adjusted \tauds{} produce bit-identical actions on every candidate in
\res{E1-NIDENT} of them. On templated medical poison the fraction of
candidates whose score falls anywhere in the decision band is
\res{E1-band-medical}: a shortcut drives score mass to the extremes, where
every threshold rule is the same rule (\res{E1-ndistinct-medical} distinct
score values per cell).
\res{E1-NIDENT-rho1} of those carry $\rhod{=}1$, where the cost-derived
rule \emph{is} the $\tau{=}0.5$ classifier; the table flags them and none
counts as evidence.

Where the score is unsaturated the cost-derived point acts: on length-matched
legal, harm per 1{,}000 queries falls from \resbare{E1-H-uniform-lm-legal} to
\resbare{E1-H-valor-lm-legal}, and on prefix-removed medical from
\resbare{E1-H-uniform-nat-medical} to \resbare{E1-H-valor-nat-medical}; the same rule
wrapped over RAGuard's and perplexity's scores reproduces the direction on
length-matched legal (\res{E2-wrapper-raguard} and \res{E2-wrapper-ppl}) at a
printed false-positive cost that rule~5 of \cref{sec:report} calls a closed
door (\cref{fig:cost}, \cref{app:grids}).

\textbf{Verdict: no.} On the benchmark the literature reports against, the
harm-blind, cost-derived and trust-adjusted rules take bit-identical actions on
every candidate, so no operating-point claim on those cells --- ours first ---
can be falsified by an experiment run on them.

\section{A Reporting Protocol for Query-Time Filters}
\label{sec:report}

Six rules, each forced by a result above.

\begin{enumerate}[leftmargin=*,itemsep=1pt]
\item \textbf{Publish the construction profile beside every reported \AUROC{}}: containment rate, length separation, and the clean band's own width, for the poison actually evaluated. Reconstruct the passage the attack inserts, not the fragment its archive stores: the difference is the result.

\item \textbf{Run at least two controls, and report the inert ones.} Which cue is load-bearing is corpus-specific, and an inert control is the evidence that the other cue carries the detection. A control that changes length needs a placebo filler: a filler with content moves the score for its own reasons.

\item \textbf{Report \FNR{} and \FPR{} at the rule you would deploy, beside \AUROC{}, and, for any detector that emits a probability, \ECE{} at the realised prevalence rather than the training base rate.} \AUROC{} alone cannot separate calibration collapse from signal collapse.

\item \textbf{Print the number of distinct score values beside every threshold-derived metric.} Where a score takes few values every threshold rule is the same rule, and no operating-point claim is falsifiable there.

\item \textbf{Compare at matched \FPR{}, or on a swept harm axis with the harm ratio stated}~\cite{drummond2006costcurves}, never at each system's own operating point. A miss rate of zero at a false-positive rate above one half is a closed door, not a detector.

\item \textbf{For every claimed evasion, report retrieval survival and payload preservation}~\cite{carlini2019evaluating}. An edit that removes a cue and then falls out of the candidate set, or that survives retrieval with the attacker's claim deleted, is a data ablation, not an attack.
\end{enumerate}

\section{Limitations and Concurrent Work}
\label{sec:discussion}

\noindent\textbf{Scope.}
Three of our four corpora carry poison regenerated from published templates, so every separability number there is an upper bound; the exception, NQ, carries the argument, and the prefix control on the attack authors' own passages (RAGuard at \AUROC{} \res{E1-AUROC-raguard-nq-published} against \res{E1-AUROC-raguard-released}) is one of the two controls everything rests on. The other, the placebo filler, withdrew a headline rather than repairing one: an earlier draft led with perplexity's length-matched collapse, which word-shuffled filler of the same length undoes (\cref{tab:placebo}).

\noindent\textbf{What the measurement does not cover.}
Seven published defenses are our re-implementations at their published rules (\cref{tab:defenses} prints nominal against realised false-positive rates); two of them, TrustRAG stage~1 and RAGDefender, also ran as released, on the poison as released (\cref{tab:their-code}), RAGDefender with its artifact's MiniLM embedder rather than the 1.5B-parameter encoder its paper reports; RAGuard, DRS, perplexity, NLI and the duplicate filter ran only as our ports. An earlier draft calibrated DRS on 199 clean documents, fewer than the 384 embedding dimensions: the covariance is then rank-deficient, its lowest-variance directions are sampling noise, the cross-fitted threshold sat on fold directions that differ from the deployed ones, and DRS blocked most of the benign corpus at a nominal 1\%. That number was ours, not DRS's, and a wider encoder would not repair it: the operative quantity is calibration documents per dimension. Every DRS cell now uses the calibration size its Algorithm~2 specifies, with the threshold on a held-out half of that set, and realises \res{E1-FPR-drs-medical} on medical and \res{E1-FPR-realised-drs} over the four template cells (withdrawn rates: \cref{tab:e1-main}'s note). The scale is 100 target questions, 40 evaluation queries per cell and corpora of 21k--34k documents, against a sub-area norm of $10^6$--$10^7$; every other fitted threshold is a quantile of the calibration split's $\le 200$ clean documents, so a 99th percentile is that sample's second-largest value, and because the split is fixed across seeds no per-seed spread contains that variance. One retriever, no reranker, no NQ length rung (the released NQ poison lies inside the clean band, overlap \res{E1-overlap-nq}), one poison rate, one trust setting: the Contriever condition, the synthetic corpus and the legitimate-update and poison-rate arms are cut, no Sybil or competing-attack condition runs~\cite{nie2026poisonfails,lee2026polymorphicsybil,zhang2025poisonarena}, and poison is never separated from a legitimate correction~\cite{wang2026purpose}. Every adaptive number is a lower bound: one attacker, tuned to our scorer and transferred unchanged to the others (\cref{sec:threat:control}), with no LLM-guided or human arm, far below the budgets of~\cite{nasr2025attacker} and without the strongest learned generator~\cite{hu2026confundo}. Generator-side harm is secondary (a 0.5B generator, substring matching, the template rung alone: \res{E9-ASR-valor}).

\noindent\textbf{Concurrent work.}
\Cref{sec:background:gap} measures the 2026 cohort's constructions rather than re-running the systems. The ranking/operating-point gap for thresholded detectors under shift is known~\cite{carlini2022lira,jacob2025promptshield,akinrele2026regime,shire2026pidsbench}; shortcut features in security benchmarks are old~\cite{arp2022dosdonts,sommer2010closedworld,pendlebury2019tesseract,geirhos2020shortcut,risse2024limits}; and the query-inclusion artifact and the prefix-removed control are concurrently observed on the attack side and in one defense (\cref{sec:background:gap}). What is new here is tying the gap to a construction cue of the poison set, the label-free profile that predicts it, and the false-positive side none of the above measures.

\section{Conclusion}
\label{sec:conclusion}

Query-time RAG poisoning filters are reported on benchmarks whose poison carries the target question verbatim, because the field's most-cited attack constructs it that way in its own released code~\cite{zou2025poisonedrag}. Of the \res{SV-n} published evaluations we surveyed, \res{SV-prefixed} run poison built exactly that way, and \res{SV-nofpr} report no false-positive rate at the point they deploy. Two numbers, computable from a poison set and a clean corpus alone, say in advance, at corpus level, which cells will saturate. On the cells that do, unrelated defenses, ours among them, report near-perfect detection at operating points no cost-derived threshold can tell apart. Delete one cue and the deployed operating point dies while the ranking survives, degraded; delete both and the ranking goes with it. The first failure is repairable, at a printed false-positive cost; the second is not.

\textbf{Until an \AUROC{} arrives with its poison set's construction profile beside it, it is a statement about how the poison was written, not about what the filter will stop.}


\section*{Open Science}
\label{sec:openscience}

Every artifact needed to reproduce every table and figure is released through a
fully anonymized repository, linked from the submission form, available within
three days of the submission deadline, frozen on 2 October 2026 and left
unedited for the remainder of review; on acceptance the same snapshot is
deposited on Zenodo with a DOI. The release contains: (i) the
scorer package (risk scorer, calibration, Bayesian trust ledger,
cost-sensitive gate, per-decision audit records) and the re-implemented
baselines under a common per-document interface, so that every measured defense
scores one identical candidate set; (ii) the four attack-template generators
(PoisonedRAG-style answer substitution, CorruptRAG-AS
framing~\cite{zhang2026corruptrag}, HijackRAG-style
instruction payloads~\cite{zhang2024hijackrag}, jamming
blockers~\cite{shafran2025jamming}), the adaptive attacks (mimicry, black-box
word substitution, HotFlip suffix~\cite{ebrahimi2018hotflip}, set-context
dilution, threshold probing) and the trust-attack scenario generators;
(iii)~the control-ladder lanes as payload-preserving edits of those generators,
every adversarial passage used here as text with its target question id, seed
and generator, the script that computes the two-number construction profile
of \cref{sec:profile}, and the measurement scripts of the \cref{tab:survey}
survey together with the released third-party artifacts they read; (iv) the
configuration that rebuilds the embedding cache for the redistributable
sub-corpora under
\texttt{all-MiniLM-L6-v2}~\cite{wang2020minilm,reimers2019sbert}, the cache
itself being a deterministic function of the released corpora; (v) scripts
that deterministically rebuild each sub-corpus from public document identifiers
(\texttt{experiments/build\_data.py}) and rerun every reported experiment at
fixed seeds (\texttt{experiments/reproduce\_all.py}); (vi) the design contract,
the prior reviews and this response; and (vii) a scrubbed snapshot of the
earlier version of the code, with identifying paths, names and repository
history removed. The audit of that earlier version cites it by file and line,
so its findings stay checkable against the snapshot. The main-table lanes take
\res{E8-total-cpu-hours} CPU-hours on the reference machine; the placebo,
DRS re-fit and ablation lanes add to that; no GPU is required.

\noindent\textbf{Datasets and licences.}
Only adversarial passages are generated; every document and query comes from a
public dataset and is redistributed only where its licence permits.
PoisonedRAG's released poison texts and retrieval
results~\cite{zou2025poisonedrag} (MIT) are included, as are the
BEIR~\cite{thakur2021beir} sub-corpora built from Natural
Questions~\cite{kwiatkowski2019nq} and HotpotQA~\cite{yang2018hotpotqa}
(CC~BY-SA~4.0) and the chunked sub-corpora from
PubMedQA~\cite{jin2019pubmedqa}, FinQA~\cite{chen2021finqa} and
TAT-QA~\cite{zhu2021tatqa} (MIT; CC~BY~4.0 on the Hugging Face mirrors) and
CUAD~\cite{hendrycks2021cuad} (CC~BY~4.0), each with its document ids and an
attribution file. MS~MARCO~\cite{nguyen2016msmarco} carries Microsoft's
research terms and no SPDX licence, so we redistribute neither its passages nor
embeddings derived from them; the release ships passage ids and a loader that
rebuilds the sub-corpus and its embedding cache. GPT-2 and the retriever
weights are fetched from their public hosts at build time, not redistributed.

\noindent\textbf{What is not released, and why.}
No trained scorer weights are released; the build script regenerates them in
\res{E0-train-minutes} minutes on CPU. Nor are E9's raw generator outputs,
beyond the substring-match verdicts and the prompts that recreate them, or any
adversarial text targeting anything but the public benchmark questions above.

\section*{LLM Usage Considerations}
\label{sec:llmusage}

\noindent\textbf{Accountability and correctness.}
LLMs were used for editorial purposes in this manuscript, and all outputs were
inspected by the authors to ensure accuracy and originality. Editorial use
means rephrasing, formatting and consistency checks; LLMs were also used as
coding assistants for the evaluation harness, the baselines and the tests.
Every reference was verified against its primary source before
inclusion; every number in the paper is produced by the released scripts from
the released data and was not generated, estimated or transcribed by a language
model. The authors take full responsibility for the text, the code and the
results.

\noindent\textbf{Transparency: LLMs in the method and in the evaluation.}
No language model is part of the scorer whose failure modes this paper
measures: it is a multilayer perceptron, with a gradient-boosted tree as the
ablation comparator, over a Beta-posterior trust term and a closed-form
decision rule. Language models appear only in the evaluation, in three places,
all open-weight and named by their Hugging Face identifiers, so that no
closed model is needed to
reproduce a result. (1)~The perplexity filter, the
RAGuard~\cite{cheng2025raguard} re-implementation and the NLI consistency
baseline score documents with GPT-2~\cite{radford2019gpt2} and a small open
cross-encoder~\cite{reimers2019sbert}, as in the original defenses. (2)~The
optional end-to-end attack-success-rate experiment (E9) uses
\texttt{Qwen/Qwen2.5-0.5B-Instruct}~\cite{qwen2025qwen25} as the generator
under attack, judged by substring match; it is a component of no defense, and
no claim depends on it. (3)~The PoisonedRAG poison
passages~\cite{zou2025poisonedrag} were generated by their authors with a
closed LLM; that text is used verbatim, assembled into the inserted passage the
way the attack's own code assembles it (\cref{sec:stack:stack}). Every other
adversarial passage here is produced by deterministic, seeded templates and gradient- or
score-based search, not by a language model.

\noindent\textbf{Responsibility.}
All data are public research datasets used under their licences, and no
personal data beyond what they already contain were processed. The evaluation
runs on a single CPU laptop, which bounds its energy footprint
(\res{E8-total-cpu-hours} CPU-hours for the main-table and control lanes) and is why
the E9 generator is a sub-billion-parameter model: the claims concern the admission decision made
before generation, which a small generator suffices to measure.

\section*{Ethical Considerations}
\label{sec:ethics}

This work is defensive. It introduces no new attack capability: every attack we
run is a published attack run as its own code
runs it (PoisonedRAG~\cite{zou2025poisonedrag}), a published attack re-implemented from
its paper (CorruptRAG~\cite{zhang2026corruptrag},
HijackRAG~\cite{zhang2024hijackrag}, jamming~\cite{shafran2025jamming},
HotFlip~\cite{ebrahimi2018hotflip}), or a standard adaptive-evaluation
technique~\cite{carlini2019evaluating,tramer2020adaptive} applied to defenses
we implement ourselves. The control ladder removes construction cues from those
attacks; it gives an attacker nothing the published generator does not. The
trust-attack scenarios instantiate the long-known taxonomy of reputation-system
attacks~\cite{hoffman2009reputation,douceur2002sybil,sun2006trustattacks} in
simulation only.

We follow the Menlo Report principles~\cite{dittrich2012menlo}. \emph{Respect
for persons}: no human subjects were involved; adjudication is by ground-truth
labels and the human-review path is simulated, not staffed.
\emph{Beneficence}: the poison generators are released only as templates bound
to the public benchmark question sets used here (NQ, HotpotQA, MS~MARCO,
PubMedQA, FinQA, CUAD),
and the generated passages are labelled adversarial and are not fluent enough
to pass as authoritative medical, financial or legal guidance outside a
benchmark harness. We release the adaptive attacks because a defense evaluated
without them is not a defense~\cite{carlini2019evaluating}. \emph{Justice}: the
deployments modelled are those where the burden of poisoning falls on end
users, and the cost-sensitive formulation makes the trade-off between missed
poison and withheld benign content explicit rather than hidden in a fixed
threshold. \emph{Respect for law and public interest}: no live system was
probed, nothing was inserted into any third-party corpus, and all datasets are
used within their licences; the medical corpus is published abstracts and the
legal corpus contracts released for research, neither containing patient or
client data.

\noindent\textbf{Dual use, and disclosure.}
The decision dial is dual-use in both directions: an operator who sets \rhod{}
or \ks{} to admit what should be blocked moves the operating point without
moving any ranking metric, which is the dissociation this paper measures under
attack. Two properties of the implementation bound that, neither a guarantee:
every decision is written to an audit record carrying \phat{}, \ks{},
\ptilde{}, \taud{}, \tauds{} and the action taken, so a shifted operating point
is visible after the fact, and \ks{} is clipped below at \kmin{}, so no
accumulated reputation moves the threshold past a stated bound. Responsible
disclosure does not arise: no third-party or live system was probed, and every
defense whose failure we report is one we re-implemented and ran ourselves. We
release the trust-attack and threshold-probing code too: both instantiate
behaviour a Kerckhoffs adversary is assumed to have already, and the threat
model (\cref{sec:threat}) puts the mitigation in authenticated source identity
and adjudicated evidence, not in secrecy.

\bibliographystyle{IEEEtran}
\bibliography{refs}

\begin{thebibliography}{100}
\providecommand{\url}[1]{#1}
\csname url@samestyle\endcsname
\providecommand{\newblock}{\relax}
\providecommand{\bibinfo}[2]{#2}
\providecommand{\BIBentrySTDinterwordspacing}{\spaceskip=0pt\relax}
\providecommand{\BIBentryALTinterwordstretchfactor}{4}
\providecommand{\BIBentryALTinterwordspacing}{\spaceskip=\fontdimen2\font plus
\BIBentryALTinterwordstretchfactor\fontdimen3\font minus
  \fontdimen4\font\relax}
\providecommand{\BIBforeignlanguage}[2]{{%
\expandafter\ifx\csname l@#1\endcsname\relax
\typeout{** WARNING: IEEEtran.bst: No hyphenation pattern has been}%
\typeout{** loaded for the language `#1'. Using the pattern for}%
\typeout{** the default language instead.}%
\else
\language=\csname l@#1\endcsname
\fi
#2}}
\providecommand{\BIBdecl}{\relax}
\BIBdecl

\bibitem{lewis2020rag}
P.~Lewis, E.~Perez, A.~Piktus, F.~Petroni, V.~Karpukhin, N.~Goyal,
  H.~K{\"u}ttler, M.~Lewis, W.-t. Yih, T.~Rockt{\"a}schel, S.~Riedel, and
  D.~Kiela, ``{Retrieval-Augmented Generation for Knowledge-Intensive NLP
  Tasks},'' in \emph{Advances in Neural Information Processing Systems
  (NeurIPS)}, vol.~33, 2020, pp. 9459--9474.

\bibitem{zou2025poisonedrag}
\BIBentryALTinterwordspacing
W.~Zou, R.~Geng, B.~Wang, and J.~Jia, ``{PoisonedRAG: Knowledge Corruption
  Attacks to Retrieval-Augmented Generation of Large Language Models},'' in
  \emph{Proc. 34th USENIX Security Symposium}.\hskip 1em plus 0.5em minus
  0.4em\relax USENIX Association, 2025, pp. 3827--3844, code and released
  poisoned texts: \url{https://github.com/sleeepeer/PoisonedRAG}. [Online].
  Available:
  \url{https://www.usenix.org/conference/usenixsecurity25/presentation/zou-poisonedrag}
\BIBentrySTDinterwordspacing

\bibitem{zhang2026corruptrag}
B.~Zhang, Y.~Chen, Z.~Liu, L.~Nie, T.~Li, Z.~Liu, and M.~Fang, ``{Practical
  Poisoning Attacks against Retrieval-Augmented Generation},'' in \emph{Proc.
  31st ACM Symposium on Access Control Models and Technologies (SACMAT)}, 2026,
  pp. 33--44, attack name CorruptRAG (CorruptRAG-AS / CorruptRAG-AK). First
  released as arXiv:2504.03957 (2025).

\bibitem{chang2025authchain}
Z.~Chang, M.~Li, X.~Jia, J.~Wang, Y.~Huang, Z.~Jiang, Y.~Liu, and Q.~Wang,
  ``{One Shot Dominance: Knowledge Poisoning Attack on Retrieval-Augmented
  Generation Systems},'' in \emph{Findings of the Association for Computational
  Linguistics: EMNLP 2025}, 2025, pp. 18\,811--18\,825, attack name AuthChain.

\bibitem{shafran2025jamming}
\BIBentryALTinterwordspacing
A.~Shafran, R.~Schuster, and V.~Shmatikov, ``{Machine Against the RAG: Jamming
  Retrieval-Augmented Generation with Blocker Documents},'' in \emph{Proc. 34th
  USENIX Security Symposium}.\hskip 1em plus 0.5em minus 0.4em\relax USENIX
  Association, 2025, pp. 3787--3806, code:
  \url{https://github.com/avitalsh/jamming_attack}. [Online]. Available:
  \url{https://www.usenix.org/conference/usenixsecurity25/presentation/shafran}
\BIBentrySTDinterwordspacing

\bibitem{chaudhari2024phantom}
\BIBentryALTinterwordspacing
H.~Chaudhari, G.~Severi, J.~Abascal, A.~Suri, M.~Jagielski, C.~A.
  Choquette-Choo, M.~Nasr, C.~Nita-Rotaru, and A.~Oprea, ``{Phantom: General
  Backdoor Attacks on Retrieval Augmented Language Generation},'' \emph{ACM
  Transactions on AI Security and Privacy}, 2026, first released as
  arXiv:2405.20485 (2024). [Online]. Available:
  \url{https://dl.acm.org/doi/10.1145/3796729}
\BIBentrySTDinterwordspacing

\bibitem{xue2024badrag}
\BIBentryALTinterwordspacing
J.~Xue, M.~Zheng, Y.~Hu, F.~Liu, X.~Chen, and Q.~Lou, ``{BadRAG: Identifying
  Vulnerabilities in Retrieval Augmented Generation of Large Language
  Models},'' 2024, arXiv preprint arXiv:2406.00083. [Online]. Available:
  \url{https://arxiv.org/abs/2406.00083}
\BIBentrySTDinterwordspacing

\bibitem{zhang2025rsb}
\BIBentryALTinterwordspacing
B.~Zhang, H.~Xin, J.~Li, D.~Zhang, M.~Fang, Z.~Liu, L.~Nie, and Z.~Liu,
  ``{Benchmarking Poisoning Attacks against Retrieval-Augmented Generation},''
  2025, arXiv preprint arXiv:2505.18543. RAG Security Bench (RSB). [Online].
  Available: \url{https://arxiv.org/abs/2505.18543}
\BIBentrySTDinterwordspacing

\bibitem{cheng2025raguard}
Z.~Cheng, J.~Sun, A.~Gao, Y.~Quan, Z.~Liu, X.~Hu, and M.~Fang, ``{Secure
  Retrieval-Augmented Generation against Poisoning Attacks},'' in \emph{Proc.
  IEEE International Conference on Big Data (BigData)}, 2025, pp. 1799--1806.

\bibitem{kim2025ragdefender}
M.~Kim, H.~Lee, and H.~Koo, ``{Rescuing the Unpoisoned: Efficient Defense
  against Knowledge Corruption Attacks on RAG Systems},'' in \emph{Proc. Annual
  Computer Security Applications Conference (ACSAC)}, 2025, pp. 1178--1192,
  system name RAGDefender. Code:
  \url{https://github.com/SecAI-Lab/RAGDefender}.

\bibitem{zhou2025trustrag}
\BIBentryALTinterwordspacing
H.~Zhou, K.-H. Lee, Z.~Zhan, Y.~Chen, Z.~Li, Z.~Wang, H.~Haddadi, and
  E.~Yilmaz, ``{TrustRAG: Enhancing Robustness and Trustworthiness in
  Retrieval-Augmented Generation},'' 2025, arXiv preprint arXiv:2501.00879.
  Code: \url{https://github.com/HuichiZhou/TrustRAG}. [Online]. Available:
  \url{https://arxiv.org/abs/2501.00879}
\BIBentrySTDinterwordspacing

\bibitem{shen2025reliabilityrag}
Z.~Shen, B.~Imana, T.~Wu, C.~Xiang, P.~Mittal, and A.~Korolova,
  ``{ReliabilityRAG: Effective and Provably Robust Defense for RAG-based
  Web-Search},'' in \emph{Advances in Neural Information Processing Systems
  (NeurIPS)}, 2025.

\bibitem{xian2025drs}
\BIBentryALTinterwordspacing
X.~Xian, T.~Wang, L.~You, and Y.~Qi, ``{Understanding Data Poisoning Attacks
  for RAG: Insights and Algorithms},'' OpenReview submission 2aL6gcFX7q,
  submitted to ICLR 2025 (unpublished), 2025, proposes the Directional Relative
  Shift (DRS) embedding-space detector. [Online]. Available:
  \url{https://openreview.net/forum?id=2aL6gcFX7q}
\BIBentrySTDinterwordspacing

\bibitem{chen2026radar}
\BIBentryALTinterwordspacing
Z.~Chen, Y.~Lyu, Y.~Liu, W.~Han, J.~Dong, C.~Shan, and T.~Tan, ``{RADAR:
  Defending RAG Dynamically against Retrieval Corruption},'' 2026, arXiv
  preprint arXiv:2605.22041. Code:
  \url{https://github.com/Etherealllllll/RADAR_code}. [Online]. Available:
  \url{https://arxiv.org/abs/2605.22041}
\BIBentrySTDinterwordspacing

\bibitem{radford2019gpt2}
\BIBentryALTinterwordspacing
A.~Radford, J.~Wu, R.~Child, D.~Luan, D.~Amodei, and I.~Sutskever, ``{Language
  Models are Unsupervised Multitask Learners},'' OpenAI technical report, 2019.
  [Online]. Available:
  \url{https://cdn.openai.com/better-language-models/language_models_are_unsupervised_multitask_learners.pdf}
\BIBentrySTDinterwordspacing

\bibitem{jain2023baselinedefenses}
\BIBentryALTinterwordspacing
N.~Jain, A.~Schwarzschild, Y.~Wen, G.~Somepalli, J.~Kirchenbauer, P.-y. Chiang,
  M.~Goldblum, A.~Saha, J.~Geiping, and T.~Goldstein, ``{Baseline Defenses for
  Adversarial Attacks Against Aligned Language Models},'' 2023, arXiv preprint
  arXiv:2309.00614. [Online]. Available: \url{https://arxiv.org/abs/2309.00614}
\BIBentrySTDinterwordspacing

\bibitem{zhang2024hijackrag}
\BIBentryALTinterwordspacing
Y.~Zhang, Q.~Li, T.~Du, X.~Zhang, X.~Zhao, Z.~Feng, and J.~Yin, ``{HijackRAG:
  Hijacking Attacks against Retrieval-Augmented Large Language Models},'' 2024,
  arXiv preprint arXiv:2410.22832. [Online]. Available:
  \url{https://arxiv.org/abs/2410.22832}
\BIBentrySTDinterwordspacing

\bibitem{cheng2024trojanrag}
\BIBentryALTinterwordspacing
P.~Cheng, Y.~Ding, T.~Ju, Z.~Wu, W.~Du, P.~Yi, Z.~Zhang, and G.~Liu,
  ``{TrojanRAG: Retrieval-Augmented Generation Can Be Backdoor Driver in Large
  Language Models},'' 2024, arXiv preprint arXiv:2405.13401. Code:
  \url{https://github.com/CTZhou-byte/TrojanRAG}. [Online]. Available:
  \url{https://arxiv.org/abs/2405.13401}
\BIBentrySTDinterwordspacing

\bibitem{zhong2023corpuspoisoning}
Z.~Zhong, Z.~Huang, A.~Wettig, and D.~Chen, ``{Poisoning Retrieval Corpora by
  Injecting Adversarial Passages},'' in \emph{Proc. Conference on Empirical
  Methods in Natural Language Processing (EMNLP)}, 2023, pp. 13\,764--13\,775,
  code: \url{https://github.com/princeton-nlp/corpus-poisoning}.

\bibitem{gong2025topicfliprag}
Y.~Gong, Z.~Chen, J.~Liu, M.~Chen, F.~Yu, W.~Lu, X.~Wang, and X.~Liu,
  ``{Topic-FlipRAG: Topic-Orientated Adversarial Opinion Manipulation Attacks
  to Retrieval-Augmented Generation Models},'' in \emph{Proc. 34th USENIX
  Security Symposium}.\hskip 1em plus 0.5em minus 0.4em\relax USENIX
  Association, 2025, pp. 3807--3826, code:
  \url{https://github.com/LauJames/Topic-FlipRAG}.

\bibitem{cho2024garag}
S.~Cho, S.~Jeong, J.~Seo, T.~Hwang, and J.~C. Park, ``{Typos that Broke the
  RAG's Back: Genetic Attack on RAG Pipeline by Simulating Documents in the
  Wild via Low-level Perturbations},'' in \emph{Findings of the Association for
  Computational Linguistics: EMNLP 2024}, 2024, pp. 2826--2844, attack name
  GARAG. Code: \url{https://github.com/zomss/GARAG}.

\bibitem{greshake2023ipi}
K.~Greshake, S.~Abdelnabi, S.~Mishra, C.~Endres, T.~Holz, and M.~Fritz, ``{Not
  what you've signed up for: Compromising Real-World LLM-Integrated
  Applications with Indirect Prompt Injection},'' in \emph{Proc. 16th ACM
  Workshop on Artificial Intelligence and Security (AISec)}, 2023, pp. 79--90.

\bibitem{abdelnabi2025taskdrift}
S.~Abdelnabi, A.~Fay, G.~Cherubin, A.~Salem, M.~Fritz, and A.~Paverd, ``{Get my
  drift? Catching LLM Task Drift with Activation Deltas},'' in \emph{Proc. IEEE
  Conference on Secure and Trustworthy Machine Learning (SaTML)}, 2025, pp.
  43--67.

\bibitem{ebrahimi2018hotflip}
J.~Ebrahimi, A.~Rao, D.~Lowd, and D.~Dou, ``{HotFlip: White-Box Adversarial
  Examples for Text Classification},'' in \emph{Proc. 56th Annual Meeting of
  the Association for Computational Linguistics (ACL), Volume 2: Short Papers},
  2018, pp. 31--36.

\bibitem{zhang2024advdecoding}
C.~Zhang, T.~Zhang, and V.~Shmatikov, ``{Adversarial Decoding: Generating
  Readable Documents for Adversarial Objectives},'' in \emph{Findings of the
  Association for Computational Linguistics: EACL 2026}, 2026, pp. 2053--2068.

\bibitem{wang2026purpose}
\BIBentryALTinterwordspacing
Z.~Wang, Y.~Zhu, M.~Dong, Y.~Lou, Y.~Li, Z.~Zhang, W.~Tong, Y.~Zhang, J.~Hua,
  and S.~Zhong, ``{PURPOSE: Poisoning Conflict Resolution in RAG via
  Proxy-Fact-Grounded Updates},'' 2026, arXiv preprint arXiv:2608.04756.
  [Online]. Available: \url{https://arxiv.org/abs/2608.04756}
\BIBentrySTDinterwordspacing

\bibitem{patil2026ragshield}
\BIBentryALTinterwordspacing
K.~Patil, ``{RAGShield: Detecting Numerical Claim Manipulation in Government
  RAG Systems},'' 2026, arXiv preprint arXiv:2604.00387. [Online]. Available:
  \url{https://arxiv.org/abs/2604.00387}
\BIBentrySTDinterwordspacing

\bibitem{hwang2025rarag}
J.~Hwang, J.~Park, H.~Park, D.~Kim, S.~Park, and J.~Ok, ``{Retrieval-Augmented
  Generation with Estimation of Source Reliability},'' in \emph{Proc.
  Conference on Empirical Methods in Natural Language Processing (EMNLP)},
  2025, pp. 34\,279--34\,303, system name RA-RAG. Code:
  \url{https://github.com/ml-postech/RA-RAG}.

\bibitem{koganti2026sourcererank}
\BIBentryALTinterwordspacing
Y.~T. Koganti and H.~G.-L. Belinchon, ``{Source-Aware Reranking for
  Retrieval-Augmented Generation: A Reliability Prior Approach},'' 2026, arXiv
  preprint arXiv:2607.22584. [Online]. Available:
  \url{https://arxiv.org/abs/2607.22584}
\BIBentrySTDinterwordspacing

\bibitem{zheng2025grada}
J.~Zheng, A.~P. Gema, G.~Hong, X.~He, P.~Minervini, Y.~Sun, and Q.~Xu,
  ``{GRADA: Graph-based Reranking against Adversarial Documents Attack},'' in
  \emph{Proc. Conference on Empirical Methods in Natural Language Processing
  (EMNLP)}, 2025, pp. 22\,244--22\,266.

\bibitem{kim2025gmtp}
S.~Kim, J.~Kim, Y.~Jeon, and G.~G. Lee, ``{Safeguarding RAG Pipelines with
  GMTP: A Gradient-based Masked Token Probability Method for Poisoned Document
  Detection},'' in \emph{Findings of the Association for Computational
  Linguistics: ACL 2025}, 2025, pp. 24\,597--24\,614, code:
  \url{https://github.com/mountinyy/GMTP}.

\bibitem{si2025seconrag}
X.~Si, M.~Zhu, S.~Qin, L.~Yu, L.~Zhang, S.~Liu, X.~Li, R.~Duan, Y.~Liu, and
  X.~Jia, ``{SeCon-RAG: A Two-Stage Semantic Filtering and Conflict-Free
  Framework for Trustworthy RAG},'' in \emph{Advances in Neural Information
  Processing Systems (NeurIPS)}, 2025.

\bibitem{edemacu2025filterrag}
K.~Edemacu, V.~M. Shashidhar, D.~Abudu, M.~Tuape, B.~Jang, and J.~W. Kim,
  ``{Defending Against Knowledge Poisoning Attacks During Retrieval-Augmented
  Generation},'' in \emph{Proc. IEEE International Conference on Information
  Reuse and Integration for Data Science (IRI)}, 2026, pp. 460--465, system
  names FilterRAG / ML-FilterRAG.

\bibitem{wang2026datafilter}
Y.~Wang, S.~Chen, R.~Alkhudair, B.~Alomair, and D.~Wagner, ``{Defending Against
  Prompt Injection with DataFilter},'' in \emph{Proc. IEEE Conference on Secure
  and Trustworthy Machine Learning (SaTML)}, 2026, first released as
  arXiv:2510.19207 (2025). Code: \url{https://github.com/yizhu-joy/DataFilter}.

\bibitem{xiang2024robustrag}
C.~Xiang, T.~Wu, Z.~Zhong, D.~Wagner, D.~Chen, and P.~Mittal, ``{Certifiably
  Robust RAG against Retrieval Corruption},'' in \emph{Proc. IEEE Conference on
  Secure and Trustworthy Machine Learning (SaTML)}, 2026, system name
  RobustRAG. First released as arXiv:2405.15556 (2024). Code:
  \url{https://github.com/inspire-group/RobustRAG}.

\bibitem{wei2025instructrag}
\BIBentryALTinterwordspacing
Z.~Wei, W.-L. Chen, and Y.~Meng, ``{InstructRAG: Instructing
  Retrieval-Augmented Generation via Self-Synthesized Rationales},'' in
  \emph{Proc. International Conference on Learning Representations (ICLR)},
  2025, code: \url{https://github.com/weizhepei/InstructRAG}. [Online].
  Available: \url{https://openreview.net/forum?id=P1qhkp8gQT}
\BIBentrySTDinterwordspacing

\bibitem{wang2025astuterag}
F.~Wang, X.~Wan, R.~Sun, J.~Chen, and S.~{\"O}. Ar{\i}k, ``{Astute RAG:
  Overcoming Imperfect Retrieval Augmentation and Knowledge Conflicts for Large
  Language Models},'' in \emph{Proc. 63rd Annual Meeting of the Association for
  Computational Linguistics (ACL)}, 2025, pp. 30\,553--30\,571.

\bibitem{ren2026dscan}
Y.~Ren, Z.~Zhao, Y.~Fu, X.~Luo, Y.-C. Chang, and C.-T. Lin, ``{When Context
  Bites: Detecting RAG Poisoning via Document-Level Attention Collapse},'' in
  \emph{Proc. 49th International ACM SIGIR Conference on Research and
  Development in Information Retrieval}, 2026, pp. 4080--4086, system name
  D-SCAN. Code: \url{https://github.com/yingtaoren/D-Scan}.

\bibitem{choudhary2026avfilter}
S.~Choudhary, N.~Palumbo, A.~Hooda, K.~D. Dvijotham, and S.~Jha, ``{Through the
  Stealth Lens: Attention-Aware Defenses Against Poisoning in RAG},'' in
  \emph{Proc. International Conference on Machine Learning (ICML)}, 2026,
  attention-variance (AV) filter. First released as arXiv:2506.04390 (2025).
  Code:
  \url{https://github.com/sarthak-choudhary/Stealthy_Attacks_Against_RAG}.

\bibitem{debenedetti2026camel}
E.~Debenedetti, I.~Shumailov, T.~Fan, J.~Hayes, N.~Carlini, D.~Fabian, C.~Kern,
  C.~Shi, A.~Terzis, and F.~Tram{\`e}r, ``{Defeating Prompt Injections by
  Design},'' in \emph{Proc. IEEE Conference on Secure and Trustworthy Machine
  Learning (SaTML)}, 2026, system name CaMeL. First released as
  arXiv:2503.18813 (2025).

\bibitem{tan2025revprag}
X.~Tan, H.~Luan, M.~Luo, X.~Sun, P.~Chen, and J.~Dai, ``{RevPRAG: Revealing
  Poisoning Attacks in Retrieval-Augmented Generation through LLM Activation
  Analysis},'' in \emph{Findings of the Association for Computational
  Linguistics: EMNLP 2025}, 2025, pp. 12\,999--13\,011.

\bibitem{zhang2025ragforensics}
B.~Zhang, H.~Xin, M.~Fang, Z.~Liu, B.~Yi, T.~Li, and Z.~Liu, ``{Traceback of
  Poisoning Attacks to Retrieval-Augmented Generation},'' in \emph{Proc. ACM
  Web Conference (WWW)}, 2025, pp. 2085--2097, system name RAGForensics.

\bibitem{zhang2026ragorigin}
B.~Zhang, H.~Xin, Y.~Chen, Z.~Liu, B.~Yi, T.~Li, L.~Nie, Z.~Liu, and M.~Fang,
  ``{Who Taught the Lie? Responsibility Attribution for Poisoned Knowledge in
  Retrieval-Augmented Generation},'' in \emph{Proc. IEEE Symposium on Security
  and Privacy (S\&P)}, 2026, pp. 4203--4222, system name RAGOrigin.

\bibitem{zeng2025raguardbench}
\BIBentryALTinterwordspacing
L.~Zeng, R.~Gupta, D.~Motwani, D.~Yang, and Y.~Zhang, ``{Worse than Zero-shot?
  A Fact-Checking Dataset for Evaluating the Robustness of RAG Against
  Misleading Retrievals},'' in \emph{Advances in Neural Information Processing
  Systems (NeurIPS)}, 2025, benchmark name RAGuard. [Online]. Available:
  \url{https://arxiv.org/abs/2502.16101}
\BIBentrySTDinterwordspacing

\bibitem{kumar2026raguardzkip}
\BIBentryALTinterwordspacing
P.~Kumar, T.~Nielson, T.~Kolhe, S.~Zala, and V.~Li, ``{RAGuard: A Layered
  Defense Framework for Retrieval-Augmented Generation Systems Against Data
  Poisoning},'' 2026, arXiv preprint arXiv:2607.26339. Workshop versions:
  NeurIPS ResponsibleFM 2025; AAAI FrontierIR 2026. Referred to as RAGuard-ZKIP
  to disambiguate from Cheng et al. [Online]. Available:
  \url{https://arxiv.org/abs/2607.26339}
\BIBentrySTDinterwordspacing

\bibitem{elkan2001costsensitive}
\BIBentryALTinterwordspacing
C.~Elkan, ``{The Foundations of Cost-Sensitive Learning},'' in \emph{Proc. 17th
  International Joint Conference on Artificial Intelligence (IJCAI)}, 2001, pp.
  973--978. [Online]. Available:
  \url{https://cseweb.ucsd.edu/~elkan/rescale.pdf}
\BIBentrySTDinterwordspacing

\bibitem{zadrozny2001costs}
B.~Zadrozny and C.~Elkan, ``{Learning and Making Decisions When Costs and
  Probabilities Are Both Unknown},'' in \emph{Proc. 7th ACM SIGKDD
  International Conference on Knowledge Discovery and Data Mining (KDD)}, 2001,
  pp. 204--213.

\bibitem{guo2017calibration}
C.~Guo, G.~Pleiss, Y.~Sun, and K.~Q. Weinberger, ``{On Calibration of Modern
  Neural Networks},'' in \emph{Proc. 34th International Conference on Machine
  Learning (ICML)}, ser. PMLR, vol.~70, 2017, pp. 1321--1330.

\bibitem{naeini2015ece}
M.~P. Naeini, G.~F. Cooper, and M.~Hauskrecht, ``{Obtaining Well Calibrated
  Probabilities Using Bayesian Binning},'' in \emph{Proc. 29th AAAI Conference
  on Artificial Intelligence}, 2015, pp. 2901--2907.

\bibitem{chow1970reject}
C.~K. Chow, ``{On Optimum Recognition Error and Reject Tradeoff},'' \emph{IEEE
  Transactions on Information Theory}, vol.~16, no.~1, pp. 41--46, 1970.

\bibitem{drummond2006costcurves}
C.~Drummond and R.~C. Holte, ``{Cost Curves: An Improved Method for Visualizing
  Classifier Performance},'' \emph{Machine Learning}, vol.~65, no.~1, pp.
  95--130, 2006.

\bibitem{androutsopoulos2000costsensitivespam}
\BIBentryALTinterwordspacing
I.~Androutsopoulos, J.~Koutsias, K.~V. Chandrinos, G.~Paliouras, and C.~D.
  Spyropoulos, ``{An Evaluation of Naive Bayesian Anti-Spam Filtering},'' in
  \emph{Proc. Workshop on Machine Learning in the New Information Age, 11th
  European Conference on Machine Learning (ECML)}, Barcelona, Spain, 2000, pp.
  9--17. [Online]. Available: \url{https://arxiv.org/abs/cs/0006013}
\BIBentrySTDinterwordspacing

\bibitem{axelsson2000baserate}
S.~Axelsson, ``{The Base-Rate Fallacy and the Difficulty of Intrusion
  Detection},'' \emph{ACM Transactions on Information and System Security},
  vol.~3, no.~3, pp. 186--205, Aug. 2000, earlier version: Proc. 6th ACM
  Conference on Computer and Communications Security (CCS), 1999, pp. 1--7,
  doi:10.1145/319709.319710.

\bibitem{provost2001robust}
F.~Provost and T.~Fawcett, ``{Robust Classification for Imprecise
  Environments},'' \emph{Machine Learning}, vol.~42, no.~3, pp. 203--231, 2001.

\bibitem{josang2002beta}
\BIBentryALTinterwordspacing
A.~J{\o}sang and R.~Ismail, ``{The Beta Reputation System},'' in \emph{Proc.
  15th Bled Electronic Commerce Conference}, Bled, Slovenia, 2002, paper 41.
  [Online]. Available: \url{https://aisel.aisnet.org/bled2002/41/}
\BIBentrySTDinterwordspacing

\bibitem{sun2006trustattacks}
Y.~L. Sun, Z.~Han, W.~Yu, and K.~J.~R. Liu, ``{Attacks on Trust Evaluation in
  Distributed Networks},'' in \emph{Proc. 40th Annual Conference on Information
  Sciences and Systems (CISS)}, 2006, pp. 1461--1466.

\bibitem{pathak2026coverage}
\BIBentryALTinterwordspacing
P.~K. Pathak and T.~K. Sharma, ``{Coverage Is Not Containment: A Fundamental
  Limit of Admission-Time Defenses Against Coordinated Poisoning of Vector
  Retrieval},'' Aug. 2026, arXiv preprint arXiv:2608.16044. [Online].
  Available: \url{https://arxiv.org/abs/2608.16044}
\BIBentrySTDinterwordspacing

\bibitem{pendlebury2019tesseract}
F.~Pendlebury, F.~Pierazzi, R.~Jordaney, J.~Kinder, and L.~Cavallaro,
  ``{TESSERACT: Eliminating Experimental Bias in Malware Classification across
  Space and Time},'' in \emph{Proc. 28th USENIX Security Symposium}.\hskip 1em
  plus 0.5em minus 0.4em\relax USENIX Association, 2019, pp. 729--746.

\bibitem{chow2026beyondtesseract}
T.~Chow, M.~D'Onghia, L.~Linhardt, Z.~Kan, D.~Arp, L.~Cavallaro, and
  F.~Pierazzi, ``{Beyond the TESSERACT: Trustworthy Dataset Curation for Sound
  Evaluations of Android Malware Classifiers},'' in \emph{Proc. IEEE Conference
  on Secure and Trustworthy Machine Learning (SaTML)}, 2026.

\bibitem{arp2022dosdonts}
\BIBentryALTinterwordspacing
D.~Arp, E.~Quiring, F.~Pendlebury, A.~Warnecke, F.~Pierazzi, C.~Wressnegger,
  L.~Cavallaro, and K.~Rieck, ``{Dos and Don'ts of Machine Learning in Computer
  Security},'' in \emph{Proc. 31st USENIX Security Symposium}.\hskip 1em plus
  0.5em minus 0.4em\relax USENIX Association, 2022, pp. 3971--3988, checklist
  at \url{https://dodo-mlsec.org/}. [Online]. Available:
  \url{https://www.usenix.org/conference/usenixsecurity22/presentation/arp}
\BIBentrySTDinterwordspacing

\bibitem{sommer2010closedworld}
R.~Sommer and V.~Paxson, ``{Outside the Closed World: On Using Machine Learning
  for Network Intrusion Detection},'' in \emph{Proc. IEEE Symposium on Security
  and Privacy (S\&P)}, 2010, pp. 305--316.

\bibitem{geirhos2020shortcut}
R.~Geirhos, J.-H. Jacobsen, C.~Michaelis, R.~Zemel, W.~Brendel, M.~Bethge, and
  F.~A. Wichmann, ``{Shortcut Learning in Deep Neural Networks},'' \emph{Nature
  Machine Intelligence}, vol.~2, no.~11, pp. 665--673, 2020.

\bibitem{risse2024limits}
\BIBentryALTinterwordspacing
N.~Risse and M.~B{\"o}hme, ``{Uncovering the Limits of Machine Learning for
  Automatic Vulnerability Detection},'' in \emph{Proc. 33rd USENIX Security
  Symposium}.\hskip 1em plus 0.5em minus 0.4em\relax USENIX Association, 2024,
  pp. 4247--4264. [Online]. Available:
  \url{https://www.usenix.org/conference/usenixsecurity24/presentation/risse}
\BIBentrySTDinterwordspacing

\bibitem{athalye2018obfuscated}
A.~Athalye, N.~Carlini, and D.~Wagner, ``{Obfuscated Gradients Give a False
  Sense of Security: Circumventing Defenses to Adversarial Examples},'' in
  \emph{Proc. 35th International Conference on Machine Learning (ICML)}, ser.
  Proceedings of Machine Learning Research, vol.~80, 2018, pp. 274--283.

\bibitem{carlini2019evaluating}
\BIBentryALTinterwordspacing
N.~Carlini, A.~Athalye, N.~Papernot, W.~Brendel, J.~Rauber, D.~Tsipras,
  I.~Goodfellow, A.~Madry, and A.~Kurakin, ``{On Evaluating Adversarial
  Robustness},'' 2019, arXiv preprint arXiv:1902.06705. Living document;
  \url{https://github.com/evaluating-adversarial-robustness/adv-eval-paper}.
  [Online]. Available: \url{https://arxiv.org/abs/1902.06705}
\BIBentrySTDinterwordspacing

\bibitem{tramer2020adaptive}
F.~Tram{\`e}r, N.~Carlini, W.~Brendel, and A.~Madry, ``{On Adaptive Attacks to
  Adversarial Example Defenses},'' in \emph{Advances in Neural Information
  Processing Systems (NeurIPS)}, vol.~33, 2020, pp. 1633--1645.

\bibitem{carlini2022lira}
N.~Carlini, S.~Chien, M.~Nasr, S.~Song, A.~Terzis, and F.~Tram{\`e}r,
  ``{Membership Inference Attacks From First Principles},'' in \emph{Proc. IEEE
  Symposium on Security and Privacy (S\&P)}, 2022, pp. 1897--1914.

\bibitem{jacob2025promptshield}
D.~Jacob, H.~Alzahrani, Z.~Hu, B.~Alomair, and D.~Wagner, ``{PromptShield:
  Deployable Detection for Prompt Injection Attacks},'' in \emph{Proc. 15th ACM
  Conference on Data and Application Security and Privacy (CODASPY)}, 2025, pp.
  341--352.

\bibitem{hand2009measuring}
D.~J. Hand, ``{Measuring Classifier Performance: A Coherent Alternative to the
  Area under the ROC Curve},'' \emph{Machine Learning}, vol.~77, no.~1, pp.
  103--123, 2009.

\bibitem{munir2026tris}
M.~B. Munir, A.~J. Ononto, N.~S. Joynab, B.~Thuraisingham, and L.~Khan,
  ``{TRIS: A Tri-Layer Retrieval Integrity Sieve Against Knowledge
  Poisoning},'' in \emph{Findings of the Association for Computational
  Linguistics: EMNLP 2026}, 2026, to appear.

\bibitem{mohanty2026trishieldrag}
\BIBentryALTinterwordspacing
S.~K. Mohanty, R.~Patel, K.~Yuvaraj, J.~Chaudhary, and D.~Singhania,
  ``{TriShieldRAG: 3 Rings, One Blind Spot in Layered Defenses for
  Retrieval-Augmented Generation},'' Jul. 2026, arXiv preprint
  arXiv:2607.23838. [Online]. Available: \url{https://arxiv.org/abs/2607.23838}
\BIBentrySTDinterwordspacing

\bibitem{xu2026ragsieve}
\BIBentryALTinterwordspacing
X.~Xu and Y.~Y. Li, ``{RAGSieve: Self-Referenced Local Contrast for
  Knowledge-Poison Detection in Retrieval-Augmented Generation},'' Aug. 2026,
  arXiv preprint arXiv:2608.13010. [Online]. Available:
  \url{https://arxiv.org/abs/2608.13010}
\BIBentrySTDinterwordspacing

\bibitem{jung2026camodocs}
\BIBentryALTinterwordspacing
J.~Jung, H.~Zheng, H.~Jang, J.~Song, B.~Chen, and J.~Lee, ``{CamoDocs: A
  Poisoning Attack Against Retrieval-Augmented Language Models Using
  Camouflaged Documents},'' Aug. 2026, arXiv preprint arXiv:2608.28389. To
  appear in Proc. EMNLP 2026. [Online]. Available:
  \url{https://arxiv.org/abs/2608.28389}
\BIBentrySTDinterwordspacing

\bibitem{wang2026refinerag}
\BIBentryALTinterwordspacing
Z.~Wang, G.~Wang, and K.~Wang, ``{RefineRAG: Word-Level Poisoning Attacks via
  Retriever-Guided Text Refinement},'' Apr. 2026, arXiv preprint
  arXiv:2604.07403. [Online]. Available: \url{https://arxiv.org/abs/2604.07403}
\BIBentrySTDinterwordspacing

\bibitem{zhang2026inceptionrag}
\BIBentryALTinterwordspacing
J.~Zhang, M.~Chen, X.~Ren, Z.~Zhang, Y.~Shu, Y.~Gao, and Z.~Zhang,
  ``{InceptionRAG: Stealthy Poisoning Attack Against Retrieval-Augmented
  Generation},'' Sep. 2026, arXiv preprint arXiv:2609.16818. To appear in Proc.
  ACM CCS 2026. [Online]. Available: \url{https://arxiv.org/abs/2609.16818}
\BIBentrySTDinterwordspacing

\bibitem{wang2026securecollarag}
Z.~Wang, D.~He, Z.~Zhang, Y.~Liu, J.~Liu, Z.~Zeng, Z.~Qin, Z.~Li, X.~Li,
  H.~Yao, J.~An, Y.~Liu, Y.~Li, Q.~Sun, X.~Liu, and L.~Zhu, ``{Combating
  Knowledge Corruption in Agent Systems: A Byzantine-Tolerant Secure
  Collaborative RAG Framework},'' in \emph{Proc. ACM Web Conference (WWW)},
  2026, pp. 2661--2672, referred to as SecureCollaRAG.

\bibitem{tan2026prarag}
X.~Tan, Y.~Zheng, C.~Huo, Y.~Zhang, Y.~Liu, H.~Luan, Z.~Yu, J.~Dai, X.~Sun, and
  P.~Chen, ``{PRA-RAG: Provably Robust Aggregation in Retrieval-Augmented
  Generation against Retrieval Corruption},'' in \emph{Findings of the
  Association for Computational Linguistics: ACL 2026}, 2026, pp.
  36\,002--36\,017.

\bibitem{hoffman2009reputation}
K.~Hoffman, D.~Zage, and C.~Nita-Rotaru, ``{A Survey of Attack and Defense
  Techniques for Reputation Systems},'' \emph{ACM Computing Surveys}, vol.~42,
  no.~1, pp. 1:1--1:31, 2009.

\bibitem{reimers2019sbert}
N.~Reimers and I.~Gurevych, ``{Sentence-BERT: Sentence Embeddings using Siamese
  BERT-Networks},'' in \emph{Proc. Conference on Empirical Methods in Natural
  Language Processing and 9th International Joint Conference on Natural
  Language Processing (EMNLP-IJCNLP)}, 2019, pp. 3980--3990.

\bibitem{wang2020minilm}
W.~Wang, F.~Wei, L.~Dong, H.~Bao, N.~Yang, and M.~Zhou, ``{MiniLM: Deep
  Self-Attention Distillation for Task-Agnostic Compression of Pre-Trained
  Transformers},'' in \emph{Advances in Neural Information Processing Systems
  (NeurIPS)}, vol.~33, 2020, pp. 5776--5788.

\bibitem{nasr2025attacker}
\BIBentryALTinterwordspacing
M.~Nasr, N.~Carlini, C.~Sitawarin, S.~V. Schulhoff, J.~Hayes, M.~Ilie,
  J.~Pluto, S.~Song, H.~Chaudhari, I.~Shumailov, A.~G. Thakurta, K.~Y. Xiao,
  A.~Terzis, and F.~Tram{\`e}r, ``{The Attacker Moves Second: Stronger Adaptive
  Attacks Bypass Defenses Against LLM Jailbreaks and Prompt Injections},'' in
  \emph{Proc. 35th USENIX Security Symposium}.\hskip 1em plus 0.5em minus
  0.4em\relax USENIX Association, 2026, pp. 1467--1486. [Online]. Available:
  \url{https://www.usenix.org/conference/usenixsecurity26/presentation/nasr}
\BIBentrySTDinterwordspacing

\bibitem{douceur2002sybil}
J.~R. Douceur, ``{The Sybil Attack},'' in \emph{Peer-to-Peer Systems: First
  International Workshop (IPTPS)}, ser. LNCS, vol. 2429.\hskip 1em plus 0.5em
  minus 0.4em\relax Springer, 2002, pp. 251--260.

\bibitem{dai2026disarmrag}
Y.~Dai, Z.~Ji, Z.~Li, K.~Li, and S.~Wang, ``{DisarmRAG: Stealthy
  Retriever-Centric Poisoning to Disable Self-Correction in Retrieval-Augmented
  Generation},'' in \emph{Proc. ACM SIGSAC Conference on Computer and
  Communications Security (CCS)}, 2026, to appear.

\bibitem{thakur2021beir}
\BIBentryALTinterwordspacing
N.~Thakur, N.~Reimers, A.~R{\"u}ckl{\'e}, A.~Srivastava, and I.~Gurevych,
  ``{BEIR: A Heterogeneous Benchmark for Zero-shot Evaluation of Information
  Retrieval Models},'' in \emph{Proc. NeurIPS Datasets and Benchmarks Track
  (Round 2)}, 2021. [Online]. Available:
  \url{https://datasets-benchmarks-proceedings.neurips.cc/paper_files/paper/2021/hash/65b9eea6e1cc6bb9f0cd2a47751a186f-Abstract-round2.html}
\BIBentrySTDinterwordspacing

\bibitem{jin2019pubmedqa}
Q.~Jin, B.~Dhingra, Z.~Liu, W.~W. Cohen, and X.~Lu, ``{PubMedQA: A Dataset for
  Biomedical Research Question Answering},'' in \emph{Proc. Conference on
  Empirical Methods in Natural Language Processing and 9th International Joint
  Conference on Natural Language Processing (EMNLP-IJCNLP)}, 2019, pp.
  2567--2577.

\bibitem{chen2021finqa}
Z.~Chen, W.~Chen, C.~Smiley, S.~Shah, I.~Borova, D.~Langdon, R.~Moussa,
  M.~Beane, T.-H. Huang, B.~Routledge, and W.~Y. Wang, ``{FinQA: A Dataset of
  Numerical Reasoning over Financial Data},'' in \emph{Proc. Conference on
  Empirical Methods in Natural Language Processing (EMNLP)}, 2021, pp.
  3697--3711.

\bibitem{zhu2021tatqa}
F.~Zhu, W.~Lei, Y.~Huang, C.~Wang, S.~Zhang, J.~Lv, F.~Feng, and T.-S. Chua,
  ``{TAT-QA: A Question Answering Benchmark on a Hybrid of Tabular and Textual
  Content in Finance},'' in \emph{Proc. 59th Annual Meeting of the Association
  for Computational Linguistics and 11th International Joint Conference on
  Natural Language Processing (ACL-IJCNLP)}, 2021, pp. 3277--3287.

\bibitem{hendrycks2021cuad}
D.~Hendrycks, C.~Burns, A.~Chen, and S.~Ball, ``{CUAD: An Expert-Annotated NLP
  Dataset for Legal Contract Review},'' in \emph{Proc. NeurIPS Datasets and
  Benchmarks Track}, 2021, data:
  \url{https://github.com/TheAtticusProject/cuad/}.

\bibitem{thornton2026chameleon}
\BIBentryALTinterwordspacing
S.~Thornton, ``{Semantic Chameleon: Corpus-Dependent Poisoning Attacks and
  Defenses in RAG Systems},'' 2026, arXiv preprint arXiv:2603.18034. Code:
  \url{https://github.com/scthornton/semantic-chameleon}. [Online]. Available:
  \url{https://arxiv.org/abs/2603.18034}
\BIBentrySTDinterwordspacing

\bibitem{nie2026poisonfails}
\BIBentryALTinterwordspacing
X.~Nie, H.~Li, S.~Wu, M.~Li, J.~Li, and W.~Jiang, ``{When Poison Fails After
  Retrieval: Revisiting Corpus Poisoning under Chunking and Reranking
  Pipelines},'' Jun. 2026, arXiv preprint arXiv:2606.11265. [Online].
  Available: \url{https://arxiv.org/abs/2606.11265}
\BIBentrySTDinterwordspacing

\bibitem{lee2026polymorphicsybil}
\BIBentryALTinterwordspacing
D.~Lee and J.~Kim, ``{A Failure-Mode Benchmark for Polymorphic Sybil Poisoning
  in RAG},'' Jul. 2026, arXiv preprint arXiv:2607.03739. [Online]. Available:
  \url{https://arxiv.org/abs/2607.03739}
\BIBentrySTDinterwordspacing

\bibitem{zhang2025poisonarena}
L.~Chen, X.~Yang, Y.~Lu, J.~Zhang, X.~Sun, Q.~Liu, S.~Wu, J.~Dong, and L.~Wang,
  ``{Uncovering Competing Poisoning Attacks in Retrieval-Augmented
  Generation},'' in \emph{Proc. 32nd ACM SIGKDD Conference on Knowledge
  Discovery and Data Mining (KDD), Vol. 2}, 2026, pp. 414--424, benchmark name
  PoisonArena. Project page: \url{https://poison-arena.github.io/}.

\bibitem{hu2026confundo}
\BIBentryALTinterwordspacing
H.~Hu, Z.~Jiang, Y.~Lyu, J.~Zhang, Y.~Liu, and K.-H. Chow, ``{Confundo:
  Learning to Generate Robust Poison for Practical RAG Systems},'' Feb. 2026,
  arXiv preprint arXiv:2602.06616. [Online]. Available:
  \url{https://arxiv.org/abs/2602.06616}
\BIBentrySTDinterwordspacing

\bibitem{akinrele2026regime}
\BIBentryALTinterwordspacing
A.~Akinrele and S.~N. Gowda, ``{Prompt Injection Detection is Regime-Dependent:
  A Deployment-Aware Evaluation with Interpretable Structural Signals},'' May
  2026, arXiv preprint arXiv:2605.26999. [Online]. Available:
  \url{https://arxiv.org/abs/2605.26999}
\BIBentrySTDinterwordspacing

\bibitem{shire2026pidsbench}
Y.~K. Shire and S.-C. Kim, ``{PIDS-Bench: Evaluating Prompt-Injection Detectors
  Under Over-Defense, Obfuscation, and Distribution Shift},'' \emph{IEEE
  Access}, vol.~14, pp. 134\,184--134\,205, 2026.

\bibitem{kwiatkowski2019nq}
T.~Kwiatkowski, J.~Palomaki, O.~Redfield, M.~Collins, A.~Parikh, C.~Alberti,
  D.~Epstein, I.~Polosukhin, J.~Devlin, K.~Lee, K.~Toutanova, L.~Jones,
  M.~Kelcey, M.-W. Chang, A.~M. Dai, J.~Uszkoreit, Q.~Le, and S.~Petrov,
  ``{Natural Questions: A Benchmark for Question Answering Research},''
  \emph{Transactions of the Association for Computational Linguistics}, vol.~7,
  pp. 452--466, 2019.

\bibitem{yang2018hotpotqa}
Z.~Yang, P.~Qi, S.~Zhang, Y.~Bengio, W.~W. Cohen, R.~Salakhutdinov, and C.~D.
  Manning, ``{HotpotQA: A Dataset for Diverse, Explainable Multi-hop Question
  Answering},'' in \emph{Proc. Conference on Empirical Methods in Natural
  Language Processing (EMNLP)}, 2018, pp. 2369--2380.

\bibitem{nguyen2016msmarco}
\BIBentryALTinterwordspacing
T.~Nguyen, M.~Rosenberg, X.~Song, J.~Gao, S.~Tiwary, R.~Majumder, and L.~Deng,
  ``{MS MARCO: A Human Generated MAchine Reading COmprehension Dataset},'' in
  \emph{Proc. Workshop on Cognitive Computation: Integrating Neural and
  Symbolic Approaches (CoCo@NIPS)}, ser. CEUR Workshop Proceedings, vol. 1773,
  2016. [Online]. Available:
  \url{https://ceur-ws.org/Vol-1773/CoCoNIPS_2016_paper9.pdf}
\BIBentrySTDinterwordspacing

\bibitem{qwen2025qwen25}
\BIBentryALTinterwordspacing
{Qwen Team}, ``{Qwen2.5 Technical Report},'' 2024, arXiv preprint
  arXiv:2412.15115. [Online]. Available: \url{https://arxiv.org/abs/2412.15115}
\BIBentrySTDinterwordspacing

\bibitem{dittrich2012menlo}
\BIBentryALTinterwordspacing
D.~Dittrich and E.~Kenneally, ``{The Menlo Report: Ethical Principles Guiding
  Information and Communication Technology Research},'' U.S. Department of
  Homeland Security, Science and Technology Directorate, Tech. Rep., Aug. 2012.
  [Online]. Available:
  \url{https://www.dhs.gov/sites/default/files/publications/CSD-MenloPrinciplesCORE-20120803_1.pdf}
\BIBentrySTDinterwordspacing

\end{thebibliography}

\appendices
\crefname{appendix}{Appendix}{Appendices}
\Crefname{appendix}{Appendix}{Appendices}
\crefalias{section}{appendix}
\balance
%
%
\section{Closed Forms for the Trust State}
\label{app:derivations}

This appendix derives the closed forms that \cref{sec:stack} quotes by
reference. They are properties of an update rule under its own
generative model. They are not guarantees, and they support no claim of
residual protection: in every condition we measure, source identity is
an exact label indicator, so no quantity below may be read as a
detection result. Every number quoted here is produced by the E3
simulation, which drives the ledger with hand-set scorer profiles
rather than with the evaluated pipeline.

\noindent\textbf{Notation.}
Notation is that of \cref{sec:stack}. A source holds state $(a,b)$ with
prior $(\pizero m,(1-\pizero)m)$, posterior mean $\pi=a/(a+b)$, and
prior-odds multiplier $\ks=\pi(1-\pizero)/\bigl((1-\pi)\pizero\bigr)$
before clipping to $[\kmin,\kmax]$. Each adjudicated event of weight
$w$ applies
\begin{equation}
a\mapsto\lambda_a a+w\,\ind{\text{poisoned}},\qquad
b\mapsto\lambda_b b+w\,\ind{\text{benign}},
\label{eq:beta-app}
\end{equation}
with forgetting indexed by the source's own events. A target
$\kappa^\star$ corresponds to the mean
$\pi^\star=\kappa^\star\pizero/(\kappa^\star\pizero+1-\pizero)$ and odds
$\omega^\star=\pi^\star/(1-\pi^\star)$; $\ks$ is increasing in $\pi$, so
$\ks\ge\kappa^\star$ iff $a\ge\omega^\star b$.

\begin{lemma}[Saturation]
\label{lem:saturation}
Under symmetric forgetting $\lambda_a=\lambda_b=\lambda$, the evidence mass
after $n$ events from $(a,b)$ is
$(a+b)\lambda^n+M(1-\lambda^n)$ with $M:=w/(1-\lambda)$, so $a+b\to M$
and reaches $95\%$ of $M$ after $n=\lceil\ln 0.05/\ln\lambda\rceil\approx
3/(1-\lambda)$ events.
\end{lemma}
\begin{proof}
Each event maps $a+b\mapsto\lambda(a+b)+w$, an affine recursion with fixed
point $M$; iterate and solve $\lambda^n=0.05$.
\end{proof}

\begin{lemma}[Poison budget and recovery under symmetric forgetting]
\label{lem:budget}
Let $\lambda_a=\lambda_b=\lambda$. From state $(a,b)$ with
$\omega^\star b>a$, the number of consecutive adjudicated poison events
after which $\ks\ge\kappa^\star$ first holds is
\begin{equation}
n^\star(a,b;\kappa^\star)=\left\lceil\frac{\ln\!\bigl(1+(1-\lambda)
(\omega^\star b-a)/w\bigr)}{-\ln\lambda}\right\rceil,
\label{eq:poison-budget}
\end{equation}
and $n^\star=0$ otherwise. From state $(a,b)$ with $a>\omega^\star b$, the
number of consecutive benign events after which $\ks\le\kappa^\star$
first holds is
\begin{equation}
n_{\mathrm{rec}}(a,b;\kappa^\star)=\left\lceil\frac{\ln\!\bigl(1+
(a-\omega^\star b)/(\omega^\star M)\bigr)}{-\ln\lambda}\right\rceil .
\label{eq:recovery}
\end{equation}
\end{lemma}
\begin{proof}
After $n$ poison events, $a_n=\lambda^n a+M(1-\lambda^n)$ and
$b_n=\lambda^n b$. The condition $a_n\ge\omega^\star b_n$ rearranges to
$\lambda^n(\omega^\star b-a+M)\le M$, i.e.\
$\lambda^n\le\bigl(1+(\omega^\star b-a)/M\bigr)^{-1}$; taking logarithms
and substituting $M=w/(1-\lambda)$ gives \eqref{eq:poison-budget}. After
$n$ benign events, $a_n=\lambda^n a$ and $b_n=\lambda^n b+M(1-\lambda^n)$;
$a_n\le\omega^\star b_n$ rearranges to
$\lambda^n(a-\omega^\star b+\omega^\star M)\le\omega^\star M$, which gives
\eqref{eq:recovery}.
\end{proof}

\noindent\textbf{The budget depends on the base rate, so the base
rate is stated with it.}
For a fully farmed source ($a\to0$, $b\to M$) the symmetric budget
\eqref{eq:poison-budget} reduces to
$\lceil\ln(1+\omega^\star)/(-\ln\lambda)\rceil$, independent of $w$. The
deployed configuration forgets asymmetrically, so the budget is instead
located by bisection on the trajectories below, which are monotone in
$n$. At the deployed calibration base rate
$\pizero=\res{E0-pi0-deployed}$ with $\kappa^\star=1$, that budget is
\res{E3-B1-nstar} adjudicated poison events (simulation). The
configuration default $\pizero=0.05$ is a hypothetical and not the
deployed value; at that setting the same computation returns
\res{E3-B1-nstar-pi005}, an order of magnitude cheaper for the attacker.
The gap is the reason no budget figure in this paper appears without its
base rate. Under $\lambda_a\neq\lambda_b$ the trajectories are
$a_n=\lambda_a^n a+M_a(1-\lambda_a^n)$ with $M_a=w/(1-\lambda_a)$ and
$b_n=\lambda_b^n b$, and the recovery count exceeds \eqref{eq:recovery}
because poison evidence decays at the slower rate $\lambda_a$.

\begin{lemma}[Steady state of an on--off source]
\label{lem:steady}
A source whose adjudicated events are poisoned with fraction $p$
(independently, unit weights) converges to
\begin{equation}
\pi_\infty=\frac{p(1-\lambda_b)}{p(1-\lambda_b)+(1-p)(1-\lambda_a)} ,
\label{eq:steady}
\end{equation}
so $\pi_\infty=p$ under symmetric forgetting and, with the defaults
$\lambda_a=0.995$, $\lambda_b=0.98$, $\pi_\infty=4p/(1+3p)$. The source
keeps $\ks<1$ iff $p<\pizero$ (symmetric) or
$p<\pizero/(4-3\pizero)$ (defaults).
\end{lemma}
\begin{proof}
In expectation each event maps $a\mapsto\lambda_a a+pw$ and
$b\mapsto\lambda_b b+(1-p)w$, with fixed points $a_\infty=pw/(1-\lambda_a)$
and $b_\infty=(1-p)w/(1-\lambda_b)$; \eqref{eq:steady} is
$a_\infty/(a_\infty+b_\infty)$. With the defaults
$(1-\lambda_b)/(1-\lambda_a)=4$. Then $\ks<1$ iff
$\pi_\infty<\pizero$; for the defaults, $4p<\pizero(1+3p)$ iff
$p(4-3\pizero)<\pizero$.
\end{proof}

Lemma~\ref{lem:steady} is read twice. For the attacker, it gives the
poison fraction at which an on--off source's steady-state mean crosses
the base rate, which asymmetric forgetting lowers by roughly a factor of
four. For the defender, an honest source whose adjudications are flipped
at rate $\epsilon$ is indistinguishable to the ledger from a source with
$p=\epsilon$, so the same crossing point moves from $\pizero$ to
$\pizero/(4-3\pizero)$: \res{E3-p} at the deployed base rate against
\res{E3-p-pi005} at the hypothetical $0.05$ (simulation, both arms). The
asymmetry that costs the attacker is the asymmetry that costs the honest
source, at the same rate and in the same direction. Every $\epsilon$
figure in this paper is simulated; the evaluated pipeline is reported at
$\epsilon=0$, and the measured trajectories are in \cref{sec:eval}.

\section{Full E1 Grids}
\label{app:grids}
\begin{figure}[tp]
\centering
\input{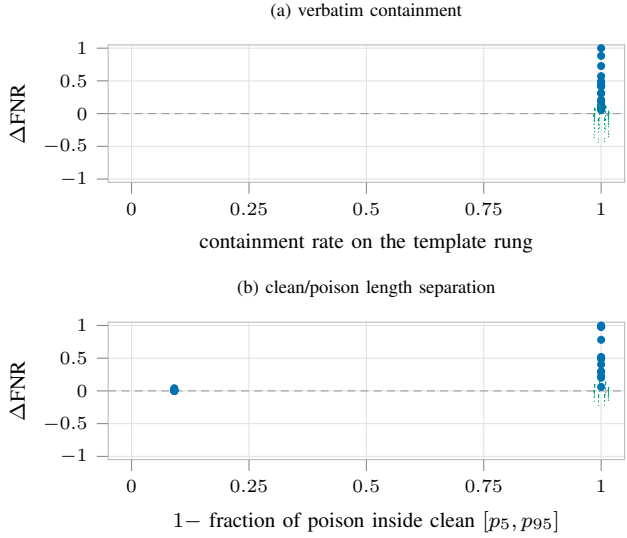}
\begin{tikzpicture}
\begin{groupplot}[group style={group size=1 by 2, vertical sep=1.85cm},
    width=0.95\columnwidth, height=3.4cm, x3base,
    xmin=-0.05, xmax=1.05, ymin=-1.05, ymax=1.05,
    xtick={0,0.25,0.5,0.75,1}, ytick={-1,-0.5,0,0.5,1},
    ylabel={$\Delta$FNR}]
\nextgroupplot[title={(a) verbatim containment}, xlabel={containment rate on the template rung}]
\draw[x3grey, densely dashed] (axis cs:-0.05,0) -- (axis cs:1.05,0);
\addplot[x3s1, only marks] table[x=x, y=y] {figures/f1_containment.dat};
\addplot[x3s3, only marks, mark=o, mark size=2.8pt]
  table[x=x, y=y] {figures/f1_containment_mis.dat};

\nextgroupplot[title={(b) clean/poison length separation}, xlabel={$1-$ fraction of poison inside clean $[p_{5},p_{95}]$}]
\draw[x3grey, densely dashed] (axis cs:-0.05,0) -- (axis cs:1.05,0);
\addplot[x3s1, only marks] table[x=x, y=y] {figures/f1_length.dat};
\addplot[x3s3, only marks, mark=o, mark size=2.8pt]
  table[x=x, y=y] {figures/f1_length_mis.dat};
\end{groupplot}
\end{tikzpicture}
\caption{\textbf{Per-point view of the construction profile.} One point per (corpus, attack, defense). Abscissa: the cue, measured on the poison set alone. Ordinate: the change in miss rate at the deployed operating point under the control that removes that cue. Circled: the \res{E1-profile-fail} of \res{E1-profile-n} points whose profiled direction is wrong at the level of a single point (the profile's claim is at corpus level, \cref{sec:profile}); their keys are in the released data.}
\label{fig:profile}
\end{figure}

\begin{figure*}[t]
\centering
\input{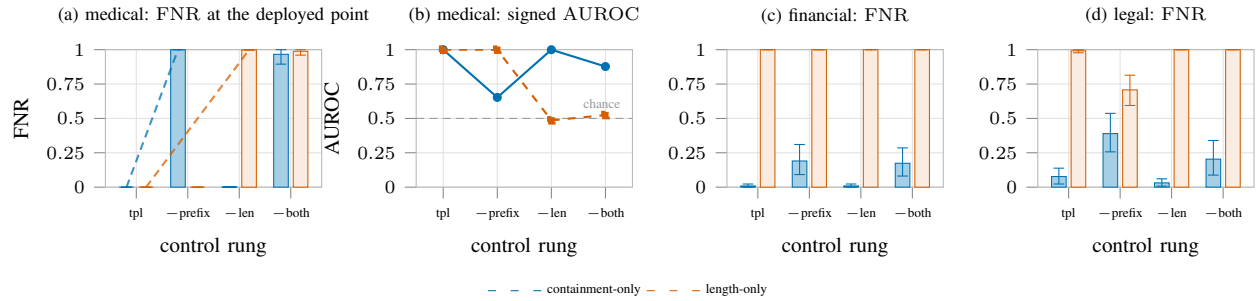}
\begin{tikzpicture}
\begin{groupplot}[group style={group size=4 by 1, horizontal sep=1.25cm},
    width=0.245\textwidth, height=3.4cm, x3base,
    xlabel={control rung}, xticklabel style={font=\tiny},
    legend style={legend columns=-1, draw=none, fill=none}]
\nextgroupplot[title={(a) medical: \FNR{} at the deployed point},
    ylabel={FNR}, x3rate, legend to name=x3f6,
    xmin=-0.6, xmax=3.6, xtick={0, 1, 2, 3}, xticklabels={tpl, $-$prefix, $-$len, $-$both}]
\addplot[x3b, fill=x3b!35, solid, ybar, bar width=5.5pt,
         error bars/.cd, y dir=both, y explicit]
  table[x=xc, y=containonly,
        y error plus expr=\thisrow{containonly_hi}-\thisrow{containonly},
        y error minus expr=\thisrow{containonly}-\thisrow{containonly_lo}]
  {figures/f6_cue_medical.dat};
\addlegendentry{containment-only}
\addplot[x3c, fill=x3c!12, solid, ybar, bar width=5.5pt,
         error bars/.cd, y dir=both, y explicit]
  table[x=xl, y=lenonly,
        y error plus expr=\thisrow{lenonly_hi}-\thisrow{lenonly},
        y error minus expr=\thisrow{lenonly}-\thisrow{lenonly_lo}]
  {figures/f6_cue_medical.dat};
\addlegendentry{length-only}
\draw[x3b, densely dashed, thick, opacity=0.75] (axis cs:-0.190,0.0000) -- (axis cs:0.810,1.0000);
\draw[x3c, densely dashed, thick, opacity=0.75] (axis cs:0.190,0.0000) -- (axis cs:2.190,0.9989);
\nextgroupplot[title={(b) medical: signed \AUROC{}}, ylabel={AUROC},
    x3rate, ytick={0,0.25,0.5,0.75,1},
    xmin=-0.5, xmax=3.5, xtick={0, 1, 2, 3}, xticklabels={tpl, $-$prefix, $-$len, $-$both}]
\draw[x3grey, densely dashed] (axis cs:-0.5,0.5) -- (axis cs:3.5,0.5);
\node[font=\tiny, x3grey, anchor=south east] at (axis cs:3.45,0.51) {chance};
\addplot[forget plot, x3s1, thick] table[x=rung, y=containonly_auroc] {figures/f6_cue_medical.dat};
\addplot[forget plot, x3s2, thick] table[x=rung, y=lenonly_auroc] {figures/f6_cue_medical.dat};
\nextgroupplot[title={(c) financial: \FNR{}}, x3rate,
    xmin=-0.6, xmax=3.6, xtick={0, 1, 2, 3}, xticklabels={tpl, $-$prefix, $-$len, $-$both}]
\addplot[forget plot, x3b, fill=x3b!35, solid, ybar, bar width=5.5pt,
         error bars/.cd, y dir=both, y explicit]
  table[x=xc, y=containonly,
        y error plus expr=\thisrow{containonly_hi}-\thisrow{containonly},
        y error minus expr=\thisrow{containonly}-\thisrow{containonly_lo}]
  {figures/f6_cue_financial.dat};
\addplot[forget plot, x3c, fill=x3c!12, solid, ybar, bar width=5.5pt,
         error bars/.cd, y dir=both, y explicit]
  table[x=xl, y=lenonly,
        y error plus expr=\thisrow{lenonly_hi}-\thisrow{lenonly},
        y error minus expr=\thisrow{lenonly}-\thisrow{lenonly_lo}]
  {figures/f6_cue_financial.dat};
\nextgroupplot[title={(d) legal: \FNR{}}, x3rate,
    xmin=-0.6, xmax=3.6, xtick={0, 1, 2, 3}, xticklabels={tpl, $-$prefix, $-$len, $-$both}]
\addplot[forget plot, x3b, fill=x3b!35, solid, ybar, bar width=5.5pt,
         error bars/.cd, y dir=both, y explicit]
  table[x=xc, y=containonly,
        y error plus expr=\thisrow{containonly_hi}-\thisrow{containonly},
        y error minus expr=\thisrow{containonly}-\thisrow{containonly_lo}]
  {figures/f6_cue_legal.dat};
\addplot[forget plot, x3c, fill=x3c!12, solid, ybar, bar width=5.5pt,
         error bars/.cd, y dir=both, y explicit]
  table[x=xl, y=lenonly,
        y error plus expr=\thisrow{lenonly_hi}-\thisrow{lenonly},
        y error minus expr=\thisrow{lenonly}-\thisrow{lenonly_lo}]
  {figures/f6_cue_legal.dat};
\end{groupplot}
\end{tikzpicture}
\par\smallskip
\pgfplotslegendfromname{x3f6}
\caption{\textbf{Each cue alone manufactures a perfect result, and each one-feature detector is destroyed by \emph{its own} control and left untouched by the other.} The detectors read one bit --- does the target question appear verbatim? --- and one scalar --- how long is the document? Both are single inputs of our own scorer's feature vector. (a)~On medical both miss nothing on the template rung; the dashed segments join each detector's template bar to the bar under the control that removes \emph{its} cue, and they cross. (b)~The same cells' \AUROC{}, \emph{signed}: below the chance line the cue is present with the opposite sign, which is why no absolute value is taken anywhere here. (c),(d)~On financial and legal the prefix control leaves containment-only standing because the two cues co-occur there; financial's length cue runs backwards and legal's clean corpus is one constant width, so neither one-feature detector is destroyed by its own control on those corpora. Whiskers as in \cref{fig:ladder}; the false-positive rates are in the released data.}
\label{fig:dissociation}
\end{figure*}

\Cref{tab:e1-delta} gives the paired cluster-bootstrap differences behind the
comparisons of \cref{sec:eval}, and \cref{tab:e1-agreement} the per-cell
decision-agreement grid whose counts \cref{sec:eval:agreement} quotes.
\Cref{fig:profile} plots the profile against the outcome one (corpus, attack,
defense) point at a time, including the points the corpus-level reading of
\cref{fig:headline} does not cover; \cref{fig:dissociation} draws the two
one-feature reference detectors of \cref{sec:profile} under every rung;
\cref{tab:e1-prefix} is the prefix control on the released NQ passages with
every defense's \AUROC{} beside the deployed point that \cref{tab:e1-main}
carries; and \cref{tab:placebo-legal} is the legal panel of the placebo table. \Cref{tab:capability} completes the survey of
\cref{tab:defenses} with the systems that expose no per-document admission
decision, and therefore cannot be scored on the shared candidate set that makes
Table~I a comparison; it carries no measurement, only the reason each was not
reproduced. All four are reproduced in the released
artifact, alongside the twelve per-cell grids and the out-of-distribution sweep,
which are artifact-only. The false-positive rates themselves are not here: they
are the second number in every cell of \cref{tab:e1-main}, in the body, because
a miss rate is not readable without them. Per-query cost is the last column of
\cref{tab:defenses}.

\noindent\textbf{Closed by design, tested as regressions.}
Domain is a property of the query, so domain laundering through document
metadata is moot (\res{E4-C1-delta-FNR}, expected zero), and attacker-declared
metadata is never a scorer input (\res{E4-A2-delta-FNR}, expected zero). Both
are run as regressions rather than reported as findings.

\begin{table}[tbp]
\centering\scriptsize\setlength{\tabcolsep}{1.5pt}
\caption{The prefix control on the attack authors' own passages: every defense on the released NQ poison with and without the question prefix the attack's code prepends.}
\label{tab:e1-prefix}
\begin{tabular}{lcccccc}
\toprule
 & \multicolumn{3}{c}{$Q{\oplus}I$, as inserted} & \multicolumn{3}{c}{$I$ only, as stored} \\
\cmidrule(lr){2-4} \cmidrule(lr){5-7}
Defense & FNR / FPR & AUROC & FNR@5\% & FNR / FPR{} & AUROC{} & FNR@5\%{} \\
\midrule
Perplexity & 1.000/0.005 & 0.773 & 0.809 & 1.000/0.005 & 0.729 & 0.866 \\
RAGuard (Cheng) & 0.191/0.065 & 0.953 & 0.216 & 0.763/0.061 & 0.758 & 0.796 \\
DRS & 0.854/0.154 & 0.539 & 0.970 & 0.930/0.164 & 0.450 & 0.995 \\
TrustRAG-S1 & 0.000/0.149 & N/A & N/A & 0.070/0.153 & N/A & N/A \\
RAGDefender & 0.045/0.025 & N/A & N/A & 0.172/0.117 & N/A & N/A \\
Dup-only & 0.678/0.005 & N/A & N/A & 0.989/0.009 & N/A & N/A \\
Ours, $\tau{=}0.5$ & 0.015/0.010 & 1.000 & 0.000 & 0.118/0.014 & 0.990 & 0.043 \\
Ours, $\tau_d$ & 0.015/0.010 & 1.000 & 0.000 & 0.118/0.014 & 0.990 & 0.043 \\
Ours, $\tau_{d,s}$ & 0.012/0.008 & 1.000 & 0.000 & 0.086/0.013 & 0.997 & 0.009 \\
length-only (ref.) & 1.000/0.000 & 0.806 & 1.000 & 1.000/0.000 & 0.864 & 1.000 \\
contain.-only (ref.) & 0.000/0.025 & 1.000 & 0.000 & 0.882/0.023 & 0.718 & 0.839 \\
NLI & 0.935/0.458 & — & — & 0.866/0.388 & — & — \\
retrieval survival & 0.99 &  &  & 0.93 &  &  \\
\bottomrule
\end{tabular}
\\[2pt]
{\footnotesize The released NQ passages of PoisonedRAG, once as its \texttt{attack.py} inserts them ($Q\oplus I$: the target question, then the released text) and once as the archive stores them ($I$ only); nothing else changes. FNR / FPR at the deployed threshold, AUROC, and the miss rate at a matched 5\% false-positive rate, point estimates over five seeds; N/A where the defense emits no score. The last row is retrieval survival with no defense.}
\end{table}

\begin{table*}[tp]
\centering\scriptsize\setlength{\tabcolsep}{3pt}
\caption{Placebo fillers for the length-matched rung on legal, whose clean corpus is a constant width: padding overshoots the clean band and installs an inverted length cue instead of removing one (\cref{sec:eval:ladder}).}
\label{tab:placebo-legal}
\begin{tabular}{lccccccccc}
\toprule
 & \multicolumn{3}{c}{clean-prose filler} & \multicolumn{3}{c}{word-shuffled filler} & \multicolumn{3}{c}{pseudo-word filler} \\
\cmidrule(lr){2-4} \cmidrule(lr){5-7} \cmidrule(lr){8-10}
Defense & FNR / FPR & AUROC & FNR@5\% & FNR / FPR{} & AUROC{} & FNR@5\%{} & FNR / FPR{}{} & AUROC{}{} & FNR@5\%{}{} \\
\midrule
Perplexity (GPT-2) & 0.775 / 0.000 & 0.878 & 0.605 & 0.247 / 0.000 & 0.948 & 0.230 & 0.280 / 0.006 & 0.942 & 0.259 \\
RAGuard (Cheng) & 0.279 / 0.116 & 0.883 & 0.321 & 0.071 / 0.115 & 0.964 & 0.071 & 0.087 / 0.119 & 0.956 & 0.087 \\
DRS & 0.857 / 0.100 & 0.688 & 0.933 & 0.922 / 0.101 & 0.649 & 0.961 & 0.479 / 0.097 & 0.836 & 0.643 \\
TrustRAG stage 1 & 0.044 / 0.206 & N/A & N/A & 0.038 / 0.185 & N/A & N/A & 0.142 / 0.199 & N/A & N/A \\
RAGDefender & 0.382 / 0.306 & N/A & N/A & 0.321 / 0.158 & N/A & N/A & 0.226 / 0.112 & N/A & N/A \\
Dup-only & 0.830 / 0.000 & N/A & N/A & 0.836 / 0.000 & N/A & N/A & 0.786 / 0.000 & N/A & N/A \\
Ours, $\tau{=}0.5$ & 0.063 / 0.041 & 0.989 & 0.053 & 0.035 / 0.042 & 0.995 & 0.025 & 0.033 / 0.037 & 0.992 & 0.030 \\
Ours, $\tau_d$ & 0.036 / 0.081 & 0.989 & 0.053 & 0.017 / 0.072 & 0.995 & 0.025 & 0.027 / 0.062 & 0.992 & 0.030 \\
Ours, $\tau_{d,s}$ ($a{=}0.3$) & 0.019 / 0.081 & 0.995 & 0.032 & 0.005 / 0.072 & 0.998 & 0.015 & 0.017 / 0.062 & 0.997 & 0.022 \\
length-only (ref.) & 1.000 / 0.000 & 0.024 & 0.992 & 1.000 / 0.000 & 0.025 & 0.988 & 1.000 / 0.000 & 0.030 & 0.991 \\
containment-only (ref.) & 0.031 / 0.025 & 0.994 & 0.022 & 0.023 / 0.022 & 0.997 & 0.016 & 0.009 / 0.023 & 0.997 & 0.009 \\
retrieval survival & 0.96 &  &  & 0.97 &  &  & 0.91 &  &  \\
\bottomrule
\end{tabular}
\\[2pt]
{\footnotesize Each column block pads the same template poison to the same draw from the clean word-length distribution with a different filler (\cref{sec:eval:ladder}). FNR / FPR at the deployed threshold, AUROC, and the miss rate at a matched 5\% false-positive rate, point estimates over five seeds; N/A where the defense emits no score. The last row is retrieval survival with no defense, the first leg of the conjunction of \cref{sec:threat}.}
\end{table*}

\begin{table*}[tp]
\centering
\caption{Query-time and generation-time RAG poisoning defenses \emph{compared by
capability only}: each exposes no per-document admission decision or no sweepable
operating point, so none can be scored on the shared candidate set of
\cref{tab:defenses}; the last row is the one per-document detector that could have
been and was not. Columns as in \cref{tab:defenses}; the last column is why each was
not reproduced.}
\label{tab:capability}
\scriptsize
\setlength{\tabcolsep}{3pt}
\begin{tabular}{@{}p{1.20in} l l p{1.28in} p{1.20in} c p{1.45in}@{}}
\toprule
Defense & Stage & Gran. & Signal & Operating-point rule & Fit? & Not reproduced because \\
\midrule
\zkip{}~\cite{kumar2026raguardzkip} & Ret+Gen & D & leave-one-out counterfactual answer change & keep iff the answer survives removal & Yes & $k{+}1$ generation passes; no per-document score \\
RobustRAG~\cite{xiang2024robustrag} & Gen & RS & isolate-then-aggregate per-document answers & certified budget, not a threshold & No & one generation pass per document \\
RADAR~\cite{chen2026radar} & PreG & RS & min-cut over an answer-agreement graph & graph cut; no per-document score & No & needs per-document answers \\
DataFilter~\cite{wang2026datafilter} & PreG & D & fine-tuned LLM data filter & model output; no exposed threshold & Yes & outside a CPU-only budget \\
RAGShield~\cite{patil2026ragshield} & Off+Ret+Gen & D, Src & numeric-claim consistency; provenance attestation & static provenance weights & No & scoped to numeric claims \\
TRIS~\cite{munir2026tris} & PreG+Gen & D, RS & cross-embedding clustering; judge; trigger filter & three fixed sieves in series & No & 16--19\,s per query as reported \\
SecureCollaRAG~\cite{wang2026securecollarag} & Ret+Gen & Src & GNN per-source credibility & Byzantine aggregation; no threshold & Yes & needs a multi-agent deployment \\
TriShieldRAG~\cite{mohanty2026trishieldrag} & Off+Ret+Gen & D, Src & ingest guard; static trust re-rank; LLM consensus & static trust weights & No & outside a CPU-only budget \\
GMTP~\cite{kim2025gmtp} & PreG & D & masked-token probability at the retriever's high-gradient tokens & threshold on the masked-token probability & No & a per-document detector we did not reproduce: the one CPU-feasible omission from \cref{tab:defenses} \\
\bottomrule
\end{tabular}
\end{table*}

\begin{table*}[tp]
\centering\scriptsize\setlength{\tabcolsep}{3pt}
\caption{E1 paired differences: every defense minus the harm-blind $\tau{=}0.5$ classifier on the same resampled query clusters, with 95\% paired cluster-bootstrap intervals. Condition: \texttt{legal/poisonedrag\_bb}.}
\label{tab:e1-delta}
\begin{tabular}{lcccc}
\toprule
Defense & FNR & FPR & CCR & $H$/1k \\
\midrule
Perplexity (GPT-2) & +0.549 [+0.418, +0.670]* & -0.059 [-0.148, +0.000] & +0.645 [+0.509, +0.778]* & +40750.000 [+29192.500, +51751.250]* \\
RAGuard (Cheng) & +0.076 [+0.028, +0.138]* & +0.049 [-0.058, +0.155] & +0.194 [+0.072, +0.339]* & +5775.000 [+2049.375, +10451.875]* \\
TrustRAG stage 1 & +0.002 [-0.004, +0.010] & +0.088 [-0.042, +0.211] & +0.007 [-0.011, +0.034] & +375.000 [-175.000, +1075.000] \\
RAGDefender & +0.337 [+0.262, +0.396]* & -0.029 [-0.127, +0.052] & +0.511 [+0.357, +0.666]* & +25075.000 [+17460.250, +32500.625]* \\
Dup-only & +0.341 [+0.290, +0.397]* & -0.059 [-0.148, +0.000] & +0.610 [+0.517, +0.707]* & +25250.000 [+21350.000, +29050.000]* \\
Ours, $\tau_d$ & -0.002 [-0.006, +0.000] & +0.020 [+0.000, +0.051] & -0.006 [-0.023, +0.000] & -100.000 [-450.000, +100.000] \\
Ours, $\tau_{d,s}$ ($a{=}0.3$) & -0.002 [-0.006, +0.000] & +0.020 [+0.000, +0.051] & -0.006 [-0.023, +0.000] & -100.000 [-450.000, +100.000] \\
Ours, $\tau_{d,s}$ ($a{=}0.1$) & -0.002 [-0.006, +0.000] & +0.020 [+0.000, +0.051] & -0.006 [-0.023, +0.000] & -100.000 [-450.000, +100.000] \\
NLI (ReliabilityRAG) & +0.791 [+0.726, +0.852]* & +0.448 [+0.335, +0.545]* & +0.972 [+0.947, +0.992]* & +60140.000 [+55638.375, +65107.000]* \\
length-only (ref.) & +0.990 [+0.976, +1.000]* & -0.059 [-0.148, +0.000] & +0.975 [+0.948, +0.996]* & +73650.000 [+65148.750, +81802.500]* \\
containment-only (ref.) & +0.075 [+0.021, +0.140]* & -0.029 [-0.074, +0.000] & +0.183 [+0.051, +0.332]* & +5525.000 [+1473.750, +10601.250]* \\
No defense & +0.998 [+0.994, +1.000]* & -0.059 [-0.148, +0.000] & +0.987 [+0.965, +1.000]* & +74250.000 [+66448.750, +82251.250]* \\
DRS & +0.814 [+0.700, +0.909]* & +0.069 [-0.057, +0.199] & +0.879 [+0.776, +0.957]* & +60825.000 [+49824.375, +71226.875]* \\
\bottomrule
\end{tabular}
\\[2pt]
{\footnotesize Each cell is the difference against Uniform threshold (tau=0.5) with a 95\% paired cluster-bootstrap interval: 1000 draws resampling the 40 query clusters (pooled over 5 seed(s)) once per draw and applying the same resample to both systems. * marks an interval that excludes zero.  Seeds vary query order, the adjudication draw and poison-template choice; the corpus split is fixed. The retrieved candidate sets are identical across all 5 seeds in financial/jamming\_blocker, legal/corruptrag\_as, legal/jamming\_blocker, medical/corruptrag\_as, medical/hijackrag\_bb, medical/jamming\_blocker, medical/poisonedrag\_bb, nq/corruptrag\_as, nq/jamming\_blocker, nq/poisonedrag\_bb (Jaccard 1.000), so the seed sd there is arithmetically forced to 0.000 and is not a sampling error. Candidate sets do move with the seed in financial/corruptrag\_as 0.995, financial/hijackrag\_bb 0.970, financial/poisonedrag\_bb 0.970, legal/hijackrag\_bb 0.914, legal/poisonedrag\_bb 0.951, nq/hijackrag\_bb 0.909 (minimum Jaccard against seed 11). Intervals are therefore paired cluster bootstraps over the evaluation queries, not seed spreads. The same table for the other 3 corpora is in the released artifact.}
\end{table*}

\begin{table*}[tp]
\centering\scriptsize\setlength{\tabcolsep}{3pt}
\caption{On the benchmark the literature reports against, three different threshold rules produce the same decisions.}
\label{tab:e1-agreement}
\begin{tabular}{lcccccc}
\toprule
Condition & cands. & in band & $\neq\tau_d$ & $\neq\tau_{d,s}$ & distinct scores & $\rho{=}1$ \\
\midrule
financial/corruptrag\_as (template) & 2000 & 0 (0.00\%) & 0 & 0 & 358 &  \\
financial/hijackrag\_bb (template) & 2000 & 0 (0.00\%) & 0 & 0 & 202 &  \\
financial/jamming\_blocker (template) & 2000 & 0 (0.00\%) & 0 & 0 & 359 &  \\
financial/poisonedrag\_bb (template) & 2000 & 0 (0.00\%) & 0 & 0 & 196 &  \\
legal/corruptrag\_as (template) & 2000 & 63 (3.15\%) & 63 & 68 & 400 &  \\
legal/hijackrag\_bb (template) & 2000 & 19 (0.95\%) & 19 & 19 & 400 &  \\
legal/jamming\_blocker (template) & 2000 & 95 (4.75\%) & 95 & 93 & 400 &  \\
legal/poisonedrag\_bb (template) & 2000 & 13 (0.65\%) & 13 & 13 & 400 &  \\
medical/corruptrag\_as (template) & 2000 & 0 (0.00\%) & 0 & 0 & 395 &  \\
medical/hijackrag\_bb (template) & 2000 & 0 (0.00\%) & 0 & 0 & 397 &  \\
medical/jamming\_blocker (template) & 2000 & 0 (0.00\%) & 0 & 0 & 396 &  \\
medical/poisonedrag\_bb (template) & 2000 & 0 (0.00\%) & 0 & 0 & 397 &  \\
nq/corruptrag\_as (template) & 2000 & 0 (0.00\%) & 0 & 4 & 400 & yes \\
nq/hijackrag\_bb (template) & 2000 & 0 (0.00\%) & 0 & 0 & 399 & yes \\
nq/jamming\_blocker (template) & 2000 & 0 (0.00\%) & 0 & 0 & 400 & yes \\
nq/poisonedrag\_bb (template) & 2000 & 0 (0.00\%) & 0 & 5 & 400 & yes \\
financial/poisonedrag\_bb (length matched) & 2000 & 0 (0.00\%) & 0 & 0 & 199 &  \\
legal/poisonedrag\_bb (length matched) & 2000 & 63 (3.15\%) & 63 & 84 & 400 &  \\
medical/poisonedrag\_bb (length matched) & 2000 & 1 (0.05\%) & 1 & 5 & 397 &  \\
financial/poisonedrag\_bb\_natural (prefix removed) & 2000 & 2 (0.10\%) & 2 & 3 & 315 &  \\
legal/poisonedrag\_bb\_natural (prefix removed) & 2000 & 36 (1.80\%) & 36 & 37 & 400 &  \\
medical/poisonedrag\_bb\_natural (prefix removed) & 2000 & 7 (0.35\%) & 7 & 7 & 399 &  \\
nq/poisonedrag\_bb\_natural (prefix removed) & 2000 & 0 (0.00\%) & 0 & 31 & 400 & yes \\
financial/poisonedrag\_bb\_natural (both removed) & 2000 & 1 (0.05\%) & 1 & 1 & 343 &  \\
legal/poisonedrag\_bb\_natural (both removed) & 2000 & 193 (9.65\%) & 193 & 265 & 400 &  \\
medical/poisonedrag\_bb\_natural (both removed) & 2000 & 0 (0.00\%) & 0 & 0 & 397 &  \\
\bottomrule
\end{tabular}
\\[2pt]
{\footnotesize Decisions counted by joining the stored per-document actions of three defenses on the same candidate set on $(\text{seed}, \text{query}, \text{document})$; nothing is re-thresholded. \emph{In band} is the number of scored candidates whose posterior lies in the closed interval between $\tau{=}0.5$ and $\tau_d$, outside which the two rules provably agree. \emph{Distinct scores} is the number of distinct values the score took: a threshold-derived metric computed on two values is not a measurement, and this column is how a reader checks that. 12 of 26 cells are bit-identical under all three rules; 2 of them have $\rho_d = 1$, where $\tau_d = 0.5$ exactly and the cost-derived rule IS the harm-blind classifier by construction -- those are a control, not evidence of inertness.}
\end{table*}

\begin{table*}[tp]
\centering\scriptsize\setlength{\tabcolsep}{3pt}
\caption{E4-D1: mimicry plus hard-label word substitution tuned against our scorer on NQ (archive-stored, prefix-less passages; \res{E4-budget-queries} oracle queries and \res{E4-budget-iters} iterations per document), then scored by every defense unchanged. Static / adaptive at each defense's deployed point; for every row but the last four this is a transfer attack (\cref{sec:threat:control}).}
\label{tab:e4-d1}
\begin{tabular}{lccccc}
\toprule
Defense & FNR & FPR & retrieved & reached & ccr \\
\midrule
No defense & 1.000 / 1.000 & 0.000 / 0.000 & 0.990 / 0.870 & 0.990 / 0.870 & 0.920 / 0.670 \\
Perplexity (GPT-2) & 1.000 / 1.000 & 0.000 / 0.000 & 0.990 / 0.870 & 0.990 / 0.870 & 0.920 / 0.670 \\
RAGuard (Cheng) & 0.162 / 0.897 & 0.119 / 0.115 & 0.990 / 0.870 & 0.160 / 0.780 & 0.154 / 0.626 \\
DRS & 0.909 / 0.943 & 0.119 / 0.124 & 0.990 / 0.870 & 0.900 / 0.820 & 0.840 / 0.657 \\
TrustRAG stage 1 & 0.000 / 0.069 & 0.198 / 0.487 & 0.990 / 0.870 & 0.000 / 0.060 & 0.000 / 0.095 \\
RAGDefender & 0.000 / 0.368 & 0.040 / 0.239 & 0.990 / 0.870 & 0.000 / 0.320 & 0.000 / 0.308 \\
NLI (ReliabilityRAG) & 0.909 / 0.690 & 0.465 / 0.354 & 0.990 / 0.870 & 0.900 / 0.600 & 0.880 / 0.574 \\
Dup-only & 0.646 / 0.954 & 0.010 / 0.009 & 0.990 / 0.870 & 0.640 / 0.830 & 0.620 / 0.650 \\
Ours, $\tau{=}0.5$ & 0.030 / 0.977 & 0.010 / 0.018 & 0.990 / 0.870 & 0.030 / 0.850 & 0.030 / 0.660 \\
Ours, $\tau_d$ & 0.030 / 0.977 & 0.010 / 0.018 & 0.990 / 0.870 & 0.030 / 0.850 & 0.030 / 0.660 \\
Ours, $\tau_{d,s}$ ($a{=}0.3$) & 0.027 / 0.966 & 0.010 / 0.018 & 0.990 / 0.870 & 0.027 / 0.840 & 0.027 / 0.650 \\
Ours (trust a=0.3 eps=0 d=0, farmed source) & 0.030 / 1.000 & 0.010 / 0.018 & 0.990 / 0.870 & 0.030 / 0.870 & 0.030 / 0.670 \\
\bottomrule
\end{tabular}
\\[2pt]
{\footnotesize Each cell is static / adaptive at the deployed operating point, a point estimate over seeds; the per-seed dispersion is in the released \texttt{e4fix\_d1\_*.json}. The defenses that are not bypassed pay for it in the FPR column: read the pair, not the miss rate alone.}
\end{table*}

\begin{figure*}[tp]
\centering
\input{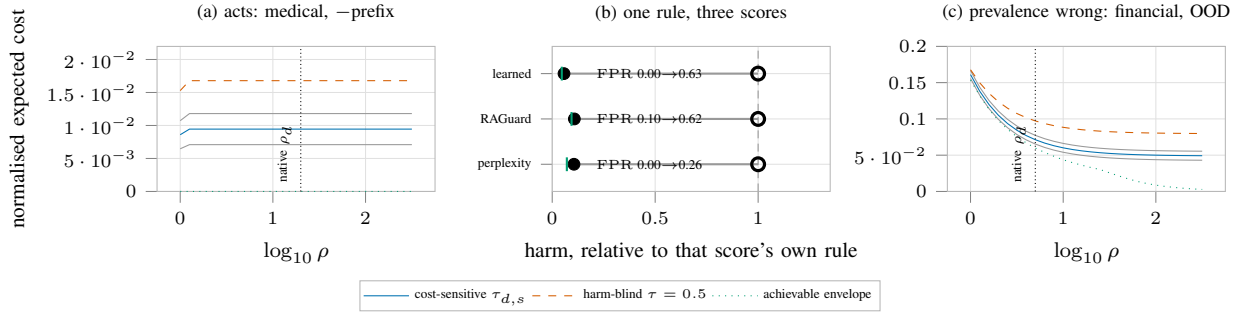}
\begin{tikzpicture}
\begin{groupplot}[group style={group size=3 by 1, horizontal sep=1.55cm},
    width=0.29\textwidth, height=3.5cm, x3base, scaled y ticks=false,
    legend style={legend columns=-1, draw=none, fill=none}]
\nextgroupplot[title={(a) acts: medical, $-$prefix}, xlabel={$\log_{10}\rho$}, ylabel={normalised expected cost}, legend to name=x3f4, x3base, ymin=0, ymax=0.021977]
\addplot[x3grey, mark=none, forget plot] table[x=log10rho, y=ours_lo] {figures/f4_acts.dat};
\addplot[x3grey, mark=none, forget plot] table[x=log10rho, y=ours_hi] {figures/f4_acts.dat};
\addplot[x3s1, mark=none] table[x=log10rho, y=ours] {figures/f4_acts.dat};
\addlegendentry{cost-sensitive $\tau_{d,s}$}
\addplot[x3s2, mark=none] table[x=log10rho, y=uniform] {figures/f4_acts.dat};
\addlegendentry{harm-blind $\tau=0.5$}
\addplot[x3s3, mark=none] table[x=log10rho, y=envelope] {figures/f4_acts.dat};
\addlegendentry{achievable envelope}
\draw[x3a, densely dotted] (axis cs:1.3010,0) -- (axis cs:1.3010,0.021977);
\node[font=\tiny, anchor=south west, rotate=90] at (axis cs:1.3010,0) {native $\rho_d$};
\nextgroupplot[title={(b) one rule, three scores},
    xlabel={harm, relative to that score's own rule},
    x3base, xmin=0, xmax=1.35, ymin=-0.6, ymax=2.6,
    ytick={0, 1, 2}, yticklabels={perplexity, RAGuard, learned},
    yticklabel style={font=\tiny}, xtick={0,0.5,1}]
\draw[x3grey, densely dashed] (axis cs:1,-0.6) -- (axis cs:1,2.6);
\addplot[x3grey, mark=none, thick, forget plot] table[x=x, y=y] {figures/f6_wrapper_link.dat};
\addplot[x3a, only marks, mark=o, mark size=2.4pt, very thick, forget plot] table[x=native, y=i] {figures/f6_wrapper.dat};
\addplot[x3a, only marks, mark=*, mark size=2.2pt, forget plot] table[x=cost, y=i] {figures/f6_wrapper.dat};
\addplot[x3d, only marks, mark=|, mark size=2.6pt, thick, forget plot] table[x=oracle, y=i] {figures/f6_wrapper.dat};
\node[font=\tiny, x3a, anchor=west] at (axis cs:0.17,0.00) {\FPR{} 0.00$\to$0.26};
\node[font=\tiny, x3a, anchor=west] at (axis cs:0.17,1.00) {\FPR{} 0.10$\to$0.62};
\node[font=\tiny, x3a, anchor=west] at (axis cs:0.17,2.00) {\FPR{} 0.00$\to$0.63};
\nextgroupplot[title={(c) prevalence wrong: financial, OOD}, xlabel={$\log_{10}\rho$},  x3base, ymin=0, ymax=0.199909]
\addplot[x3grey, mark=none, forget plot] table[x=log10rho, y=ours_lo] {figures/f4_loss.dat};
\addplot[x3grey, mark=none, forget plot] table[x=log10rho, y=ours_hi] {figures/f4_loss.dat};
\addplot[forget plot, x3s1, mark=none] table[x=log10rho, y=ours] {figures/f4_loss.dat};
\addplot[forget plot, x3s2, mark=none] table[x=log10rho, y=uniform] {figures/f4_loss.dat};
\addplot[forget plot, x3s3, mark=none] table[x=log10rho, y=envelope] {figures/f4_loss.dat};
\draw[x3a, densely dotted] (axis cs:0.6990,0) -- (axis cs:0.6990,0.199909);
\node[font=\tiny, anchor=south west, rotate=90] at (axis cs:0.6990,0) {native $\rho_d$};
\end{groupplot}
\end{tikzpicture}
\par\smallskip
\pgfplotslegendfromname{x3f4}
\caption{\textbf{A cost-derived operating point is a lever only where the score is unsaturated.} (a)~Normalised expected cost against $\log_{10}\rho$ where the score is unsaturated and the rule acts; shading is the $\rho$ interval in which the cost-sensitive and the harm-blind $\tau=0.5$ rule decide \emph{bit-identically}. On the cell the literature reports against that interval covers the whole axis and every trace is identically zero, which is why that cell is a number in \cref{sec:eval:operating} and not a panel (\texttt{f4\_inert.dat}). (b)~The same rule fitted out of fold over three score families that share no code path, length-matched medical: $\circ$ each score's own deployed rule (harm $1.0$ by construction), $\bullet$ the cost-derived rule, $|$ the best that score's thresholds allow, with the false-positive rate each gain costs printed beside it. (c)~The cross-domain lane, where the calibration prevalence is wrong; dotted rule, the native $\rho_d$. Note the per-panel vertical scale.}
\label{fig:cost}
\end{figure*}

\noindent\textbf{The defense-aware attacker.}\label{sec:eval:operating}
\Cref{tab:e4-d1} is the query-based attacker of \cref{sec:eval:ladder}: the
mimicry rung followed by hard-label word substitution tuned against our scorer
on NQ (archive-stored, prefix-less passages; \res{E4-budget-queries} oracle
queries and \res{E4-budget-iters} iterations per document), then scored by
every defense unchanged. Each cell is static / adaptive at that defense's
deployed point; for every row but our own the number is a transfer attack and
a lower bound~\cite{tramer2020adaptive}. The DRS row is the pre-fix port and
is footnoted as such.

\noindent\textbf{Where the operating point acts, and where it hurts.}
\Cref{fig:cost} sweeps normalised expected cost against $\log_{10}\rhod$ on the
control lanes, where the score is unsaturated, wrapping one rule over three
unrelated scores: ours, the chunk score of
RAGuard-Cheng~\cite{cheng2025raguard} and GPT-2
perplexity~\cite{radford2019gpt2}. Two existence proofs. On length-matched
legal, harm per 1{,}000 queries falls from \resbare{E1-H-uniform-lm-legal} to
\resbare{E1-H-valor-lm-legal} at a miss rate of \resbare{E1-FNR-uniform-lm-legal}
against \resbare{E1-FNR-valor-lm-legal}; on prefix-removed medical, from
\resbare{E1-H-uniform-nat-medical} to \resbare{E1-H-valor-nat-medical} at
\resbare{E1-FNR-uniform-nat-medical} against \resbare{E1-FNR-valor-nat-medical}. The
same rule wrapped over the two scores we did not build reproduces the direction
on length-matched legal (\res{E2-wrapper-raguard} for the chunk score,
\res{E2-wrapper-ppl} for perplexity, in harm per 1{,}000 queries), which is the
evidence that the effect belongs to the decision layer and not to our features.
Both wrappers buy it on the clean side, at the false-positive cost
\cref{fig:cost}(b) prints, because realised prevalence on that cell makes
aggressive blocking cost-optimal; and unlike \taud{} the wrapper needs a
labelled calibration set.

The negative half is not on the harm axis. On the financial
out-of-distribution condition (the scorer calibrated at a prevalence it does not
meet) the rule still lowers harm (\res{E1-OOD-HARM}) but pays
\res{E1-OOD-dFPR} points of false-positive rate for it, and a deployment
pricing a false block differently from our \rhod{} reads that trade the other
way. The prevalence correction is not the repair either: harm at \pizero{}
against \pione{} runs \res{E2-pi1-harm}, so re-anchoring the prior recovers
nothing and on legal costs more. Nor is the cost-derived point the achievable
optimum at the native harm ratio: proportionality regret, the harm of the
cost-derived threshold minus the least harm any threshold reaches at the same
ratio, is \resbare{E2-regret-financial} on financial and \resbare{E2-regret-legal} on
legal against \resbare{E2-regret-medical} on medical, where every threshold is the
same one. The failure is assumption~A1, not the threshold formula. The decision
layer is a lever everywhere except the benchmark: unsaturated scores give it
room, the literature's own cells give it none, and a wrong calibration
prevalence makes it pay for its gain in benign blocks.

\FloatBarrier
\clearpage
\section{Prior Reviews and Response}

\label{app:prior-reviews}

An earlier version of this work was submitted to HICSS-60 and rejected. That
round produced two reviews and no rebuttal; both are reproduced here in full
and there were no others. \cref{app:map} maps every distinct criticism to the
part of this submission that addresses it; \cref{app:verbatim} gives the
reviews. A code audit against the reviews confirmed every
point raised and found further defects the reviewers could not see from the
manuscript: an untrained head used in ranking, a reliability signal derived
from the filter's own decisions, a baseline fit on the evaluation corpus with
ground-truth labels, and hard-coded strings from the authors' own attack
generator. \emph{All numerical results of the earlier version are withdrawn};
this submission reports only results from the rewritten pipeline on public
corpora with published attacks. Experiment identifiers E1--E9 are those of
\cref{sec:eval}, except E3, a trust simulation whose closed forms are
\cref{app:derivations} and whose remaining arms are artifact-only; E5 and E6
were planned, cut for compute, and named as limitations
(\cref{sec:discussion}).

\subsection{How each criticism is addressed}
\label{app:map}

\footnotesize
\noindent Status codes: \textbf{A} addressed; \textbf{N} addressed with a
narrowed claim; \textbf{R} criticised component or metric retired, or result
withdrawn; \textbf{O} out of scope, stated as a limitation.

\smallskip\noindent\textbf{Reviewer~1.}
\begin{itemize}[leftmargin=3.0em,labelwidth=2.6em,labelsep=0.4em,align=left,nosep,itemsep=1pt]
\item[R1.0] Methodology not coherently specified; evaluation does not support the claims. Restated as a cost-sensitive decision problem with an explicit objective, \cref{prop:bayes} and stated assumptions, one of which is measured to fail; every claim is tied to an experiment. \cref{sec:stack}, \cref{sec:eval}, \cref{tab:defenses}. [A]
\item[R1.1] Pass rule inverted ($\theta$ falls with risk yet documents pass when $R\le 1-\theta$). An exposition error: the legacy code blocked when $R\ge\theta_d$ and was stricter for higher-risk domains; but $\theta_d$ spanned only $0.450$--$0.474$ across domains, so the mechanism was nearly inert. Both are replaced by the Bayes threshold $\tauds=1/(1+\rhod\ks)$, monotone in $\rhod$ and $\ks$ (\cref{prop:bayes}, \cref{eq:tauds}); trust now enters the threshold. E2 (\cref{sec:eval:operating}) measures the effect of $\rhod$ with the scorer fixed. [A]
\item[R1.2] Combined score $S$ mixes benignness with impact and relevance; its role in filtering is unexplained. $S$ only reranked passed documents and is removed; the sole decision variable is $\ptilde$ (\cref{sec:stack:decision}). [R]
\item[R1.3] Three heads $R,I,D$ but the loss covers only $R,I$. The $D$ head was never trained and the $I$ label was a function of the poison label; both removed. Single-output calibrated scorer (\cref{sec:stack:system}); ablation in E7 (\cref{sec:eval:mechanism}). [A]
\item[R1.4] Query never reaches the encoder. Confirmed; the scorer is now query-conditioned, $\phat_i=f(x_i,q)$, and near-verbatim query containment is one of its features (\cref{sec:stack:decision}, \cref{sec:stack:system}); its feature groups are ablated in E7 (\cref{tab:e7-ablation}). [A]
\item[R1.5] Reliability signal $r$ never defined. It was the filter's own decision, and no trust update preceded any reported number. Replaced by an exogenous adjudication ledger (\cref{sec:stack:system}); the channel is in the threat model as C4 (\cref{sec:threat:adversary}); the corrupted-ledger and merged-channel arms are E3 B6 and the trust term itself is ablated in E7 (\cref{tab:e7-ablation}); every E3 number is simulation, and in these conditions no trust number is a detection result (\cref{app:derivations}). [A]
\item[R1.6] GSQ has no recall term. Retired; alert precision/recall, escalation volume, time-to-flag, false-demotion and recovery are reported separately (E3). [R]
\item[R1.7] DPS ``ideal'' vs.\ ``actual'' thresholds not distinguished. Retired; proportionality is judged by $\HPR_d$ against the oracle-optimal threshold and by cost curves (E2). [R]
\item[R1.8] TCA omits the calibration weight. Retired; ECE (two binnings), Brier decomposition and reliability diagrams are reported separately (E1, E3). [R]
\item[R1.9] Baselines not apt; F1$=0$ reflects an interface mismatch. One filter interface for every defense over the same candidate set; detector metrics only for label-emitting systems, N/A otherwise (\cref{sec:stack:stack}, E1). [A]
\item[R1.10] Document-level screening baselines would be more appropriate. Same-task baselines re-implemented (perplexity, RAGuard, DRS, TrustRAG-S1, RAGDefender, NLI consistency, uniform-threshold classifier); systems needing a GPU or generator internals appear in \cref{tab:defenses} only. E1, E4 (\cref{sec:eval:ladder}). [A]
\item[R1.11] Adversarial training reached DR$=1$, PPR$=0$, beating the proposal. Cause: that baseline was fit on the evaluation corpus with ground-truth labels. Result withdrawn; the fair version (disjoint training split, uniform $\tau=0.5$) is the cost-blind baseline in E1 and E7. [R]
\end{itemize}

\smallskip\noindent\textbf{Reviewer~2.}
\begin{itemize}[leftmargin=3.0em,labelwidth=2.6em,labelsep=0.4em,align=left,nosep,itemsep=1pt]
\item[R2.1] Section~2 is not a comparative related-work review. Rewritten by intervention stage, granularity, attack class, source reliability, consistency, attribution and certification, including every work named (\cref{sec:background:defenses}, \cref{tab:defenses}). [A]
\item[R2.2] Unsupported ``first/only'' claims. All removed; query-time per-document filtering is credited to prior work (\cref{sec:intro}, \cref{sec:background:gap}). [A]
\item[R2.3] Concise comparison table; discuss the six named papers. \Cref{tab:defenses} carries stage, granularity, signal, operating-point rule, whether anything is fitted, nominal against realised \FPR{} and per-query cost; attack class and the post-generation and attribution stages are covered in the prose of \cref{sec:background:attacks}, \cref{sec:background:defenses} rather than as columns. Of the six named works, RAGuard, DRS and the ReliabilityRAG-style NLI filter are reproduced as baselines (E1, \cref{tab:e4-d1}); RAGForensics (post-generation attribution) and D-SCAN (generator-attention filtering) lie outside the per-document admission class and are discussed, not reproduced (\cref{sec:background:defenses}); RSB~\cite{zhang2025rsb} is cited for the finding that motivates the adaptive lane (\cref{sec:intro}). [A]
\item[R2.4] The contribution is the integration, not filtering per se. Overtaken rather than adopted: no system-level contribution is claimed now. The composition---cost-sensitive threshold, Beta trust, per-document decision, audit record---is credited to prior work as classical instrumentation (\cref{sec:background:lineage}) and used as an \emph{instrument} (\cref{sec:stack}); the three contributions are the construction profile, the two failure modes with their protocol, and a both-signs measurement of the decision layer (\cref{sec:intro}). [R]
\item[R2.5] Figure~1 uninformative. Removed, not redrawn: the paper no longer proposes the system it depicted, and the adversary entry points it carried are enumerated in \cref{sec:threat}. The figures that replace it (\cref{fig:profile,fig:ladder,fig:roc,fig:cost}) each state a measured claim. [A]
\item[R2.6] Table~1 should span both columns. \Cref{tab:defenses} is full width. [A]
\item[R2.7] Reconcile the pass rule with Eq.~10; confirm equations match the code. See R1.1; equations are transcribed from the decision module, whose unit tests compare the closed form with brute-force harm minimisation (\cref{sec:stack:decision}). [A]
\item[R2.8] Direction of $S$; relevance head absent from the loss; score weights unreported. Components removed; single-purpose notation (\cref{sec:stack}). [R]
\item[R2.9] If $r$ is the model's own decision the system reinforces its errors; if human, describe the workflow; how is compromise detected? Trust updates are stated to be non-automatic; the escalate--adjudicate--ledger workflow is described (\cref{sec:stack:system}); the self-reinforcement the review predicts is the E3 B6 merged-channel arm, in which every honest source is eventually demoted; time-to-demotion and the attacker's poison budget are E3 B1 (\cref{app:derivations}), cold start and the long-history insider E3 B7. Every one of these is simulation; every pipeline result is reported at $\varepsilon=0$ (\cref{sec:threat:adversary}). [A]
\item[R2.10] Malicious modification vs.\ legitimate correction: provenance, versioning, authentication, corroboration, cold start, compromised trusted sources, reversal, superseded documents. Source identity is bound to an authenticated channel (\cref{sec:threat:adversary}); contradiction is evidence routed to \ESCALATE, never adjudication; cold start $\kappa=1$; the trust floor bounds insiders; reversal recomputes the ledger. The legitimate-update benchmark (E5) is cut for compute and named as a limitation (\cref{sec:discussion}). Content-only signals cannot separate identical texts and the paper says so (\cref{sec:discussion}). [N]
\item[R2.11] Separate architectural capability from security performance. \Cref{tab:defenses} (capability) vs.\ E1 (cross-system outcomes; detector metrics only where defined). [A]
\item[R2.12] Table~2/Figure~3 misleading; F1 should be N/A. Figure removed, table withdrawn (see R1.11); E1 uses N/A. [R]
\item[R2.13] Financial/legal gains within noise; reframe around common outcomes. Every comparison is a paired cluster-bootstrap CI on the difference against a named comparator; per-seed spread is never reported as a precision claim, because candidate sets are seed-invariant; the instruction-injection and jamming conditions are retained and reported as is, in the attack set of \cref{sec:stack:stack} and in every cell of \cref{tab:e1-profile,tab:e1-agreement}. [A]
\item[R2.14] DR$=1-$PPR for every system. Renamed \emph{context contamination rate}; detector metrics are class-conditional and counted over the candidate set (\cref{sec:stack:metrics}). [A]
\item[R2.15] Proportionality confounded with domain difficulty. Domain is no longer a treatment. $\rhod$ is swept \emph{within} one corpus over stored scores, with corpus, candidate sets, attack and scorer all fixed, so the harm ratio is the only thing that moves (E2, \cref{fig:cost}); the cross-corpus $\rhod$ swap is reported separately, and it is a loss (\cref{sec:eval:operating}). [A]
\item[R2.16] Risk weights unjustified; applications within a sector differ. Replaced by one harm ratio $\rhod$ per deployment, a governance input labelled illustrative and always swept over $\log_{10}\rho\in[0,2.5]$ (\cref{sec:stack:decision}, E2, \cref{sec:discussion}). [N]
\item[R2.17] Narrow to ``partial support in the medical condition''. Earlier conclusions withdrawn; every prediction is scoped to corpus level with its per-point failure count printed (\cref{sec:profile:limits}, \cref{sec:eval:prediction}), and no claim is stated on a rung the paper calls unfalsifiable. [A]
\item[R2.18] DPS circular. Retired; $\HPR_d$ is oracle-referenced (E2). [R]
\item[R2.19] TCA parameters unreported. Retired; see R1.8. [R]
\item[R2.20] GSQ constituents undefined; GSQ$=1$ misread as recall. Retired; see R1.6. [R]
\item[R2.21] Table~3/Figure~4 duplicate. Both gone; no figure duplicates a table. [A]
\item[R2.22] Vector-database manipulation unevaluated. Removed from the defended classes; index/retriever tampering stated out of scope (\cref{sec:threat:adversary}). [O]
\item[R2.23] Unintended 5\%$\to$8\% poison rate needs evidence. Root cause (append rather than substitute) fixed; realised rate logged; the dedicated poison-rate sweep (E6) is cut for compute and named as a limitation (\cref{sec:discussion}); one poison rate is used throughout, and the injection budget is instead read across attack families at $N{=}1$ and $N{=}5$, where it decides which defense class wins (\cref{sec:eval:mechanism}); primary metrics are prevalence-invariant, with an odds correction (\cref{sec:stack:decision}). [A]
\item[R2.24] Table~4 operating guidance not derived from results. Removed; operating points are read off cost curves and escalation is Chow's rule with an explicit review cost (\cref{sec:stack:decision}). [R]
\item[R2.25] Synthetic corpus, few conditions and trials. Four real corpora (NQ/PoisonedRAG over BEIR, PubMedQA, FinQA with TAT-QA, CUAD), published attacks, five seeds on the main grid --- three on the E7 ablation, one on the secondary E9, stated wherever they appear --- and an out-of-distribution training regime; the synthetic corpus is dropped, and the single retriever and the 40 evaluation queries per cell are stated as limitations (\cref{sec:stack:stack}, \cref{sec:discussion}). [A]
\item[R2.26] Frame as an exploratory prototype; no superiority or deployment-readiness claims. Abstract, \cref{sec:intro} and \cref{sec:discussion} rewritten; adaptive-attack success is reported even when high. [A]
\end{itemize}

\smallskip\noindent\textbf{Reviewer~2, ``Suggested revisions'' S1--S16.} The sixteen closing bullets restate the items above and are answered by them:

\smallskip\noindent\begin{tabular}{@{}l@{\ }l@{\qquad}l@{\ }l@{}}
S1 & R2.1, R2.3, R2.4 & S9 & R1.9, R2.11 \\
S2 & R2.2, R2.4 & S10 & R1.11, R2.12 \\
S3 & R2.10 & S11 & R1.6--R1.8, R2.18--R2.20 \\
S4 & R2.5 & S12 & R2.14 \\
S5 & R2.6 & S13 & R2.21 \\
S6 & R1.1, R2.7 & S14 & R1.1, R2.15--R2.17 \\
S7 & R1.2--R1.4, R2.8 & S15 & R2.24 \\
S8 & R1.5, R2.9 & S16 & R2.26 \\
\end{tabular}
\normalsize

\subsection{The reviews}
\label{app:verbatim}

As the call for papers requires, the two reviews are reproduced below
anonymized but otherwise unedited and complete: the text and scores are
exactly as received. The only omission is the programme chairs' cover e-mail,
a form rejection notice containing no review content; no chair or reviewer
names appeared in the original.
Nothing is excerpted, summarised or reordered, and apart from the redaction
recorded immediately below, the only changes are typesetting
ones---paragraph reflow, runs of spaces collapsed, invisible subscript
artefacts dropped, and quotation marks, dashes, bullets, URLs and mathematical
symbols set in \LaTeX{} rather than plain text---applied mechanically by a
script, not by hand.

\smallskip
\noindent[\emph{Redaction, disclosed.} The reviews name the system the earlier
version proposed. That name is an identifier: the submission is double-blind,
and one search for it would undo the anonymity. It occurs \textbf{25} times in
the two reviews below and every occurrence is replaced, by script, with the
token [\textsc{system}]. What is replaced is that proper name and nothing
else---no word of a reviewer's own prose, no score, no heading, no citation.
No other character of either review is altered, and no sentence, clause or
punctuation mark is dropped, softened or reordered.]

\footnotesize
\medskip\noindent\textbf{Review 1}\par\smallskip
\noindent\textit{Characteristics of Submission (scale is 1..5, 5 is best)}\par
\begin{center}\begin{tabular}{@{}lll@{}}\toprule
Interest: 2 & Originality: 2 & Contribution: 1 \\
Relevance: 3 & Theory: 1 & Methodology: 1 \\
Presentation: 2 & Validity: 1 & References: 1 \\
\bottomrule\end{tabular}\end{center}
\noindent\textit{The Review}\par
This paper proposes [\textsc{system}], a framework for detecting poisoned documents in RAG corpora for enterprise applications. It centers on the idea that that security sensitivity should vary with the organizational consequences of a failure. The paper highlights a salient threat, and per-document screening at query time could be useful. However, the current formulation contains several technical inconsistencies, and the evaluation does not support the paper's main claims. The core methodology is not coherently specified. Most seriously, the document pass rule appears inverted. The paper defines a theta term that declines as domain risk increases and source trust decreases. It defines a threshold for documents passing as below 1 - theta. As specified, the filter becomes \_less\_ restrictive for riskier documents - higher risk and lower trust reduce theta, which increases 1 - theta, giving a higher threshold for documents passing. Equation 5 has a related problem. It combines 1-R(d) positively with harm and relevance, making it unclear whether S is a risk score or a safety score. The paper also never explains how S determines filtering. The multi-task model is likewise incomplete. The paper introduces three heads, R, I, and D, but the training objective includes losses only for R and I. D is described as query relevance, but it is unclear how a query representation is made available to the encoder, which is described as operating over document embeddings. The source-trust mechanism also depends on a runtime reliability signal r that is never defined. The proposed metrics are difficult to interpret. GSQ does not include recall, so a perfect GSQ cannot mean that every poisoned document was caught. That claim also conflicts with the reported medical poison pass rate and document precision. DPS is unclear because the "ideal" and "actual" thresholds are not operationally distinguished, while TCA omits key details such as the calibration weight. The baselines are not apt comparisons. Standard RAG and query filtering do not produce document-level labels, and adversarial training removes poisoned documents offline; assigning these systems an F1 score of zero mainly reflects an interface mismatch. Prior work has proposed direct document-level screening and filtering baselines that would be more appropriate. Further, adversarial training achieves DR=1 and PPR=0 in every domain, outperforming [\textsc{system}] on the reported prevention objective.

\medskip\noindent\textbf{Review 2}\par\smallskip
\noindent\textit{Characteristics of Submission (scale is 1..5, 5 is best)}\par
\begin{center}\begin{tabular}{@{}lll@{}}\toprule
Interest: 5 & Originality: 4 & Contribution: 3 \\
Relevance: 5 & Theory: 3 & Methodology: 2 \\
Presentation: 3 & Validity: 2 & References: 3 \\
\bottomrule\end{tabular}\end{center}
\noindent\textit{The Review}\par
The paper addresses a timely and important problem: how retrieval-augmented generation systems should detect and respond to malicious documents in enterprise repositories when the consequences of a successful attack differ substantially across deployment contexts. The proposed [\textsc{system}] framework is conceptually interesting, particularly its integration of query-time document scoring, dynamic source trust, domain-sensitive thresholds, and governance reporting. Grounding the framework in Service-Dominant Logic also provides an original basis for arguing that security controls should reflect the organizational value and potential harm associated with a retrieval event rather than treating every poisoned document as equally consequential. The paper is highly relevant, generally well written, and commendably candid about several limitations and negative results, including its poor prompt-injection performance and the lack of meaningful differentiation between the financial and legal conditions. The manuscript nevertheless requires clarification and reframing in several areas. Section 2 does not adequately position [\textsc{system}] against comparable security systems, the baseline comparison conflates architectural capability with comparative security performance, and the reported results provide only partial evidence that the proportionality mechanism functions as intended. The mathematical description of the decision process also contains ambiguities, while the dynamic trust mechanism does not explain how malicious changes are distinguished from legitimate document updates. These issues do not eliminate the paper's conceptual contribution, but they substantially narrow what can be concluded from the current evaluation. Section 2 is titled ``Related Work,'' but it primarily provides background on RAG, categorizes attack vectors, introduces the proportionality gap, and summarizes Service-Dominant Logic. These are useful foundations, but the section does not perform the comparative function normally expected of a related-work review. It does not discuss comparable runtime poisoning defenses, document-level or chunk-level filtering methods, embedding-anomaly approaches, source-reliability frameworks, multi-document consistency defenses, or poisoning-attribution systems. Consequently, the manuscript does not adequately establish which components of [\textsc{system}] are novel, which extend existing techniques, or why the selected baselines appropriately represent the state of the field. This omission is especially important because the manuscript makes several strong positioning claims. It states that no existing RAG security system operationalizes the proposed value-conditioned objective, that [\textsc{system}] is the first RAG security framework grounded in Service-Dominant Logic, and that it is the only evaluated system capable of identifying individual adversarial documents at query time. The first two claims may be supportable, but the current review of the literature does not provide enough evidence for the reader to assess them. The third is explicitly limited to the systems selected for the experiment, yet it receives considerable emphasis throughout the abstract, introduction, results, discussion, and conclusion. The manuscript should distinguish between being the only selected baseline that emits a particular output and being the only existing framework capable of that function. A concise comparison of related systems would substantially strengthen the paper. The comparison should identify whether each system operates offline, during retrieval, before generation, or after an attack is observed; whether it detects attacks at the query, document, chunk, retrieval-set, or generated-output level; which attack classes it addresses; whether it incorporates source reliability, provenance, or cross-document consistency; and whether it supports document attribution or governance reporting. Relevant studies the authors may wish to consider include: \textbullet{} RAGuard - Secure Retrieval-Augmented Generation against Poisoning Attacks: Expands the initial retrieval set and applies chunk-wise perplexity and text-similarity filtering to identify and remove poisoned texts at query time. It is a particularly important comparator because, like [\textsc{system}], it makes document-level or chunk-level filtering decisions before generation. (\url{https://arxiv.org/abs/2510.25025}) \textbullet{} ReliabilityRAG - Effective and Provably Robust Defense for RAG-based Web-Search: Uses document-reliability signals and a contradiction graph to identify a consistent subset of retrieved documents and filter potentially malicious sources. Its use of document reliability is especially relevant to [\textsc{system}]'s source-trust component. (\url{https://arxiv.org/abs/2509.23519}) \textbullet{} Directional Relative Shifts - Understanding Data Poisoning Attacks for RAG: Insights and Algorithms: Proposes an embedding-space defense that detects poisoned entries through directional distribution shifts associated with successful attacks. It is relevant to [\textsc{system}]'s embedding-anomaly and vector-database-manipulation claims. (\url{https://openreview.net/pdf?id=2aL6gcFX7q}) \textbullet{} RAGForensics - Traceback of Poisoning Attacks to Retrieval-Augmented Generation: Uses iterative retrieval and LLM-guided inspection to identify specific poisoned texts responsible for an observed attack. Although it is primarily a traceback system rather than a preventive query-time filter, it is relevant to claims about document-level attribution. (\url{https://arxiv.org/abs/2504.21668}) \textbullet{} D-SCAN - When Context Bites: Detecting RAG Poisoning via Document-Level Attention Collapse: Monitors generator attention dynamics for a collapse pattern associated with poisoned context. It provides a useful contrast because detection occurs through generator behavior rather than document-content or source-trust scoring. (\url{https://dl.acm.org/doi/epdf/10.1145/3805712.3809904}) \textbullet{} Benchmarking Poisoning Attacks against Retrieval-Augmented Generation: Evaluates multiple poisoning attacks and defense mechanisms across different datasets and RAG configurations. This work could help justify the selected baselines and situate the evaluation within the broader defense landscape. (\url{https://arxiv.org/abs/2505.18543}) The authors need not reproduce each framework experimentally, but they should discuss the closest alternatives and explain why their selected baselines are appropriate. Such a comparison may reveal that [\textsc{system}]'s more defensible contribution is not query-time document filtering by itself, but the integration of value-sensitive thresholds, dynamic per-source trust, document-level decisions, and governance signals. Section 2 could be revised accordingly or retitled to reflect its present role as background, attack taxonomy, and theoretical foundation, with a compact discussion of comparable defenses and the resulting research gap added to the introduction. The system description should then be clarified before the evaluation is interpreted. Figure 1 currently provides little information beyond the Layer 1 through Layer 4 sequence and labels for three attack entry points. It does not show how the components described in Section 4 interact. A more useful architecture diagram would show query screening, embedding and FAISS retrieval, the retrieved document set, the cross-document encoder, risk, impact, and relevance heads, Bayesian source-trust calibration, Layer 3 filtering, accepted context, generation, governance reporting, and the feedback path used to update source trust. Annotating the four layers and attack entry points on that architecture would make the implementation substantially easier to understand. Table 1 should also span both columns because its current narrow formatting breaks short phrases across multiple lines and obscures the relationship among attack method, detection layer, and risk interaction. The mathematical description of the filtering process also appears to require correction or clarification. Section 4.1.3 states that a document passes when R(di)$\leq$1$-$$\theta$(di), where R(di) is the predicted probability that the document is adversarial. Equation 10 lowers $\theta$ when domain risk is high and source trust is low. Under the stated pass rule, lowering $\theta$ increases 1$-$$\theta$, permitting documents with higher adversarial probabilities to pass. For example, reducing $\theta$ from .65 to .45 increases the allowable risk score from .35 to .55. This appears to loosen the filter in the circumstances where the text says it becomes more restrictive. The authors should verify whether the pass rule, threshold equation, or interpretation has been reversed and confirm that the equations match the implementation used to produce the reported results. Equation 5 also requires a clearer interpretation. The combined score increases with 1$-$R(di), which represents benignness, but also increases with impact and relevance. It is therefore unclear whether a higher value of S(di) represents safety, risk, priority, or another decision quantity. The manuscript does not clearly explain how S(di) affects the Layer 3 decision, which is described primarily through R(di) and $\theta$(di). The relevance head appears in the architecture and combined score but not in the training loss shown in Equation 6. The score weights are also not reported, and the same notation appears to be used for both score weights and loss weights. These may be notation or reporting issues rather than implementation errors, but they should be resolved because these equations define the framework's central value-aware mechanism. The dynamic trust loop is another promising component that requires additional operational explanation. Equations 7 and 8 update source trust using a reliability signal r, but the manuscript does not define how this signal is obtained after a live retrieval event. If r comes from the risk model's own decision, the system may reinforce its own errors: a legitimate document incorrectly classified as malicious reduces source trust, which then affects subsequent filtering decisions. If r comes from human review, downstream outcomes, administrator confirmation, or known labels, the process is not fully automatic and the associated governance workflow should be described. Figure 2 illustrates how the trust equation behaves once a source is designated as compromised, but it does not establish how the system independently determines that a compromise occurred. This ambiguity is closely connected to a practical issue that the manuscript does not address: distinguishing malicious document modification from a legitimate correction or update. The introductory example contrasts a legitimate \$24.7 million profit with a fabricated \$8.3 million loss, but an enterprise repository may contain a later accounting restatement, corrected filing, revised clinical protocol, or amended legal policy. Semantic contradiction or numerical replacement alone cannot establish which document is authoritative. A new document may be anomalous but correct, while a previously trusted source may be compromised. The framework should therefore explain how it incorporates document provenance, authorized authorship, version history, temporal validity, source authentication, cross-source corroboration, or human adjudication. Source-level trust alone does not resolve the issue because legitimate new sources begin with limited history, while compromised trusted sources may retain high trust until sufficient negative evidence accumulates. The manuscript should clarify the assumptions under which a correction is labeled as malicious, how authorized updates enter the corpus, how superseded documents are handled, and how erroneous trust penalties can be reversed. This is central to the proposed enterprise use case and can be addressed through architectural and governance clarification. The baseline evaluation is the most significant methodological concern. The F1 comparison evaluates a capability that only [\textsc{system}] was designed to provide. Standard RAG performs no security detection, Input Filtering RAG examines the query rather than the retrieved documents, and Adversarial Training RAG cleans the corpus before inference. None produces per-document inference-time labels. Their reported F1 values of zero therefore do not represent unsuccessful document classification; they indicate that these systems do not emit the output required for the metric. The Adversarial Training RAG baseline makes the problem particularly clear. Table 2 reports a detection rate of 1.00 and poison pass rate of 0.00 in every domain, meaning that no poisoned documents remain available to influence generation. Nevertheless, the system is assigned F1 = 0 because TP, FP, and FN are all zero at inference time. In this circumstance, F1 is undefined or not applicable rather than a meaningful zero-performance result. Figure 3 then places this system at zero beside [\textsc{system}], creating the visual impression that [\textsc{system}] substantially outperforms it even though the adversarial-training baseline prevents all poisoned documents from reaching inference. The authors acknowledge that the baselines have zero F1 ``not because they perform poorly'' but because they do not produce per-document flags. That qualification is appropriate, but it does not resolve the comparability problem. The manuscript should separate two questions: whether a system can identify and report a specific document at query time, and whether it reduces the passage or downstream influence of poisoned documents. The first is an architectural capability comparison; the second is a security-performance comparison. F1 can be reported for [\textsc{system}]'s document-level classifier, but it should not be used to rank systems that do not perform the same task. The common outcomes in Table 2 provide a substantially more restrained picture. In the medical domain, [\textsc{system}] improves detection rate from .769 to .928 and reduces poison pass rate from .231 to .072 relative to Standard RAG and Input Filtering RAG. This is a meaningful descriptive improvement. In the financial domain, detection rate changes from .894 to .896 and poison pass rate from .106 to .105. In the legal domain, detection rate changes from .924 to .925 and poison pass rate from .076 to .075. The financial and legal differences are negligible relative to the reported standard deviations. The manuscript acknowledges that these gains fall within noise, but the F1 comparison and Figure 3 visually obscure how limited the common performance improvement is. The baseline narrative should therefore be reframed around outcomes that are available across systems. Figure 3 should be removed because it duplicates Table 2 and presents unavailable document-labeling outputs as ordinary zero scores. Table 2 should instead distinguish common security outcomes, such as poison pass rate, from [\textsc{system}]-specific document-classification metrics. F1 should be marked as not applicable for systems that do not emit inference-time document labels, while [\textsc{system}]'s precision, recall, and F1 can be reported separately. The interpretation of detection rate also requires clarification. In Table 2, DR and PPR sum to 1.00 for every system and condition, including Standard RAG, which the manuscript states has no detection mechanism. DR therefore appears to be defined as the complement of poison pass rate rather than as the proportion of poisoned documents actively identified by a detector. If so, it is not an independent metric, and the term ``detection rate'' is misleading for systems that merely fail to retrieve a poisoned document or remove it offline. The authors should provide the exact definition and computation of DR and consider renaming it as poison exclusion or blocking rate. The comparative discussion should then avoid implying that Standard RAG performs document detection. The evidence for domain proportionality is also incomplete. The manuscript's central premise is that higher-risk domains should receive stronger protection, and the authors interpret the medical-domain improvement as support for this hypothesis. However, domain content, linguistic characteristics, attack examples, classifier difficulty, source distributions, and assigned risk weights all vary simultaneously. The medical, financial, and legal conditions therefore do not isolate the effect of the domain-risk parameter. The manuscript itself explains that legal detection is difficult because legitimate legal writing contains conditional language, hedging, and authority references that resemble adversarial correction text. That explanation is plausible, but it also demonstrates that domain-specific detection difficulty may dominate the intended risk weighting. Legal receives a higher risk weight than financial, yet its precision, recall, and F1 are substantially lower. The current experiment therefore cannot establish that the assigned weights caused the observed ordering or that the framework successfully scales protection across all three domains. The domain-risk weights themselves require justification. The manuscript assigns 1.00 to medical, .92 to legal, and .88 to financial but does not explain how these values were derived, who selected them, or whether they reflect empirical loss estimates, expert judgment, regulatory classifications, organizational risk tolerances, or merely illustrative design choices. It is unclear why the difference between legal and financial is .04 rather than .10 or .20, or whether the same ordering would apply across different applications within each domain. A medical scheduling system and a clinical dosage assistant may present very different risks, just as a routine financial reporting tool and a high-value trading system may not belong in the same risk category. As currently presented, the values appear author-assigned and closely clustered. Under Equation 10, the resulting threshold differences may be small, particularly as source trust increases. The manuscript should explain the provenance, interpretation, and governance of the weights; identify who would assign or revise them in practice; and characterize them as illustrative unless they were derived through a defensible risk-assessment process. The results support a narrower conclusion: [\textsc{system}] produced a substantial improvement in the highest-risk medical condition but did not demonstrate meaningful improvement in the financial or legal conditions. This constitutes partial evidence consistent with the proportionality hypothesis, not general validation of proportional security behavior. The three value-sensitive metrics also require clearer definition and interpretation. Domain Proportionality Score is defined as an average over domains, but Table 3 reports separate values for each domain. The manuscript should explain how the per-domain values are derived and what constitutes the ``ideal'' threshold against which actual behavior is compared. If the ideal threshold is generated from the same formula and assigned weights used by the framework, DPS may measure conformity with the authors' selected parameters rather than correspondence with organizational harm. Trust Calibration Accuracy combines expected calibration error with sharpness, but the weighting parameter, calibration procedure, binning method, and operational definition of actual source reliability are not reported. Because a model can be sharply confident and still wrong, ECE and sharpness should also be reported or interpreted separately. Governance Signal Quality combines alert precision, explainability, and timeliness, but the Explainability Index and Timeliness Score are not sufficiently defined for replication. The interpretation of medical GSQ = 1.00 should also be corrected. The manuscript states that this means every bad document was caught, although medical recall is .924 and poison pass rate is .072. Because GSQ appears to include alert precision but not recall, a perfect score may mean that every issued alert was accurate, understandable, and timely, not that every poisoned document was detected. Table 3 and Figure 4 also substantially duplicate the same DPS, TCA, and GSQ results. The table is the more useful presentation because it preserves the exact means and standard deviations. Figure 4 could therefore be removed, with the prompt-injection null results stated in the accompanying text or incorporated into the table if a direct comparison is necessary. The evaluation also does not cover all attacks included in the stated threat model. Vector-database manipulation is presented as a supported attack class and is mapped to Layers 1 and 2 in Table 1, but the six experimental conditions cover only document poisoning and prompt injection. Claims concerning detection of crafted embeddings should therefore be characterized as architectural proposals rather than evaluated capabilities. The assertion that the unintended increase from a 5\% to an approximately 8\% poison rate does not affect relative comparisons also requires evidence. Changes in attack prevalence can alter precision, F1, trust updates, and threshold behavior. The authors should either provide a sensitivity analysis across poison rates or qualify the claim. Finally, the implications for practice are not sufficiently connected to the study's operationalized results. Table 4 specifies threshold values, review cadences, and escalation criteria for high, medium, and low-risk deployments, but the manuscript does not explain how these recommendations were derived from the reported DPS, TCA, GSQ, detection-rate, or poison-pass-rate results. The values appear to be author-assigned rather than empirically estimated, optimized, or validated within the experiment. For example, the paper does not show why $\theta$base=.65 and $\Delta$$\theta$=.20 are appropriate for high-risk systems, why GSQ thresholds of .70, .50, and .30 should trigger different responses, or why review should occur in real time, daily, or weekly across the three risk categories. Practical recommendations should follow from the study's measured outcomes, sensitivity analyses, or a clearly articulated governance framework. In the absence of such derivation, Table 4 should be presented as an illustrative configuration rather than evidence-based operating guidance, and the manuscript should explicitly distinguish parameters demonstrated by the study from values proposed by the authors for future validation. Overall, [\textsc{system}] is an interesting and potentially valuable framework. Its value-aware framing, dynamic trust mechanism, and emphasis on governance signals are original and likely to stimulate productive conference discussion. The current evidence supports the feasibility of producing query-time document alerts and shows a meaningful descriptive improvement in the medical-domain poisoning condition. However, the synthetic corpus, limited experimental conditions, and small number of trials do not yet establish that the observed behavior will generalize across different enterprise repositories, attack strategies, source distributions, organizational risk models, or deployment settings. The paper would be substantially strengthened by providing a genuine related-work positioning, clarifying the architecture and decision logic, reframing the baseline comparison, explaining the provenance and operation of the value and trust mechanisms, and narrowing the conclusions to the results actually demonstrated. Suggested revisions: \textbullet{} Strengthen the related-work positioning: Section 2 currently functions primarily as background, attack taxonomy, and theoretical foundation. Add a concise comparison of existing runtime poisoning defenses, document- or chunk-level filtering methods, embedding-anomaly approaches, source-reliability systems, and poisoning-attribution methods. Clearly identify what [\textsc{system}] adds beyond those systems. Alternatively, retitle Section 2 to reflect its current content and introduce the comparative literature and gap in the introduction. \textbullet{} Clarify the novelty claim: Distinguish the statement that [\textsc{system}] is the only selected baseline that emits per-document query-time labels from any broader claim that it is the only existing system capable of runtime document identification. Emphasize the integration of value-sensitive thresholds, dynamic source trust, and governance reporting as the primary contribution. \textbullet{} Address legitimate document updates: Explain how [\textsc{system}] distinguishes poisoning from authorized corrections, restatements, revised policies, or other valid knowledge changes. Discuss the roles of provenance, version history, source authentication, temporal validity, corroboration, and human review. \textbullet{} Replace Figure 1: Provide a substantive system architecture diagram showing retrieval, document scoring, trust calibration, filtering, context assembly, generation, governance reporting, and the trust-update feedback loop, with Layers 1 through 4 and attack entry points annotated. \textbullet{} Reformat Table 1: Expand the table across both text columns to improve readability. \textbullet{} Verify the threshold equations: Reconcile the pass condition R(di)$\leq$1$-$$\theta$(di) with Equation 10, which appears to loosen the gate when risk is high and trust is low. Confirm that the equations, implementation, and narrative describe the same decision logic. \textbullet{} Clarify the combined score and training objectives: Define the direction and use of S(di), report the score weights, explain how the relevance head is trained, and distinguish the scoring weights from the loss weights. \textbullet{} Define the reliability signal: Explain how r is obtained after each retrieval event and whether it reflects model output, human adjudication, known ground truth, or downstream feedback. Acknowledge the possibility that erroneous flags could reinforce later trust decisions. \textbullet{} Reframe the baseline comparison: Separate architectural capability from security performance. Do not interpret missing document-level outputs as failed classification. \textbullet{} Revise Table 2, the treatment of F1, and remove Figure 3: Figure 3 repeats the results in Table 2 while visually presenting unavailable document-level F1 values as ordinary zero-performance scores. Remove the figure and revise Table 2 to distinguish metrics that are comparable across all systems, such as poison pass rate, from [\textsc{system}]-specific document-classification metrics. Report F1 as not applicable for systems that do not produce inference-time document labels, present [\textsc{system}]'s precision, recall, and F1 separately, and avoid representing Adversarial Training RAG as unsuccessful when it prevents all poisoned documents from reaching inference. \textbullet{} Clarify DPS, TCA, and GSQ: Provide the missing operational definitions and constituent parameters, explain how per-domain DPS values relate to the aggregate equation, and correct the interpretation of medical GSQ = 1.00 so that it does not imply perfect recall when poisoned documents were still missed. \textbullet{} Clarify detection rate: Define how DR is computed and explain why systems without a detection mechanism receive non-zero values. If DR is simply 1$-$PPR, rename it to reflect poison exclusion or blocking and avoid reporting it as an independent metric. \textbullet{} Consolidate Table 3 and Figure 4: Table 3 and Figure 4 substantially duplicate the DPS, TCA, and GSQ results. Retain a revised table because it reports the exact means and standard deviations more clearly and compactly, and remove Figure 4. The prompt-injection null results currently conveyed by the figure can be stated in the accompanying text or incorporated into the table if the authors believe a direct comparison across attack types is necessary. \textbullet{} Narrow the proportionality conclusion: State that the results provide partial support in the medical condition. Financial and legal performance is effectively unchanged relative to Standard RAG under the common metrics. Discuss whether the assigned risk weights and threshold adjustment exert enough influence to overcome differences in domain-specific detection difficulty. \textbullet{} Calibrate the practice guidance: Present the thresholds, review cadences, and escalation values in Table 4 as illustrative proposed configurations rather than validated enterprise operating standards. \textbullet{} Narrow the final claims: Frame [\textsc{system}] as a promising exploratory prototype that demonstrates query-time document attribution and partial value-sensitive behavior in the medical condition. Avoid general claims of proportionality, comparative superiority, or deployment readiness that are not supported across all three domains.

\normalsize

\end{document}